\documentclass[aps,prb,10pt, twocolumn, superscriptaddress, 
]{revtex4-1}

\pdfoutput=1
\usepackage[utf8]{inputenc}
\usepackage{amsmath}
\usepackage{wasysym}
\usepackage{amsfonts}
\usepackage{braket}
\usepackage{tensor}
\usepackage{amssymb}
\usepackage{mathbbol}
\usepackage{bm,bbm}
\usepackage{graphicx}
\usepackage{titlesec}
\usepackage[titletoc,toc,title]{appendix}
\usepackage{tikz}
\usepackage{tikz-network}
\usepackage{amsthm}
\theoremstyle{definition}

\usetikzlibrary{decorations.pathreplacing}
\usetikzlibrary{plotmarks}
\usetikzlibrary{shapes.misc}
\usetikzlibrary{decorations.pathmorphing}

\usetikzlibrary{patterns,decorations.pathreplacing}
\usepackage[colorlinks=true,linkcolor=blue, citecolor=blue, urlcolor=blue, bookmarks]{hyperref}
\usetikzlibrary{patterns}
\tikzset{cross/.style={cross out, draw=black, minimum size=2*(#1-\pgflinewidth), inner sep=0pt, outer sep=0pt},
	cross/.default={1pt}}

\usepackage{color}
\definecolor{myred}{RGB}{232,102,102}
\definecolor{myblue}{RGB}{187,187,255}
\definecolor{myviolet}{RGB}{210,145,178}
\definecolor{myvioletc}{RGB}{45,110,77}
\definecolor{mygreen}{RGB}{34,139,34}
\definecolor{myorange}{RGB}{255,165,0}
\definecolor{OliveGreen}{RGB}{85,107,47}
\definecolor{NavyBlue}{RGB}{0,0,128}

\def\be{\begin{equation}}
\def\ee{\end{equation}}
\def\bea{\begin{eqnarray}}
\def\eea{\end{eqnarray}}

\newtheorem{Theorem}{Theorem}
\newtheorem{Lemma}{Lemma}

\begin{document}

\title{ Exact dynamics in dual-unitary quantum circuits}

\author{Lorenzo Piroli}
\affiliation{Max-Planck-Institut f\"ur Quantenoptik, Hans-Kopfermann-Str.~1, 85748 Garching, Germany.}
\affiliation{Munich Center for Quantum Science and Technology, Schellingstraße 4, 80799 M\"unchen, Germany}

\author{Bruno Bertini}
\affiliation{Department of Physics, Faculty of Mathematics and Physics, University of Ljubljana, Jadranska 19, SI-1000 Ljubljana, Slovenia}

\author{J. Ignacio Cirac}
\affiliation{Max-Planck-Institut f\"ur Quantenoptik, Hans-Kopfermann-Str.~1, 85748 Garching, Germany.}
\affiliation{Munich Center for Quantum Science and Technology, Schellingstraße 4, 80799 M\"unchen, Germany}

\author{Toma\v{z} Prosen}
\affiliation{Department of Physics, Faculty of Mathematics and Physics, University of Ljubljana, Jadranska 19, SI-1000 Ljubljana, Slovenia}

\begin{abstract}
	
We consider the class of dual-unitary quantum circuits in $1+1$ dimensions and introduce a notion of ``solvable'' matrix product states (MPSs), defined by a specific condition which allows us to tackle their time evolution analytically. We provide a classification of the latter, showing that they include certain MPSs of arbitrary bond dimension, and study analytically different aspects of their dynamics. For these initial states, we show that while any subsystem of size $\ell$ reaches infinite temperature after a time $t\propto \ell$, irrespective of the presence of conserved quantities, the light-cone of two-point correlation functions displays qualitatively different features depending on the ergodicity of the quantum circuit, defined by the behavior of infinite-temperature dynamical correlation functions. Furthermore, we study the entanglement spreading from such solvable initial states, providing a closed formula for the time evolution of the entanglement entropy of a connected block. This generalizes recent results obtained in the context of the self-dual kicked Ising model. By comparison, we also consider a family of non-solvable initial mixed states depending on one real parameter $\beta$, which,  as $\beta$ is varied from zero to infinity, interpolate between the infinite temperature density matrix and arbitrary initial pure product states. We study analytically their dynamics for small values of $\beta$, and highlight the differences from the case of solvable MPSs.

\end{abstract}

\maketitle


\section{Introduction}
\label{sec:intro}

{The extensive study of isolated quantum matter out of equilibrium carried out in the last two decades reminded us, once again, of how tremendously complex the quantum many-body dynamics can be~\cite{polkovnikov2011colloquium, Gogolin_2016, eisertreview}}. Even though the past decade has witnessed the development of powerful numerical techniques based on matrix product states~\cite{perez2006matrix} (MPSs) that are able to determine, quite generally, the dynamics of quantum many-body systems in one spatial dimension~\cite{white2004real,daley2004time,vidal2007classical,orus2008infinite,banuls2009matrix,schollwock2011density}, these methods are usually limited to small or intermediate time scales~\cite{schollwock2011density}. This is due to the generic linear growth of the entanglement entropy~\cite{calabrese2005evolution}, which is a major obstacle for the MPS-representation of the time evolving state. For this reason, it is of great interest to find instances where the many-body dynamics can be solved exactly, allowing for an analytic study of interesting physical phenomena, such as the emergence of thermalization~\cite{rigol2008thermalization}.

Integrable models provide a natural arena to search for such solvable examples~\cite{calabrese2016introduction,essler2016quench} and allowed for great progress in the characterization of large-time properties of many-body systems out of equilibrium~\cite{essler2016quench,vidmar2016generalized,caux2016quench,ilievski2016quasilocal}. The computation of the full dynamics, however, remains a challenge. In particular, while an impressive number of results have been derived in theories that can be mapped onto free fermionic systems (see Ref.~[\onlinecite{essler2016quench}] for a comprehensive review), only a few special cases exist where analytic computations could be done  in the presence of genuine interactions~\cite{iyer_exact_2013,liu_quench_2014,bertini2014quantum,de2015relaxation,delfino2014quantum,*delfino2017theory,cubero2016planar,van_den_berg_separation_2016,moca2017hybrid,alba_entanglement_2017,alba_entanglement_2018,piroli2018non,klobas2019time}. Furthermore, integrable models are by their own nature very special, and can not be representative of generic systems, which are expected to display qualitatively different features.

Recently, increasing attention has been devoted to the class of random unitary quantum circuits, which provide an alternative theoretical laboratory for the study of the many-body dynamics~\cite{znidaric_optimal_2007,znidaric_exact_2008,brown_convergence_2010,shenker_stringy_2015,nahum2017quantum,nahum2018operator,chan2018solution,chan2018spectral,khemani2018operator,vonKeyserlingk2018operator,rakovszky2018diffusive,sunderhauf2018localization,xu_locality_2018,gharibyan_onset_2018,zhou_operator_2019}. The main appeal of these systems is that they represent minimally structured dynamical models where analytic results can be obtained beyond the realm of integrability. One typically considers a set of finite-dimensional Hilbert spaces sequentially updated by unitary gates, which are chosen randomly out of a suitable probability distribution. Analytic predictions for physical quantities are then obtained after averaging over disorder realizations~\cite{chan2018solution,chan2018spectral,khemani2018operator,vonKeyserlingk2018operator,rakovszky2018diffusive,sunderhauf2018localization}. This approach allows one to obtain exact results for quantities notoriously hard to compute, such as out-of-time-ordered correlators (OTOCs) ~\cite{nahum2018operator,vonKeyserlingk2018operator}, operator-space entanglement entropies~\cite{zanardi_entanglement_2001,prosen_operator_2007,jonay_coarse-grained_2018} and other measures of quantum chaos such as the so-called tripartite mutual information~\cite{hosur_chaos_2016,sunderhauf_quantum_2019}. Of particular interest for our work are the settings (sometimes called \emph{local} random unitary quantum circuits) where the unitary gates couple only neighboring sites, simulating the  ``local dynamics" of generic many-body quantum systems~\cite{nahum2017quantum,nahum2018operator,chan2018solution,chan2018spectral,khemani2018operator,vonKeyserlingk2018operator,rakovszky2018diffusive,sunderhauf2018localization}. One can still wonder, however, whether the presence of disorder averages gives rise to qualitative differences when compared to clean, homogeneous systems.

In this respect, an interesting class of quantum circuits without disorder, called ``dual-unitary", was recently introduced~\cite{bertini2019exact}, for which several dynamical features could be investigated analytically. These systems implement a dynamics in which at each time step the configuration of the system is updated by applying a product of identical unitary gates on neighboring sites. Furthermore, as a defining feature, these gates remain unitary under a reshuffling of their indices. Remarkably, this class was shown to include instances of unitary dynamics both integrable (\emph{i.e.} with local conservation laws) and chaotic, providing a rare example where the differences between the two can be analyzed to a high degree of control.

Previous works on these circuits focused on the study of infinite-temperature dynamical two-point functions~\cite{bertini2019exact}, and on the growth of the operator-space entanglement entropy~\cite{bertini2019operator,bertini2019operator_II}. It is then natural to wonder whether the class of dual-unitary circuits can also provide solvable models for the quantum dynamics arising from given initial states. This question represents the main motivation for the present work. 

For one particular dual-unitary circuit, corresponding to the self-dual point of the kicked Ising model~\cite{prosen2002general,prosen2002ruelle,prosen2007chaos, akila_particle-time_2016,bertiniSDKI}, this problem has been already partially addressed in Ref.~[\onlinecite{bertini2019entanglement}], where it was shown that for two special families of initial product states the growth of bipartite entanglement entropy could be computed exactly. In this paper we show that there exists a much broader family of ``solvable'' initial states, for which the dynamics can be tackled analytically, irrespective of the choice of the dual-unitary gates. This class includes MPSs of arbitrary bond dimension, and allows for the exact computation of several quantities beyond the growth of bipartite entanglement entropy, including the spreading of two-point correlation functions, and the thermalization time of finite subsystems. In this work we will focus on the case of spatially homogeneous systems, with a ``Floquet-like'' time evolution in which the same product of unitary gates is applied periodically in time. We stress, however, that our constructions also work for circuits with explicit space-time modulations, and can be used to obtain analytic results also in that case.

The rest of this manuscript is organized as follows. In Sec.~\ref{sec:dual_unitary_circuits} we introduce the local quantum circuits considered in this work, while in Sec.~\ref{sec:solvable_states} we define the class of solvable initial states with respect to the dual-unitary dynamics. These are classified in Sec.~\ref{sec:classification} and \ref{sec:explicit_examples}, while their time evolution is studied in Sec.~\ref{sec:time_ev}. The dynamics arising from a set of non-solvable states is studied for comparison in Sec.~\ref{sec:non_solvable_states}, while our conclusions are reported in Sec.~\ref{sec:conclusions}. Finally, the most technical aspects of our study are consigned to two appendices.

\section{The dual-unitary dynamics}
\label{sec:dual_unitary_circuits}

We consider a chain of $2L$ sites, each one associated with a $d$-level system. The corresponding Hilbert space is  thus $\mathcal{H}_{2L}=h_1\otimes \cdots \otimes h_{2L}$, where $h_{j}\simeq \mathbb{C}^{d}$. In the following, we will denote the corresponding basis vectors by $\ket{j}$, $j=1\,,\ldots d$. We are interested in local quantum circuits, which implement a particular periodically driven unitary evolution. Specifically, given the initial state $\ket{\Psi_0}$ we study the quantum dynamics defined by
\bea
\ket{\Psi_{2t+1}}&=&\mathcal{U}_-\ket{\Psi_{2t}}\,,\label{eq:ev_rule_even}\\
\ket{\Psi_{2t+2}}&=&\mathcal{U}_+\ket{\Psi_{2t+1}}\label{eq:ev_rule_odd}\,,
\eea
where $t\in \mathbb{N}$, while
\bea
\mathcal{U}_{-} &=&U_{2,3} U_{4,5} \cdots U_{2 L-2,2 L-1} U_{2 L, 1}\,,\label{eq:u_m}\\
\mathcal{U}_{+} &=&U_{1,2} U_{3,4} \cdots U_{2 L-1,2 L}\,.\label{eq:u_p}
\eea
Here $U_{j,k}$ are two-site unitary operators acting on the product of local spaces $h_{j}\otimes h_k$, and periodic boundary conditions are implemented in Eq.~\eqref{eq:u_m}.  We will mainly consider the case of infinite systems, namely we will study the above dynamics in the limit $L\to\infty$.

We focus on the special class of \emph{dual-unitary} quantum circuits recently introduced in Ref.~[\onlinecite{bertini2019exact}]. These circuits are defined as follows. Let $U$ be a unitary gate, and define $\tilde{U}$ as the operator given by the following reshuffling of indices
\be
\langle k|\otimes\langle\ell|\tilde{U}| i\rangle \otimes| j\rangle=\langle j|\otimes\langle\ell|U| i\rangle \otimes| k\rangle\,.
\ee
We say that $\tilde{U}$ is the dual gate of $U$. Then, dual-unitary circuits are defined by local unitary gates  $U$ such  that $\tilde{U}$ is also unitary, namely
\bea
U U^{\dagger}&=&U^{\dagger} U=\mathbb{1}\,, \label{eq:unitarity}\\
\tilde{U} \tilde{U}^{\dagger}&=&\tilde{U}^{\dagger} \tilde{U}=\mathbb{1}\,.\label{eq:dual_unitarity}
\eea
This definition is reminiscent of those of perfect tensors and $2$-unitary matrices, introduced in Refs.~[\onlinecite{pastawski2015holographic}] and~[\onlinecite{Goyeneche2015absolutely}] respectively. The former are defined as tensors which are isometric operators for any bipartition of their indices into two subsets (not necessarily of the same size). The latter are matrices, acting on the tensor product of two qudits, that remain unitary for any bipartition of indices into two pairs. We note, however, that dual-unitary gates are generally neither $2$-unitary matrices nor perfect tensors, since they only correspond to unitary operators for two special bipartitions of the indices. Finally, we mention that the partitioning of indices in the definition of dual-unitarity is the same appearing in the one of the cross-norm and realignment criteria for separability~\cite{rudolph_properties_2003,rudolph_further_2005, horodecki2006separability}.

The dual-unitary condition can be naturally expressed using the customary tensor-network graphical representation of quantum circuits. In this language, matrix elements of local operators are depicted as boxes with a number of incoming and outgoing legs. To each leg corresponds an index associated with one of the local sites on which the local operator acts on. In particular, for the operators $U$ and $U^\dagger$ we have the representation
\be
\begin{tikzpicture}[baseline=(current  bounding  box.center), scale=0.8]
\node at (1.5,0.35) {\scalebox{1.2}{$,$}};
\node at (5.5,0.35) {\scalebox{1.2}{$.$}};
\draw [fill=myred] (0.15, 0.2) rectangle (0.85,0.8);
\foreach \y in {0}{
	\foreach \x in {0.} { 
		\draw[=latex] (\x-0.15,-0.1+\y)  -- (\x+0.15,0.2+\y);
	}
}
\foreach \y in {0.9}{
	\foreach \x in {1.} { 
		\draw[=latex] (\x-0.15,-0.1+\y)  -- (\x+0.15,0.2+\y);
	}
}
\foreach \y in {0.9}{
	\foreach \x in {0} { 
		\draw[=latex] (\x-0.15,0.2+\y)  -- (\x+0.15,-0.1+\y);
	}
}
\foreach \y in {0.}{
	\foreach \x in {1} { 
		\draw[=latex] (\x-0.15,0.2+\y)  -- (\x+0.15,-0.1+\y);
	}
}
\node at (-1,0.5) {$U_{i,j}^{k,l}=$};
\node at (-0.35,-0.4) {$i$};
\node at (1.3,-0.4) {$j$};
\node at (-0.35,1.4) {$k$};
\node at (1.3,1.4) {$l$};
\foreach \x in {4} { 
	\node at (-1+\x,0.5) {$\left(U^\dagger\right)_{i,j}^{k,l}=$};
	\node at (-0.35+\x,-0.4) {$i$};
	\node at (1.3+\x,-0.4) {$j$};
	\node at (-0.35+\x,1.4) {$k$};
	\node at (1.3+\x,1.4) {$l$};
	\draw [fill=myblue] (0.15+\x, 0.2) rectangle (0.85+\x,0.8);
}
\foreach \y in {0}{
	\foreach \x in {4} { 
		\draw[=latex] (\x-0.15,-0.1+\y)  -- (\x+0.15,0.2+\y);
	}
}
\foreach \y in {0.9}{
	\foreach \x in {5} { 
		\draw[=latex] (\x-0.15,-0.1+\y)  -- (\x+0.15,0.2+\y);
	}
}
\foreach \y in {0.9}{
	\foreach \x in {4} { 
		\draw[=latex] (\x-0.15,0.2+\y)  -- (\x+0.15,-0.1+\y);
	}
}
\foreach \y in {0.}{
	\foreach \x in {5} { 
		\draw[=latex] (\x-0.15,0.2+\y)  -- (\x+0.15,-0.1+\y);
	}
}
\label{eq:u_tensors}
\end{tikzpicture}
\ee
When legs of different operators are joined together it is understood that one should sum over the index of the corresponding local space. Finally, an explicit label for the legs can be omitted when it does not generate confusion. Using this  notation, Eq.~\eqref{eq:unitarity} can be rewritten as
\be
\begin{tikzpicture}[baseline=(current  bounding  box.center), scale=0.6]
\foreach \z in {0}{
	\draw[=latex, ] (0.18+\z,0.2) arc(90:270: 0.1 and 0.45);
	\draw[=latex, ] (0.82+\z,0.2) arc(90:-90: 0.1 and 0.45);
	\draw [fill=myblue, ] (0.15+\z, 0.2) rectangle (0.85+\z,0.8);
	\foreach \y in {0.9}{
		\foreach \x in {1} { 
			\draw[=latex] (\x-0.15,-0.1+\y)  -- (\x+0.15,0.2+\y);
		}
	}
	\foreach \y in {0.9}{
		\foreach \x in {0} { 
			\draw[=latex] (\x-0.15,0.2+\y)  -- (\x+0.15,-0.1+\y);
		}
	}
	\foreach \y in {-1.5} { 
		\draw [fill=myred, ] (0.15+\z, 0.2+\y) rectangle (0.85+\z,0.8+\y);
	}
	
	\foreach \y in {-1.5}{
		\foreach \x in {0.} { 
			\draw[=latex] (\x-0.15,-0.1+\y)  -- (\x+0.15,0.2+\y);
		}
	}
	\foreach \y in {0.9}{
		\foreach \x in {1.} { 
			\draw[=latex] (\x-0.15,-0.1+\y)  -- (\x+0.15,0.2+\y);
		}
	}
	\foreach \y in {-1.5}{
		\foreach \x in {1} { 
			\draw[=latex] (\x-0.15,0.2+\y)  -- (\x+0.15,-0.1+\y);
		}
	}
}
\node at (2.0,-0.25) {\scalebox{1.2}{$=$}};
\node at (5.0,-0.25) {\scalebox{1.2}{$=$}};
\node at (8.0,-0.25) {\scalebox{1.2}{$\,,$}};
\foreach \z in {3}{
	\draw[=latex, ] (0.18+\z,0.2) arc(90:270: 0.1 and 0.45);
	\draw[=latex, ] (0.82+\z,0.2) arc(90:-90: 0.1 and 0.45);
	\draw [fill=myred] (0.15+\z, 0.2) rectangle (0.85+\z,0.8);
	\foreach \y in {0.9}{
		\foreach \x in {1} { 
			\draw[=latex] (\x-0.15,-0.1+\y)  -- (\x+0.15,0.2+\y);
		}
	}
	\foreach \y in {0.9}{
		\foreach \x in {0} { 
			\draw[=latex] (\x-0.15+\z,0.2+\y)  -- (\x+0.15+\z,-0.1+\y);
		}
	}
	\foreach \y in {-1.5} { 
		\draw [fill=myblue] (0.15+\z, 0.2+\y) rectangle (0.85+\z,0.8+\y);
	}
	
	\foreach \y in {-1.5}{
		\foreach \x in {0.} { 
			\draw[=latex] (\x-0.15+\z,-0.1+\y)  -- (\x+0.15+\z,0.2+\y);
		}
	}
	\foreach \y in {0.9}{
		\foreach \x in {1.} { 
			\draw[=latex] (\x-0.15+\z,-0.1+\y)  -- (\x+0.15+\z,0.2+\y);
		}
	}
	\foreach \y in {-1.5}{
		\foreach \x in {1} { 
			\draw[=latex] (\x-0.15+\z,0.2+\y)  -- (\x+0.15+\z,-0.1+\y);
		}
	}
	
}
\foreach \z in {6}{
	\foreach \y in {0.9}{
		\foreach \x in {1} { 
			\draw[=latex] (\x-0.15,-0.1+\y)  -- (\x+0.15,0.2+\y);
		}
	}
	\foreach \y in {0.9}{
		\foreach \x in {0} { 
			\draw[=latex] (\x-0.15+\z,0.2+\y)  -- (\x+0.15+\z,-0.1+\y);
		}
	}
	\foreach \y in {-1.5}{
		\foreach \x in {0.} { 
			\draw[=latex] (\x-0.15+\z,-0.1+\y)  -- (\x+0.15+\z,0.2+\y);
		}
	}
	\foreach \y in {0.9}{
		\foreach \x in {1.} { 
			\draw[=latex] (\x-0.15+\z,-0.1+\y)  -- (\x+0.15+\z,0.2+\y);
		}
	}
	\foreach \y in {-1.5}{
		\foreach \x in {1} { 
			\draw[=latex] (\x-0.15+\z,0.2+\y)  -- (\x+0.15+\z,-0.1+\y);
		}
	}
	\foreach \y in {-1.5}{
		\foreach \x in {1} { 
			\draw[=latex] (\x-0.15+\z,+0.2+\y)  -- (\x-0.15+\z,2.3+\y);
		}
		\foreach \x in {0.3} { 
			\draw[=latex] (\x-0.15+\z,+0.2+\y)  -- (\x-0.15+\z,2.3+\y);
		}
	}
}
\end{tikzpicture}
\label{eq:unitarity_pic}
\ee
while Eq.~\eqref{eq:dual_unitarity} reads
\be
\begin{tikzpicture}[baseline=(current  bounding  box.center), scale=0.6]
\foreach \z in {0}{
	\draw[=latex, ] (-0.2+\z,0.8) arc(90:270: 0.2 and 1.05);
	\draw[=latex, ] (0.18+\z,0.2) arc(90:270: 0.1 and 0.45);
	\draw [fill=myblue, ] (0.15+\z, 0.2) rectangle (0.85+\z,0.8);
	\foreach \y in {0.9}{
		\foreach \x in {1.} { 
			\draw[=latex, ] (\x-0.15+\z,-0.1+\y)  -- (\x+0.25+\z,-0.1+\y);
		}
	}
	\foreach \y in {0.9}{
		\foreach \x in {0} { 
			\draw[=latex, ] (\x-0.2+\z,-0.1+\y)  -- (\x+0.15+\z,-0.1+\y);
		}
	}
	\foreach \y in {0.}{
		\foreach \x in {1} { 
			\draw[=latex, ] (\x-0.15+\z,0.2+\y)  -- (\x+0.25+\z,0.2+\y);
		}
	}
	\foreach \y in {-1.5} { 
		\draw [fill=myred, ] (0.15+\z, 0.2+\y) rectangle (0.85+\z,0.8+\y);
	}
	\foreach \y in {-1.5}{
		\foreach \x in {0} { 
			\draw[=latex, ] (\x-0.2+\z,0.2+\y)  -- (\x+0.15+\z,0.2+\y);
		}
	}
	\foreach \y in {-0.6}{
		\foreach \x in {1} { 
			\draw[=latex, ] (\x-0.15+\z,-0.1+\y)  -- (\x+0.25+\z,-0.1+\y);
		}
	}
	\foreach \y in {-1.5}{
		\foreach \x in {1} { 
			\draw[=latex, ] (\x-0.15+\z,0.2+\y)  -- (\x+0.25+\z,0.2+\y);
		}
	}
}
\node at (2.0,-0.25) {\scalebox{1.2}{$=$}};
\node at (5.,-0.25) {\scalebox{1.2}{$,$}};
\foreach \z in {3}{
	\draw[=latex, ] (-0.2+\z,0.8) arc(90:270: 0.2 and 1.05);
	\draw[=latex, ] (0.18+\z,0.2) arc(90:270: 0.1 and 0.45);
	\foreach \y in {0.9}{
		\foreach \x in {1.} { 
			\draw[=latex, ] (\x-0.95+\z,-0.1+\y)  -- (\x+0.25+\z,-0.1+\y);
		}
	}
	\foreach \y in {0.9}{
		\foreach \x in {0} { 
			\draw[=latex, ] (\x-0.2+\z,-0.1+\y)  -- (\x+0.15+\z,-0.1+\y);
		}
	}
	\foreach \y in {0.}{
		\foreach \x in {1} { 
			\draw[=latex, ] (\x-0.82+\z,0.2+\y)  -- (\x+0.25+\z,0.2+\y);
		}
	}
	\foreach \y in {-1.5}{
		\foreach \x in {0} { 
			\draw[=latex, ] (\x-0.2+\z,0.2+\y)  -- (\x+0.15+\z,0.2+\y);
		}
	}
	\foreach \y in {-0.6}{
		\foreach \x in {1} { 
			\draw[=latex, ] (\x-0.82+\z,-0.1+\y)  -- (\x+0.25+\z,-0.1+\y);
		}
	}
	\foreach \y in {-1.5}{
		\foreach \x in {1} { 
			\draw[=latex, ] (\x-0.95+\z,0.2+\y)  -- (\x+0.25+\z,0.2+\y);
		}
	}
}
\foreach \z in {6.5}{
	\draw[=latex, ] (1.15+\z,0.8) arc(90:-90: 0.2 and 1.05);
	\draw[=latex, ] (0.82+\z,0.2) arc(90:-90: 0.1 and 0.45);
	\draw [fill=myblue, ] (0.15+\z, 0.2) rectangle (0.85+\z,0.8);
	\foreach \y in {0.9}{
		\foreach \x in {1.} { 
			\draw[=latex, ] (\x-0.15+\z,-0.1+\y)  -- (\x+0.15+\z,-0.1+\y);
		}
	}
	\foreach \y in {0.9}{
		\foreach \x in {0} { 
			\draw[=latex, ] (\x-0.25+\z,-0.1+\y)  -- (\x+0.15+\z,-0.1+\y);
		}
	}
	\foreach \y in {0.}{
		\foreach \x in {0.} { 
			\draw[=latex, ] (\x-0.25+\z,0.2+\y)  -- (\x+0.25+\z,0.2+\y);
		}
	}
	\foreach \y in {-1.5} { 
		\draw [fill=myred, ] (0.15+\z, 0.2+\y) rectangle (0.85+\z,0.8+\y);
	}
	\foreach \y in {-1.5}{
		\foreach \x in {0} { 
			\draw[=latex, ] (\x-0.25+\z,0.2+\y)  -- (\x+0.15+\z,0.2+\y);
		}
	}
	\foreach \y in {-0.6}{
		\foreach \x in {0} { 
			\draw[=latex, ] (\x-0.25+\z,-0.1+\y)  -- (\x+0.15+\z,-0.1+\y);
		}
	}
	\foreach \y in {-1.5}{
		\foreach \x in {1} { 
			\draw[=latex, ] (\x-0.15+\z,0.2+\y)  -- (\x+0.15+\z,0.2+\y);
		}
	}
}
\node at (8.5,-0.25) {\scalebox{1.2}{$=$}};
\node at (11.3,-0.25) {\scalebox{1.2}{$,$}};
\foreach \z in {9.5}{
	\draw[=latex, ] (1.15+\z,0.8) arc(90:-90: 0.2 and 1.05);
	\draw[=latex, ] (0.82+\z,0.2) arc(90:-90: 0.1 and 0.45);
	\foreach \y in {0.9}{
		\foreach \x in {1.} { 
			\draw[=latex, ] (\x-0.95+\z,-0.1+\y)  -- (\x+0.15+\z,-0.1+\y);
		}
	}
	\foreach \y in {0.9}{
		\foreach \x in {0} { 
			\draw[=latex, ] (\x-0.25+\z,-0.1+\y)  -- (\x+0.15+\z,-0.1+\y);
		}
	}
	\foreach \y in {0.}{
		\foreach \x in {0.} { 
			\draw[=latex, ] (\x-0.25+\z,0.2+\y)  -- (\x+0.82+\z,0.2+\y);
		}
	}
	\foreach \y in {-1.5}{
		\foreach \x in {0} { 
			\draw[=latex, ] (\x-0.25+\z,0.2+\y)  -- (\x+0.95+\z,0.2+\y);
		}
	}
	\foreach \y in {-0.6}{
		\foreach \x in {0} { 
			\draw[=latex, ] (\x-0.25+\z,-0.1+\y)  -- (\x+0.82+\z,-0.1+\y);
		}
	}
	\foreach \y in {-1.5}{
		\foreach \x in {1} { 
			\draw[=latex, ] (\x-0.15+\z,0.2+\y)  -- (\x+0.15+\z,0.2+\y);
		}
	}
}
\end{tikzpicture}
\label{eq:dual_unitarity_pic}
\ee
where continuous solid lines represent the identity operator.

In the case of qubits, \emph{i.e.} circuits with local dimension $d=2$, it was shown in Ref.~[\onlinecite{bertini2019exact}] that the most general dual-unitary gate reads
\be
U=e^{i \phi} (u_+ \otimes u_-)\cdot V[J]\cdot (v_-\otimes v_+)\,,
\label{eq:dualunitaryU}
\ee
where $\phi, J \in \mathbb R$, $u_\pm,v_\pm\in {\rm SU}(2)$ and 
\be
\!\!\!V[J]\!=\! \exp\!\left[\!-i\left(\frac{\pi}{4} \sigma^x\otimes\sigma^x \!+\! \frac{\pi}{4} \sigma^y\otimes\sigma^y\!+\! J \sigma^z\otimes\sigma^z\right)\right]\!\!.
\label{eq:v_mat_dual}
\ee
Furthermore, it was shown that this family includes both integrable~\cite{gritsev_integrable_2017,vanicat_integrable_2018,ljubotina_ballistic_2019} and non-integrable cases. In particular, it contains a full parameter line of the integrable trotterized XXZ chain~\cite{vanicat_integrable_2018,ljubotina_ballistic_2019}
\be
U_{\rm XXZ}[J]=V[J]\,,
\label{eq:gate_xxz}
\ee
and a quantum circuit representation of the self-dual kicked Ising (SDKI) model
\be
U_{\rm SDKI }[h]=e^{-i h \sigma^{z}} e^{i \frac{\pi}{4} \sigma^{x}} \otimes e^{i \frac{\pi}{4} \sigma^{x}} \cdot \tilde{V}[0] \cdot e^{-i h \sigma^{z}} \otimes \mathbb{1}\,,
\label{eq:gate_SKDI}
\ee
with
\be
\tilde{V}[0]=e^{-i \frac{\pi}{4} \sigma^{y}} \otimes e^{-i \frac{\pi}{4} \sigma^{y}} \cdot V[0] \cdot e^{i \frac{\pi}{4} \sigma^{y}} \otimes e^{i \frac{\pi}{4} \sigma^{y}}\,.
\ee
We recall that the dynamics defined by the gate~\eqref{eq:gate_SKDI} is integrable for $h=0$, while it is chaotic otherwise~\cite{bertini2019exact}, and, accordingly, in the latter case its spectral form factor is described by Random Matrix Theory~\cite{bertiniSDKI}.

\section{The solvable initial states}
\label{sec:solvable_states}

Even for dual-unitary quantum circuits, the computation of the time-evolution from arbitrary initial states appears to be extremely hard. Still, in the special case corresponding to the self-dual kicked Ising Floquet dynamics, it was found that the evolution of the bipartite R\'enyi entropies could be computed exactly for a particular family of product states~\cite{bertini2019entanglement}. This result relied on a specific mathematical property of the latter, which were called ``separating''. In this section we see that a similar logic can be followed for arbitrary dual-unitary circuits; in particular, we show how one can introduce a natural notion of ``solvability'' for a given initial state, and how this relates to the ``separating'' property in the special case of Ref.~[\onlinecite{bertini2019entanglement}]. 

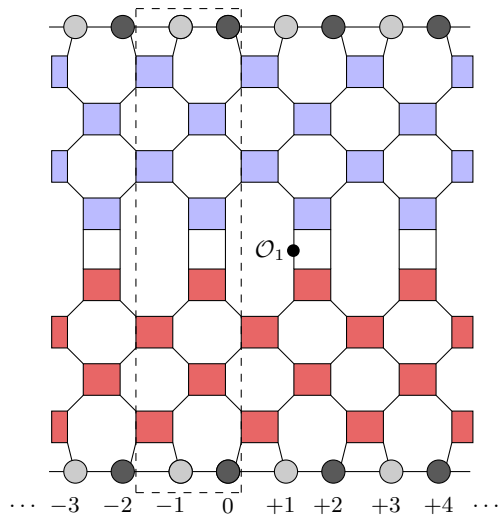
\begin{figure}
	\begin{center}
		\begin{tikzpicture}[scale=0.7]
		\draw[=latex] (0.5,-0.35)  -- (8.5,-0.35);
		\draw[=latex] (0.5,8.1)  -- (8.5,8.1);
		\foreach \x in {0.85,2.85,4.85,6.85} { 
			\draw[=latex] (\x+0.1,-0.2)  -- (\x,0.2);
			\draw[=latex] (\x,7.75-0.2)  -- (\x+0.1,7.75+0.2);
		}
		\foreach \x in {2.85-0.7,4.85-0.7,6.85-0.7,8.85-0.7} { 
			\draw[=latex] (\x-0.1,-0.2)  -- (\x,0.2);
			\draw[=latex] (\x,7.75-0.2)  -- (\x-0.1,7.75+0.2);
		}
		\foreach \y in {0,1.8} {
			\foreach \x in {2,4.,...,6.} { 
				\draw [fill=myred] (\x +0.15, 0.2+\y) rectangle (\x+0.85,0.8+\y);
			}
			\draw [fill=myred] (0.55, 0.2+\y) rectangle (0.85,0.8+\y);
			\draw [fill=myred] (8.15, 0.2+\y) rectangle (8.55,0.8+\y);
		}
		\foreach \y in {0,1.8} {
			\foreach \x in {1.,3.,...,7.} { 
				\draw [fill=myred] (\x +0.15,1.1+\y) rectangle (\x+0.85,1.7+\y);
			}
		}
		\foreach \y in {4.5+0.45,6.3+0.45} {
			\foreach \x in {2.,4.,...,6.} { 
				\draw [fill=myblue] (\x +0.15, 0.2+\y) rectangle (\x+0.85,0.8+\y);
			}
			\draw [fill=myblue] (0.55, 0.2+\y) rectangle (0.85,0.8+\y);
			\draw [fill=myblue] (8.15, 0.2+\y) rectangle (8.55,0.8+\y);
		}
		\foreach \y in {2.7+0.45,4.5+0.45} {
			\foreach \x in {1.,3.,...,7.} { 
				\draw [fill=myblue] (\x +0.15,1.1+\y) rectangle (\x+0.85,1.7+\y);
			}
		}
		\foreach \y in {1.8,4.5+0.45,6.3+0.45}{
			\foreach \x in {2.,4.,6.,8.} { 
				\draw[=latex] (\x-0.15,-0.1+\y)  -- (\x+0.15,0.2+\y);
			}
		}
		\foreach \y in {0.9,2.7,5.4+0.45}{
			\foreach \x in {2.,4.,6.,8.} { 
				\draw[=latex] (\x-0.15,0.2+\y)  -- (\x+0.15,-0.1+\y);
			}
		}
		\foreach \y in {1.8,4.5+0.45,6.3+0.45}{
			\foreach \x in {1,3.,5.,7.} { 
				\draw[=latex] (\x-0.15,0.2+\y)  -- (\x+0.15,-0.1+\y);
			}
		}
		\foreach \y in {0.9,2.7,5.4+0.45}{
			\foreach \x in {1,3.,5.,7.} { 
				\draw[=latex] (\x-0.15,-0.1+\y)  -- (\x+0.15,0.2+\y);
			}
		}
		\foreach \y in {-1.45,7.0} {
			\draw[fill=white!80!black] (0.85 +0.15,1.1+\y) circle (.22);
			\draw[fill=white!35!black] (1.75 +0.15,1.1+\y) circle (.22);
			\draw[fill=white!80!black] (2.85 +0.15,1.1+\y) circle (.22);
			\draw[fill=white!35!black] (3.75 +0.15,1.1+\y) circle (.22);
			\draw[fill=white!80!black] (2.85 +0.15,1.1+\y) circle (.22);
			\draw[fill=white!35!black] (3.75 +0.15,1.1+\y) circle (.22);
			\draw[fill=white!80!black] (4.85 +0.15,1.1+\y) circle (.22);
			\draw[fill=white!35!black] (5.75 +0.15,1.1+\y) circle (.22);
			\draw[fill=white!80!black] (6.85 +0.15,1.1+\y) circle (.22);
			\draw[fill=white!35!black] (7.75 +0.15,1.1+\y) circle (.22);
		}
		\foreach \x in {1,3.,5.,7.} { 
			\draw[=latex] (\x+0.15,3.5)  -- (\x+0.15,3.8+0.45);
		}
		\foreach \x in {2,4.,6.,8.} { 
			\draw[=latex] (\x-0.15,3.5)  -- (\x-0.15,3.8+0.45);
		}
		\draw[=latex,dashed] (2+0.15,-0.75)  -- (2+0.15,8.45);
		\draw[=latex,dashed] (4+0.15,-0.75)  -- (4+0.15,8.45);
		\draw[=latex,dashed] (2+0.15,-0.75)  -- (4+0.15,-0.75);
		\draw[=latex,dashed] (2+0.15,8.45)  -- (4+0.15,8.45);
		\draw[thick, fill=black] (5.22-0.08,3.85) circle (0.1cm); 
		\node at (4.7,3.85) {$\mathcal{O}_1$};
		\node at (0.,-1) {\scalebox{0.95}{$\ldots$}};
		\node at (0.8,-1) {\scalebox{0.95}{$-3$}};
		\node at (1.8,-1) {\scalebox{0.95}{$-2$}};
		\node at (2.8,-1) {\scalebox{0.95}{$-1$}};
		\node at (3.9,-1) {\scalebox{0.95}{$0$}};
		\node at (4.9,-1) {\scalebox{0.95}{$+1$}};
		\node at (5.8,-1) {\scalebox{0.95}{$+2$}};
		\node at (6.9,-1) {\scalebox{0.95}{$+3$}};
		\node at (7.9,-1) {\scalebox{0.95}{$+4$}};
		\node at (8.8,-1) {\scalebox{0.95}{$\ldots$}};
		\end{tikzpicture}
	\end{center}
	\caption{Pictorial representation of a time-dependent one-point correlation function. In the figure, an initial two-site shift invariant MPS $\ket{\Psi^L_0}$ is time-evolved by applying $t=4$ layers of unitary gates. The operator $\mathcal{O}_1$, localized at $j=1$, is represented by a small black dot, while black dashed lines enclose the transverse transfer matrix $E(t)$, \emph{cf.} Eq.~\eqref{eq:transfer_mat_E}.}
	\label{fig:quantum_circuit}
\end{figure}

We consider a generic initial state $\ket{\Psi^L_0}$ in the form of an MPS 
\be
|\Psi^L_0\rangle=\sum_{i_{1}, \ldots, i_{2L}=1}^{d} \operatorname{tr}\left(A_{1}^{i_{1}} \cdots A_{2L}^{i_{2L}}\right)\left|i_{1}, \ldots, i_{2L}\right\rangle\,,
\ee
where $A_{j}^{i_{j}}$ are matrices of dimensions $\chi_j\times \chi_{j+1}$. Since we are interested in the limit of infinite system sizes, it is natural to restrict to initial states that are invariant under translation of $p$ sites, where $p$ is some integer number. For reasons that will become clear later, we choose in the following $p=2$, which also contains the class of translationally invariant states. In this case, we can rewrite
\be
|\Psi^L_0\rangle=\!\!\!\!\!\!\sum_{i_{1}, \ldots, i_{2L}=1}^{d} \!\!\!\!\!\operatorname{tr}\left(A^{i_{1}}B^{i_{2}} \cdots A^{i_{2L-1}}B^{i_{2L}}\right)\left|i_{1}, \ldots, i_{2L}\right\rangle\,,
\label{eq:two_site_MPS}
\ee
so that now $\ket{\Psi^L_0}$ only depends on two sets of matrices $\{A^{i}\}_{i=1}^d$ and $\{B^{i}\}_{i=1}^d$ of dimensions $\chi\times \chi^\prime$ and $\chi^\prime \times \chi$ respectively. Finally, we consider MPSs that are normalized in the thermodynamic limit, namely
\be
\lim_{L\to\infty }\braket{\Psi^L_0|\Psi^L_0}=1\,.
\label{eq:normalization}
\ee
In analogy with the previous section, we can make use of a standard graphical notation and represent the individual tensors as
\be
\begin{tikzpicture}[scale=0.8]
\foreach \x in {0}{
	\draw[=latex, ] (0.6+\x,0.8)  -- (0.6+\x,1.4);
	\draw[=latex, ] (0.+\x,0.6)  -- (0.6+\x,0.6);
	\draw[=latex, ] (0.6+\x,0.6)  -- (1.2+\x,0.6);
	\draw[fill=white!80!black] (0.6+\x,0.6) circle (.22);
	\node at (-1.6+\x,0.6) {$\left(A^{i}\right)_{j,k}=$};
	\node at (1.7+\x,0.6) {$,$};
	\node at (-0.25+\x,0.6) {$j$};
	\node at (1.35+\x,0.6) {$k$};
	\node at (0.6+\x,1.65) {$i$};
}
\foreach \x in {5}{
	\draw[=latex, ] (0.6+\x,0.8)  -- (0.6+\x,1.4);
	\draw[=latex, ] (0.+\x,0.6)  -- (0.6+\x,0.6);
	\draw[=latex, ] (0.6+\x,0.6)  -- (1.2+\x,0.6);
	\draw[fill=white!35!black] (0.6+\x,0.6) circle (.22);
	\node at (-1.6+\x,0.6) {$\left(B^{i}\right)_{j,k}=$};
	\node at (1.7+\x,0.6) {$,$};
	\node at (-0.25+\x,0.6) {$j$};
	\node at (1.35+\x,0.6) {$k$};
	\node at (0.6+\x,1.65) {$i$};
}
\label{eq:mps_tensors}
\end{tikzpicture}
\ee
and
\be
\begin{tikzpicture}[scale=0.8]
\foreach \x in {0}{
	\draw[=latex, ] (0.6+\x,0.8)  -- (0.6+\x,-0.3);
	\draw[=latex, ] (0.+\x,0.6)  -- (0.6+\x,0.6);
	\draw[=latex, ] (0.6+\x,0.6)  -- (1.2+\x,0.6);
	\draw[fill=white!80!black] (0.6+\x,0.6) circle (.22);
	\node at (-1.6+\x,0.6) {$\left(A^{i}\right)^\ast_{j,k}=$};
	\node at (1.7+\x,0.6) {$,$};
	\node at (-0.25+\x,0.6) {$j$};
	\node at (1.35+\x,0.6) {$k$};
	\node at (0.6+\x,-0.6) {$i$};
}
\foreach \x in {5}{
	\draw[=latex, ] (0.6+\x,0.8)  -- (0.6+\x,-0.3);
	\draw[=latex, ] (0.+\x,0.6)  -- (0.6+\x,0.6);
	\draw[=latex, ] (0.6+\x,0.6)  -- (1.2+\x,0.6);
	\draw[fill=white!35!black] (0.6+\x,0.6) circle (.22);
	\node at (-1.6+\x,0.6) {$\left(B^{i}\right)^\ast_{j,k}=$};
	\node at (1.7+\x,0.6) {$,$};
	\node at (-0.25+\x,0.6) {$j$};
	\node at (1.35+\x,0.6) {$k$};
	\node at (0.6+\x,-0.6) {$i$};
}
\end{tikzpicture}
\label{eq:bra_tensors}
\ee	
so that MPSs are represented by a sequence of circles connected by lines, with additional out-coming legs corresponding to the physical local spaces. 

To isolate the special property that the MPSs \eqref{eq:two_site_MPS} should have in order to generate an exactly-solvable dynamics, it is instructive to consider the computation of the time evolution of one-point functions. Specifically, we consider $\braket{\Psi^L_t|\mathcal{O}_{1}|\Psi^L_t}$, where $\mathcal{O}_{1}$ is an arbitrary local operator acting on site $j=1$. A pictorial representation of this expectation value is reported in Fig.~\ref{fig:quantum_circuit}, where we employed the graphical notations introduced above. Borrowing standard ideas from the literature on tensor networks~\cite{perez2006matrix,banuls2009matrix,muller-hermes_tensor_2012}, one can write
\be
\lim_{L\to\infty}\braket{\Psi^L_t|\mathcal{O}_1|\Psi^L_t}=\lim_{k\to\infty}{\rm tr}\left[E^k(t)E_{\mathcal{O}_1}(t)E^{k}(t)\right]\,,
\label{eq:expectation_value}
\ee
where $E(t)$ and $E_{\mathcal{O}_1}(t)$ are appropriate transfer matrices acting on the tensor product of $2t+2$ local sites along the ``transverse direction''. There are several (in general inequivalent) ways to define these operators. For the purpose of the present discussion, it is useful to define $E(t)$ and $E_{\mathcal{O}_1}(t)$ directly in terms of the elementary tensors in Eqs.~\eqref{eq:u_tensors}, \eqref{eq:mps_tensors} and \eqref{eq:bra_tensors}. In particular, using a graphical notation and focusing for concreteness on the case where $t$ is even, we can define
\begin{equation}
\label{eq:transfer_mat_E}
\begin{tikzpicture}[baseline=(current  bounding  box.center), scale=0.5]
\node at (1.3,3.85) {$E(t)=$};
\node at (6.8,3.85) {$\,,$};
\node at (14.8,3.85) {$\,.$};
\node at (6.8+2.25,3.85) {$E_{\mathcal{O}_1}(t)=$};
\draw[thick, fill=black] (10.22-0.08+1.5,3.85) circle (0.1cm); 
\node at (11,3.85) {$\mathcal{O}_1$};
\foreach \t in {1,8.5}{
	\node at (4+1.1+\t,-1.3) {\scalebox{0.75}{$1$}};
	\node at (4+1.1+\t,-0.6) {\scalebox{0.75}{$2$}};
	\node at (4+1.1+\t,0.25) {\scalebox{0.75}{$3$}};
	\node at (4+1.1+\t,1.2) {\scalebox{0.95}{$\vdots$}};
	\node at (4+1.4+\t,2.9) {\scalebox{0.75}{$t+1$}};
	\node at (4+1.4+\t,4.9) {\scalebox{0.75}{$t+2$}};
	\node at (4+1.1+\t,6.8) {\scalebox{0.95}{$\vdots$}};
	\node at (4.2+1.0+\t,7.5) {\scalebox{0.75}{$2t$}};
	\node at (4.5+1.0+\t,8.3) {\scalebox{0.75}{$2t+1$}};
	\node at (4.5+1.0+\t,8.9) {\scalebox{0.75}{$2t+2$}};
	\foreach \y in {-0.8} { 
		\foreach \x in {2.85} { 
			\draw[=latex] (\x+0.1+\t,-0.2+\y)  -- (\x+\t,0.2+\y);
		}
		\foreach \x in {2.15} { 
		\draw[=latex] (\x+\t-0.55,0.2+\y)  -- (\x+\t,0.2+\y);
		}
		\draw[=latex] (3.9+\t,-0.2+\y)  -- (3.9+\t,0.2+\y);
		\draw[=latex] (3.9+\t,0.2+\y)  -- (4.7+\t,0.2+\y);
		\draw[=latex] (1.6+\t,-0.35+\y)  -- (4.7+\t,-0.35+\y);
		\draw[fill=white!80!black] (2.85 +0.15+\t,1.1+\y-1.45) circle (.22);
		\draw[fill=white!35!black] (3.75 +0.15+\t,1.1+\y-1.45) circle (.22);
		\foreach \x in {2} { 
		\draw [fill=myred] (\x +0.15+\t, 0.2+\y) rectangle (\x+0.85+\t,0.8+\y);
		}
		\foreach \x in {3} { 
		\draw [fill=myred] (\x +0.15+\t,1.1+\y) rectangle (\x+0.85+\t,1.7+\y);
		}
		\draw[=latex] (1.6+\t,0.8+\y)  -- (2.4+\t,0.8+\y);
		\draw[=latex] (3.5+\t,\y-0.6+0.9+0.8)  -- (4.7+\t,\y-0.6+0.9+0.8);
		\draw[=latex] (3.5+\t,\y+0.9+0.8)  -- (4.7+\t,\y+0.9+0.8);
		\draw[=latex] (3-0.15+\t,-0.1+0.9 +\y)  -- (3+0.15+\t,0.2+0.9 +\y);
	}
	\foreach \y in {0.8} { 
		\foreach \x in {2.85} { 
			\draw[=latex] (\x+\t,7.75-0.2+\y)  -- (\x+0.1+\t,7.75+0.2+\y);
		}
		\foreach \x in {2.15} { 
		\draw[=latex] (\x+\t,7.75-0.2+\y)  -- (\x-0.55+\t,7.75-0.2+\y);
		}
		\draw[=latex] (3.9+\t,7.75-0.2+\y)  -- (3.9+\t,7.75+0.25+\y);
		\draw[=latex] (3.9+\t,7.75-0.2+\y)  -- (4.7+\t,7.75-0.2+\y);
		\draw[=latex] (1.6+\t,8.1+\y)  -- (4.7+\t,8.1+\y);
		\draw[fill=white!80!black] (2.85+\t +0.15,1.1+\y+7.0) circle (.22);
		\draw[fill=white!35!black] (3.75 +0.15+\t,1.1+\y+7.0) circle (.22);
		\foreach \x in {2} { 
			\draw [fill=myblue] (\x +0.15+\t, 0.2+6.3+0.45+\y) rectangle (\x+0.85+\t,0.8+6.3+0.45+\y);
		}
		\foreach \x in {3} { 
			\draw [fill=myblue] (\x +0.15+\t,1.1+4.5+0.45+\y) rectangle (\x+0.85+\t,1.7+4.5+0.45+\y);
		}
		\draw[=latex] (1.6+\t,\y-0.6+7.1+0.45)  -- (2.4+\t,\y-0.6++7.1+0.45);
	 	\draw[=latex] (3-0.15+\t,0.2+\y+6.3+0.45)  -- (3+0.15+\t,-0.1+\y+6.3+0.45); 
	 	\draw[=latex] (3.5+\t,\y-0.6+0.9+5.3+0.45)  -- (4.7+\t,\y-0.6+0.9+5.3+0.45);
	 	\draw[=latex] (3.5+\t,\y+0.9+5.3+0.45)  -- (4.7+\t,\y+0.9+5.3+0.45);
	}
	\foreach \y in {1.8} {
		\foreach \x in {2} { 
			\draw [fill=myred] (\x +0.15+\t, 0.2+\y) rectangle (\x+0.85+\t,0.8+\y);
			\foreach \x in {3} { 
			\draw [fill=myred] (\x +0.15+\t,1.1+\y) rectangle (\x+0.85+\t,1.7+\y);
			}
		}
	}
	\foreach \y in {4.5+0.45} {
		\foreach \x in {2} { 
			\draw [fill=myblue] (\x +0.15+\t, 0.2+\y) rectangle (\x+0.85+\t,0.8+\y);
		}
	}
	\foreach \y in {2.7+0.45} {
		\draw [fill=myblue] (3.15+\t,1.1+\y) rectangle (3.85+\t,1.7+\y);
	}
	\foreach \y in {1.8,4.5+0.45}{
		\draw[=latex] (3-0.15+\t,0.2+\y)  -- (3.15+\t,-0.1+\y);
	}
	\foreach \y in {2.7,5.4+0.45}{
		\draw[=latex] (3-0.15+\t,-0.1+\y)  -- (3.15+\t,0.2+\y);
	}
	\foreach \x in {3.} { 
		\draw[=latex] (\x+0.15+\t,3.5)  -- (\x+0.15+\t,3.8+0.45);
	}
	\foreach \x in {4.} { 
		\draw[=latex] (\x-0.15+\t,3.5)  -- (\x-0.15+\t,3.8+0.45);
	}
	\foreach \y in {2.6,5.3+0.45}{
		\draw[=latex] (1.6+\t,\y-0.6)  -- (2.4+\t,\y-0.6);
		\draw[=latex] (1.6+\t,\y)  -- (2.4+\t,\y);
	}
	\foreach \y in {2.6,4.1+0.45}{
		\draw[=latex] (3.5+\t,	\y-0.6+0.9)  -- (4.7+\t,\y-0.6+0.9);
	}
	\foreach \y in {0.3,5.85}{
	\draw[=latex,dashed] (3.15+\t,0.25+\y)  -- (3.15+\t,1.4+\y);
	}
}
\end{tikzpicture}
\end{equation}
Here the right out-coming $2t+2$ legs correspond to the input space on which $E(t)$ and $E_{\mathcal{O}_1}(t)$ act on. The validity of Eq.~\eqref{eq:expectation_value} is straightforwardly established. Indeed, one can simply note that the graphical representation for ${\rm tr}\left[E^k(t)E_{\mathcal{O}_1}(t)E^k(t)\right]$, which is obtained by placing $2k+1$ transfer matrices side by side, is the same as for the expectation value of $\braket{\Psi_t|\mathcal{O}_1|\Psi_t}$ in a chain of $4k+2$ sites (where periodic boundary conditions are implemented). 

Next, suppose that the largest eigenvalue $\lambda_0$ of $E(t)$ is non-degenerate; more precisely, suppose that its algebraic multiplicity (namely, the number of diagonal elements in the Jordan form of $E(t)$ that are equal to $\lambda_0$) is $1$ and that there are no other eigenvalues with the same absolute value. Then, for large~$L$
\be
\braket{\Psi^L_t|\Psi^L_t}= {\rm tr}\left[E(t)^{L}\right]\simeq \lambda^{L}_0\,,
\ee
where we used that the length of the system is $2L$. Since $\braket{\Psi^L_t|\Psi^L_t}=\braket{\Psi^L_0|\Psi^L_0}$, Eq.~\eqref{eq:normalization} implies $\lambda_0=1$.  In turn, this yields
\be
\lim_{L\to\infty}\langle\Psi^L_t|\mathcal{O}_1| \Psi^L_t\rangle=\left\langle L\left|E_{\mathcal{O}_1}(t)\right| R\right\rangle\,,
\label{eq:transverse_contraction}
\ee
where we denoted by $\ket{L}$ and $\ket{R}$ the left and right eigenstates of $E(t)$ associated with $\lambda_0$, with the normalization $\braket{L|R}=1$.  In general, the evaluation of Eq.~\eqref{eq:transverse_contraction} can only be done numerically for small times. However, for dual-unitary circuits there exist a class of states for which $\ket{L}$ and $\ket{R}$ can be determined exactly.

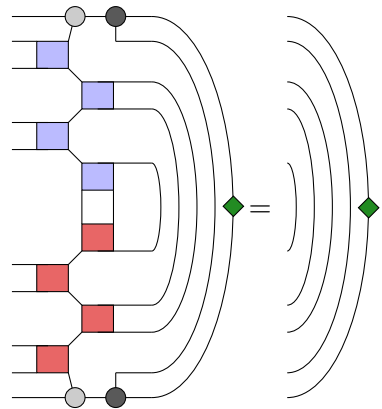
\begin{figure}
	\begin{center}
		\begin{tikzpicture}[scale=0.6]
		\draw (4.7,8.10) arc(90:-90: 1.8 and 4.225);
		\draw (4.7,7.55) arc(90:-90: 1.4 and 3.675);
		\draw (4.7,6.65) arc(90:-90: 1.0 and 2.775);
		\draw (4.7,6.05) arc(90:-90: 0.6 and 2.175);
		\draw (4.7,4.85) arc(90:-90: 0.2 and 0.975);
		\foreach \x in {2.85} { 
			\draw[=latex] (\x+0.1,-0.2)  -- (\x,0.2);
			\draw[=latex] (\x,7.75-0.2)  -- (\x+0.1,7.75+0.2);
		}
		\foreach \x in {2.15} { 
			\draw[=latex] (\x-0.55,0.2)  -- (\x,0.2);
			\draw[=latex] (\x,7.75-0.2)  -- (\x-0.55,7.75-0.2);
		}
		\draw[=latex] (3.9,-0.2)  -- (3.9,0.2);
		\draw[=latex] (3.9,0.2)  -- (4.7,0.2);
		\draw[=latex] (3.9,7.75-0.2)  -- (3.9,7.75+0.25);
		\draw[=latex] (3.9,7.75-0.2)  -- (4.7,7.75-0.2);
		\draw[=latex] (1.6,-0.35)  -- (4.7,-0.35);
		\draw[=latex] (1.6,8.1)  -- (4.7,8.1);	
		\foreach \y in {0,1.8} {
			\foreach \x in {2} { 
				\draw [fill=myred] (\x +0.15, 0.2+\y) rectangle (\x+0.85,0.8+\y);
			}
		}
		\foreach \y in {0,1.8} {
			\foreach \x in {3} { 
				\draw [fill=myred] (\x +0.15,1.1+\y) rectangle (\x+0.85,1.7+\y);
			}
		}
		\foreach \y in {4.5+0.45,6.3+0.45} {
			\foreach \x in {2} { 
				\draw [fill=myblue] (\x +0.15, 0.2+\y) rectangle (\x+0.85,0.8+\y);
			}
		}
		\foreach \y in {2.7+0.45,4.5+0.45} {
			\foreach \x in {3} { 
				\draw [fill=myblue] (\x +0.15,1.1+\y) rectangle (\x+0.85,1.7+\y);
			}
		}
		\foreach \y in {1.8,4.5+0.45,6.3+0.45}{
			\foreach \x in {3} { 
				\draw[=latex] (\x-0.15,0.2+\y)  -- (\x+0.15,-0.1+\y);
			}
		}
		\foreach \y in {0.9,2.7,5.4+0.45}{
			\foreach \x in {3.} { 
				\draw[=latex] (\x-0.15,-0.1+\y)  -- (\x+0.15,0.2+\y);
			}
		}
		\foreach \y in {-1.45,7.0} {
			\draw[fill=white!80!black] (2.85 +0.15,1.1+\y) circle (.22);
			\draw[fill=white!35!black] (3.75 +0.15,1.1+\y) circle (.22);
		}
		\foreach \x in {3.} { 
			\draw[=latex] (\x+0.15,3.5)  -- (\x+0.15,3.8+0.45);
		}
		\foreach \x in {4.} { 
			\draw[=latex] (\x-0.15,3.5)  -- (\x-0.15,3.8+0.45);
		}
		\foreach \y in {2.6,5.3+0.45}{
			\draw[=latex] (1.6,\y-0.6)  -- (2.4,\y-0.6);
			\draw[=latex] (1.6,\y)  -- (2.4,\y);
		}
		\foreach \y in {0.8}{
			\draw[=latex] (1.6,\y)  -- (2.4,\y);
		}
		\foreach \y in {7.1+0.45}{
			\draw[=latex] (1.6,\y-0.6)  -- (2.4,\y-0.6);
		}
		\foreach \y in {0.8,5.3+0.45}{
			\draw[=latex] (3.5,\y-0.6+0.9)  -- (4.7,\y-0.6+0.9);
			\draw[=latex] (3.5,\y+0.9)  -- (4.7,\y+0.9);
		}
		\foreach \y in {2.6,4.1+0.45}{
			\draw[=latex] (3.5,	\y-0.6+0.9)  -- (4.7,\y-0.6+0.9);
		}
		\node at (7.1,3.8) {\scalebox{1.2}{$=$}};
		\draw (7.7,8.10) arc(90:-90: 1.8 and 4.225);
		\draw (7.7,7.55) arc(90:-90: 1.4 and 3.675);
		\draw (7.7,6.65) arc(90:-90: 1.0 and 2.775);
		\draw (7.7,6.05) arc(90:-90: 0.6 and 2.175);
		\draw (7.7,4.85) arc(90:-90: 0.2 and 0.975);
		\draw [fill=mygreen, rotate around={45:(6.5-0.05,3.85)},scale=0.8] (0+8-0.05,0+4.6) rectangle (0.4+8-0.05,0.4+4.6);
		\draw [fill=mygreen, rotate around={45:(6.5+3-0.02,3.85)},scale=0.8] (0+8+3.7-0.02,0+4.6) rectangle (0.4+8+3.7-0.02,0.4+4.6);
		\end{tikzpicture}
	\end{center}
	\caption{Graphical representation of the transfer matrix $E(t)$ acting on the state $\ket{R}$ defined in Eq.~\eqref{eq:right_eigenstate}, for $t=4$. By making repeated use of the diagrams in Eq.~\eqref{eq:dual_unitarity_pic}, and finally using Eq.~\eqref{eq:solvability_pic}, one directly obtains $E(t)\ket{R}=\ket{R}$.}
	\label{fig:transfer_matrix}
\end{figure}

Consider in particular an initial two-site shift invariant MPS, as defined in Eq.~\eqref{eq:two_site_MPS}, and suppose that there exists a $\chi$-dimensional matrix $S$ such that 
\be
\sum_{k=1}^d\left(A^iB^k\right) S \left(A^jB^k\right)^\dagger =\frac{1}{d}\delta_{i,j}S\,.
 \label{eq:solvability_cond}
\ee
Then one can show that
\bea
|R\rangle=\bigotimes _{k=2}^{t+1}\left(\frac{1}{\sqrt{d}}\sum_{j=1}^d\ket{j}_k\otimes \ket{j}_{2t-k+3}\right)\nonumber\\
\otimes \left(\sum_{\alpha,\beta=1}^\chi S_{\beta,\alpha}\ket{\beta}_1\otimes \ket{\alpha}_{2t+2} \right)\,,
\label{eq:right_eigenstate}
\eea
is a right eigenstate of $E(t)$ with eigenvalue $\lambda_0=1$. Here $S_{\alpha,\beta}=\braket{\alpha|S|\beta}$ are the matrix elements of $S$ in the basis $\{\ket{\alpha}\}_{\alpha=1}^\chi$ of the auxiliary space associated with the initial MPS. The proof can be carried out graphically by noting that Eq.~\eqref{eq:solvability_cond} can be represented as
\be
\begin{tikzpicture}[baseline=(current  bounding  box.center), scale=0.7]
\begin{scope}[xscale=-1,xshift=-4cm]
\draw[=latex, ] (0.8,2.05) arc(90:270: 0.27 and 1.025);
\draw[=latex] (1,2.1)  -- (2,2.1);
\draw[=latex] (1,0.05)  -- (2,0.05);
\draw[=latex] (1,0.1)  -- (1,2.1);
\draw[=latex] (2,0.1)  -- (2.2,0.8);
\draw[=latex] (2.2,1.4)  -- (2,2.1);
\foreach \y in {-1,1.0} {
	\draw[=latex] (2,1.1+\y)  -- (2.7,1.1+\y) ;
	\draw[fill=white!35!black] (1,1.1+\y) circle (.22);
	\draw[fill=white!80!black] (2,1.1+\y) circle (.22);	
	\draw [fill=mygreen, rotate around={45:(0.4,1.4)},scale=0.8] (0.6,1.2) rectangle (0.2,1.6);
}
\end{scope}
\node at (4.5,1.) {\scalebox{1.2}{$=\frac{1}{d}$}};
\begin{scope}[xscale=-1,xshift=-11cm]
\draw[=latex, ] (4.4,2.1) arc(90:270: 0.27 and 1.025);
\draw[=latex, ] (4.7,1.8) arc(90:270: 0.27 and 0.725);
\draw[=latex] (3.4+1,2.1)  -- (3.6+2,2.1);
\draw[=latex] (3.4+1,0.05)  -- (3.6+2,0.05);
\draw[=latex] (2+3.4,0.05+.3)  -- (2.2+3.4,0.35+.3);
\draw[=latex] (2.2+3.4,1.8-.3)  -- (2+3.4,2.1-.3);

\draw[=latex] (3.4+1.3,2.1-0.3)  -- (3.4+2,2.1-0.3);
\draw[=latex] (3.4+1.3,0.05+0.3)  -- (3.4+2,0.05+0.3);
\draw [fill=mygreen, rotate around={45:(0.4+3.6,1.4)},scale=0.8] (0.6+4.5,1.2) rectangle (0.2+4.5,1.6);
\end{scope}
\node at (7.5,1.) {,};
\end{tikzpicture}
\label{eq:solvability_pic}
\ee
where we introduced the following notation
\be
\begin{tikzpicture}[baseline=(current  bounding  box.center), scale=0.7]
\node at (5,1.) {\scalebox{1.2}{$S_{\beta,\alpha}=$}};
\node at (6.4,2.3) {\scalebox{1.}{$\alpha$}};
\node at (6.4,-0.2) {\scalebox{1.}{$\beta$}};
\begin{scope}[xscale=-1,xshift=-11cm]
\draw[=latex, ] (4.4,2.1) arc(90:270: 0.27 and 1.025);
\draw [fill=mygreen, rotate around={45:(0.4+3.6,1.4)},scale=0.8] (0.6+4.5,1.2) rectangle (0.2+4.5,1.6);
\end{scope}
\node at (7.5,1.) {.};
\end{tikzpicture}
\ee
Indeed, as shown in Fig.~\ref{fig:transfer_matrix}, this allows one to compute the action of $E(t)$ on  $\ket{R}$ by making use of the diagrams in Eqs.~\eqref{eq:dual_unitarity_pic} and ~\eqref{eq:solvability_pic}.

The above discussion motivates us to introduce the notion of \emph{solvable} initial states for the quantum dynamics, and take Eq.~\eqref{eq:solvability_pic} as a defining property of solvability. A priori, however, this condition alone is not sufficient to guarantee the uniqueness of the leading eigenvalue, which is necessary, for instance, to obtain Eq.~\eqref{eq:transverse_contraction}. Accordingly, we say that a two-site shift invariant MPS [as defined in Eq.~\eqref{eq:two_site_MPS}] is solvable with respect to the class of dual-unitary quantum circuits if the following two conditions are satisfied:
\begin{itemize}
	\item[{\bf C1}.] The transfer matrix $E(t)$ has a unique eigenvalue $\lambda_0$ with largest absolute value, $\lambda_0=1$ and $\lambda_0$ has algebraic multiplicity $1$ $\forall t\in \mathbb{N}$; \label{cond:unique_eigen}
	\item[{\bf C2}.] There exists a non-zero $\chi$-dimensional matrix $S$ satisfying Eq.~\eqref{eq:solvability_cond}\,.
\end{itemize}
As we will see in the following, condition C1 could be removed. Indeed, as it will be clear from our derivations, the sum of $k$ MPSs satisfying C1 and C2 gives us another MPS whose transfer matrix $E(t)$ has $k$ known eigenvectors with maximum absolute value, and for which analytic results could be derived. However, condition C1 allows us to classify the most ``elementary" solvable states, from which all the others can be built simply out of linear superposition.

It is worth to discuss a connection between the present work and the findings of Refs.~[\onlinecite{banuls2009matrix},\onlinecite{muller-hermes_tensor_2012}], where a \emph{folding technique} to contract infinite tensor networks was introduced. Indeed, it can be seen that the solvable initial states are such that the leading right eigenvector $\ket{R}$ of $E(t)$ is a product state in the folded tensor network introduced in [\onlinecite{banuls2009matrix}], in analogy with what happens from different initial states in the toy model studied in Ref.~[\onlinecite{muller-hermes_tensor_2012}]. Note, however, that the latter was manifestly non-interacting, while the dual-unitary circuits generally implement a chaotic time evolution~\cite{bertini2019operator_II}, making the dynamics of solvable initial states non-trivial.

We also note that the logic of the present paper is similar to that of Refs.~[\onlinecite{piroli2017integrable}],~[\onlinecite{pozsgay2019integrable}], where it was shown that for any Bethe-Ansatz integrable Hamiltonian it is always possible to find ``integrable'' MPSs for which the quantum dynamics can be tackled analytically. In that case, these MPSs were defined as the initial states for which the transverse transfer matrix $E(t)$ (obtained after a discretization of time through an appropriate Trotterization procedure~\cite{vanicat_integrable_2018}) becomes Bethe-Ansatz integrable, so that its eigenvectors can be determined exactly. In our case, we stress that the transfer matrix $E(t)$ corresponding to solvable MPSs is in general not Bethe-Ansatz integrable and, accordingly, we only have access to the eigenstate with largest absolute value.

Finally, before leaving this section, we recall that there is a close relation between the above definition of solvable and that of ``separating'' initial states, as defined in Ref.~[\onlinecite{bertini2019entanglement}]. This is reported in Appendix~\ref{sec:solvability_vs_separability}, to which we refer the interested reader.

\section{Classification of the solvable initial states}
\label{sec:classification}

 In this section, we proceed by providing a complete classification of the states that can be described by solvable MPSs in the thermodynamic limit. This is achieved by Theorem~\ref{th:classification}, which is stated below. The proof of the latter, which is based on well established techniques in quantum information and tensor network theory~\cite{perez2006matrix,sanz2010quantum,cirac2017matrix}, is rather technical and the interested reader can find it in Appendix~\ref{sec:proof_theorem_1}. In the rest of this section we only provide the definitions that are needed in order to present its statement.
 
We begin by introducing the following parametrization for two-site shift invariant MPSs
\bea
&&|\Psi^L_0\left(\mathcal{M}\right)\rangle= \label{eq:two_site_MPS_new}
\\
&&\sum_{\{i_{j}\}}^{d} \operatorname{tr}\left(\mathcal{M}^{(i_1,i_2)} \mathcal{M}^{(i_3,i_4)} \cdots \mathcal{M}^{(i_{2L-1},i_{2L})}\right)\left|i_{1}, \ldots, i_{2L}\right\rangle\,,
\nonumber
\eea 
where $\{\mathcal{M}^{(i,j)}\}_{i,j=1}^d$ is a single set of $\chi$-dimensional matrices, which encode all of the information stored in the sets $\{A^j\}_{j=1}^d$ and $\{B^j\}_{j=1}^d$. We note that the solvability condition~\eqref{eq:solvability_cond} can be written in terms of the tensors $\mathcal{M}^{(i,j)}$ as
\be
\sum_{k=1}^d \mathcal{M}^{(i,k)} S \left(\mathcal{M}^{(j,k)}\right)^\dagger =\frac{1}{d}\delta_{i,j}S\,. 
\label{eq:solvability}
\ee

Next, we provide two additional definitions that are needed in order to state our main result. First, let $\{\ket{\Phi^{L}_0}\}_{L}$ be a class of states defined on systems of increasing (even) sizes. We say that $\{\ket{\Phi^L_0}\}_{L}$ is equivalent to the class of two-site shift invariant MPSs $\{\ket{\Psi_0^L(\mathcal{M})}\}_{L}$ if all local correlation functions coincide in the thermodynamic limit, namely if $\forall R\in \mathbb{N}$
\be
\lim_{L\to \infty }\braket{\Phi^L_0|\mathcal{O}_{R}|\Phi^L_0}=\lim_{L\to \infty }\braket{\Psi_0^L(\mathcal{M})|\mathcal{O}_{R}|\Psi_0^L(\mathcal{M})}\,,
\label{eq:equivalence}
\ee
where $\mathcal{O}_R$ is any observable acting non-trivially only on a finite product of local Hilbert spaces $h_{j_1}\otimes \cdots \otimes h_{j_R}$. In this case, we also say that $\ket{\Phi^L_0}$ and $\ket{\Psi_0^L(\mathcal{M})}$ are equivalent in the thermodynamic limit.

Second, we recall the well-known notion of injectivity~\cite{perez2006matrix}: we say that a two-site shift invariant MPS $\ket{\Psi_0^L(\mathcal{M})}$ is \emph{injective} if the linear map 
\bea
\Gamma_L: X\mapsto \sum_{\{i_{j}\}}^{d} \operatorname{tr}\left(X\mathcal{M}^{(i_1,i_2)} \cdots\right. \nonumber\\
\left. \cdots\mathcal{M}^{(i_{2L-1},i_{2L})}\right)\left|i_{1}, \ldots, i_{2L}\right\rangle\,,
\eea
is injective. It can be proven~\cite{perez2006matrix} that if $\ket{\Psi_0^L(\mathcal{M})}$ is injective for $L>0$, so is for $L^\prime>L$. This means that we can define the class $\{\ket{\Psi_0^L(\mathcal{M})}\}_{L}$ to be injective if $\ket{\Psi_0^{\bar L}(\mathcal{M})}$ is injective for $\bar L$ sufficiently large.

Using the above definitions, we are now in a position to state one of our main results, \emph{i.e.} the following theorem.

\begin{Theorem}
	\label{th:classification}
	\emph{The state $\ket{\Phi^L_0}$ is equivalent to a solvable MPS $\ket{\Psi^L_0(\mathcal{M})}$ satisfying conditions $C1$ and $C2$ if and only if $\ket{\Phi^L_0}$ is equivalent to a two-site shift invariant MPS $\ket{\Psi^L_0(\mathcal{N})}$ which is injective (for $L$ sufficiently large) and such that}
	\be
	\sum_{k=1}^d \mathcal{N}^{(i,k)}\left(\mathcal{N}^{(j,k)}\right)^\dagger =\frac{1}{d}\delta_{i,j}\,. 
	\label{eq:solvability_identity}
	\ee
\end{Theorem}
As we already mentioned, the proof of this theorem is rather technical, and is therefore reported in Appendix~\ref{sec:proof_theorem_1}.

At this point, it is important to note that Eq.~\eqref{eq:solvability_identity} also allows us to write down the left eigenvector of the transfer matrix $E(t)$. Indeed, it is not difficult to show that Eq.~\eqref{eq:solvability_identity} implies
\be
\sum_{k=1}^d \left(\mathcal{N}^{(i,k)}\right)^\dagger  \mathcal{N}^{(j,k)} =\frac{1}{d}\delta_{i,j}\,,
\label{eq:solvability_identity_left}
\ee
from which it follows that right and left eigenvectors are equal, namely
\bea
\ket{L}=|R\rangle=\bigotimes _{k=2}^{t+1}\left(\frac{1}{\sqrt{d}}\sum_{j=1}^d\ket{j}_k\otimes \ket{j}_{2t-k+3}\right)\nonumber\\
\otimes \left(\frac{1}{\sqrt{\chi}}\sum_{\alpha=1}^\chi \ket{\alpha}_1\otimes \ket{\alpha}_{2t+2} \right)\,.
\label{eq:right_left_eigenstate}
\eea
Altogether, Theorem~\ref{th:classification} provides us with a useful criterion to construct solvable MPSs, as we now explain. Given the tensors $\mathcal{N}^{(i,j)}$, we begin by defining the matrix $W(\mathcal{N})$ acting on the tensor product $\mathbb{C}^d\otimes \mathbb{C}^\chi$ via
\bea
\bra{i}\otimes \bra{\alpha} W(\mathcal{N}) \ket{j}\otimes \ket{\beta}=\left[\mathcal{N}^{(i,j)}\right]_{\alpha,\beta}\,,
\label{eq:W_matrix}
\eea
where $ i,j=1,\ldots d$, $\alpha,\beta=1,\ldots \chi$, so that Eq.~\eqref{eq:solvability_identity} can be straightforwardly rewritten as
\be
W(\mathcal{N})\left[W(\mathcal{N})\right]^{\dagger}=\mathbb{1}\,.
\label{eq:unitarity_matrix}
\ee
This equation is particularly useful: when combined with Theorem~\ref{th:classification} it tells us that solvable states can be completely parametrized by matrices $W\in {\rm{End}}(\mathbb{C}^d\otimes \mathbb{C}^\chi)$  that are \emph{unitary}. Next, note that if $\ket{\Psi^L_0(\mathcal{N})}$ is an MPS satisfying Eq.~\eqref{eq:solvability_identity}, then the same is true for the MPS
\be
\ket{\Psi^{\prime L}_0}=\left(\prod_{j=1}^{L}\left[u_{2j}v_{2j-1}\right]\right)\ket{\Psi^L_0(\mathcal{N})}\,,
\ee
where $u_j,v_j\in U(d)$ are arbitrary unitary operators acting on the local Hilbert space $h_j$. This means that solvable MPSs can be classified up to products of local unitaries. On the level of the matrix $W(\mathcal N)$, this transformation reads as
\be
W(\mathcal N)\longmapsto (v_{2j-1}^\dag\otimes\mathbbm{1}_\chi) W(\mathcal N) (u_{2j}\otimes \mathbbm{1}_\chi)\,,
\label{eq:Wmapping}
\ee
where $\mathbbm{1}_\chi$ is the identity on $\mathbb{C}^\chi$. 

In the next section we see how the above consideration can be used to construct explicitly solvable MPSs.

\section{Qubit systems: explicit solutions}
\label{sec:explicit_examples}

In this section we address the explicit construction of solvable MPSs for the simplest case of a qubit system, corresponding to $d=2$. In particular, we provide formulae for solvable MPSs with bond dimensions $\chi=1$ and $\chi=2$. We recall that here $\chi$ denotes the bond dimension of the MPS $\ket{\Psi_0^L(\mathcal{N})}$ obtained by grouping together two sites, namely it is the dimension of the matrices $\mathcal{N}^{(i,j)}$.

\subsection{Bond dimension $\chi$=1}
\label{sec:bond_chi1}

The simplest example of solvable MPSs is given by product states, corresponding to bond dimension $\chi=1$. In this case, up to products of local unitaries, it is always possible to choose $W(\mathcal N)\propto\mathbbm{1}_2$, which leads to the following explicit form
\be
\ket{\Psi^L_0}=\frac{1}{\sqrt{2^{2L}}}\bigotimes_{k=1}^{L}\left(\ket{1,1}_{2k-1,2k}+\ket{2,2}_{2k-1,2k}\right)\,.
\label{eq:chi_1_MPS}
\ee
On the other hand, it is trivial to verify that the state $\ket{\Psi^L_0}$ defined above is injective.  Putting all together, we obtain a single solvable MPS with bond dimension $\chi=1$. 

\subsection{Bond dimension $\chi$=2}

The case of bond dimension $\chi=2$ is more interesting. Now the unitary matrix $W$ defined in Eq.~\eqref{eq:W_matrix} acts on the tensor product $\mathbb{C}^2\otimes \mathbb{C}^2$. Then, we can use the following known parametrization~\cite{kraus2001optimal} for $W\in U(4)$, which is complete up to a global irrelevant phase:
\be
W=\left(z \otimes v\right) V\left[\{K_{j}\}_{j=1}^3\right]\left( t \otimes u\right)\,,
\ee
where $t,u,v,z\in SU(2)$,  while
\bea
V\left[\{K_{j}\}_{j=1}^3\right]&=&\exp \left[-i\left(K_{1} \sigma^{x} \otimes \sigma^{x}\right.\right.\nonumber\\
&+&\left.\left. K_{2} \sigma^{y} \otimes \sigma^{y}+K_{3} \sigma^{z} \otimes \sigma^{z}\right)\right]\,,
\eea
with $K_j\in \mathbb{R}$. By performing the matrix exponential $V\left[\{K_{j}\}_{j=1}^3\right]$, and rearranging the indices as in Eq.~\eqref{eq:W_matrix}, we arrive at the following general parametrization (up to products of local unitaries)
\bea
\mathcal{N}^{(1,1)}
&=&
v
\begin{bmatrix}
	e^{-iK_3}\cos K_- & 0 \\
	0 & 		e^{iK_3}\cos K_+  \\
\end{bmatrix}
u\,, \label{eq:n_11}\\
\mathcal{N}^{(1,2)}
&=&
v
\begin{bmatrix}
	0 &-ie^{-iK_3}\sin K_-  \\
	-ie^{iK_3}\sin K_+ & 0  \\
\end{bmatrix}
u,\\
\mathcal{N}^{(2,1)}
&=&
v
\begin{bmatrix}
	0 &-ie^{iK_3}\sin K_+   \\
	-ie^{-iK_3}\sin K_-  & 0  \\
\end{bmatrix}
u,\\
\mathcal{N}^{(2,2)}
&=&
v
\begin{bmatrix}
	e^{iK_3}\cos K_+  & 0 \\
	0 & e^{-iK_3}\cos K_- \\
\end{bmatrix}
u\,,\label{eq:n_22}
\eea
where $u,v\in SU(2)$ and $K_{\pm}=K_{1}\pm K_2\in \mathbb{R}$. 
Finally, the MPS $\ket{\Psi_0^L(\mathcal{N})}$ is injective, except for a set of measure zero in the space of parameters $\{K_{j}\}_{j=1}^3$. This simply follows from the fact that for generic choices of $\{K_{j}\}_{j=1}^3$ the operators $\mathcal{N}^{(i,j)}$ span the whole set of $2\times 2$ matrices.

\begin{figure}
	\begin{center}
		\begin{tikzpicture}[scale=0.7]
		\draw[=latex] (0.7+-0.35,0.5)  -- (0.7+-0.35,4.);
		\draw[=latex] (4.5+-0.35,0.5)  -- (4.5+-0.35,4.);
		\draw[=latex] (0,-0.25)  -- (4.5,-0.25);
		\draw[=latex] (0,4.8)  -- (4.5,4.8);
		\draw[=latex] (0.3,0.25)  -- (0.8,-0.25);
		\draw[=latex] (-1,1.45)  -- (0.,0.45);
		\draw[=latex] (0.3,4.25)  -- (0.8,4.85);
		\draw[=latex] (-0.75,3.3)  -- (-0.2,3.85);
		\draw[=latex] (3.75,-0.25)  -- (4.2,0.25);
		\draw[=latex] (4.5,0.45)  -- (5.5,1.45);
		\draw[=latex] (3.75,4.85)  -- (4.2,4.25);
		\draw[=latex] (4.6,3.95)  -- (5.4,3.15);
		\draw[=latex] (-1.35,1.5)  -- (-1.35,3);
		\draw[=latex] (-0.65,1.5)  -- (-0.65,3);
		\draw[=latex] (5.15,1.5)  -- (5.15,3);
		\draw[=latex] (5.85,1.5)  -- (5.85,3);
		%
		\draw[=latex] (1.7,-0.25)  -- (1.7,4.8);
		\draw[=latex] (2.8,-0.25)  -- (2.8,4.8);
		%
		\draw[=latex, ] (-1.35,3.4) arc(90:270: 0.4 and 1.145);
		\draw[=latex, ] (5.85,1.115) arc(-90:90: 0.4 and 1.145);
		\draw[=latex, ] (-2.4,4.29) arc(90:270: 0.4 and 2.04);
		\draw[=latex, ] (6.9,0.215) arc(-90:90: 0.4 and 2.04);
		\draw[=latex] (4.7,4.3)  -- (6.9,4.3);
		\draw[=latex] (4.7,0.2)  -- (6.9,0.2);
		\draw[=latex] (-2.4,0.2)  -- (0,0.2);
		\draw[=latex] (-2.4,4.3)  -- (0,4.3);
		\draw[=latex] (-3.,-0.25)  -- (0,-0.25);
		\draw[=latex] (-3.,4.8)  -- (0,4.8);
		\draw[=latex, ] (-3,4.8) arc(90:270: 0.4 and 2.52);
		\draw[=latex, ] (7.5,-0.25) arc(-90:90: 0.4 and 2.52);
		\draw[=latex] (4.2,4.8)  -- (7.5,4.8);
		\draw[=latex] (4.2,-0.25)  -- (7.5,-0.25);
		\foreach \y in {-0.25,4.8} {
			\draw[fill=white!80!black] (0.8,+\y) circle (.22);
			\draw[fill=white!35!black] (1.7,+\y) circle (.22);
			\draw[fill=white!80!black] (2.8,+\y) circle (.22);
			\draw[fill=white!35!black] (3.7,+\y) circle (.22);
			\draw[fill=white!80!black] (2.8,+\y) circle (.22);
			\draw[fill=white!35!black] (3.7,+\y) circle (.22);
		}
		\foreach \y in {0} {
			\foreach \x in {-0.5,4} { 
				\draw [fill=myred] (\x +0.15, 0.2+\y) rectangle (\x+0.85,0.8+\y);
			}
		}
		\foreach \y in {0} {
			\foreach \x in {-1.5,5.} { 
				\draw [fill=myred] (\x +0.15,1.1+\y) rectangle (\x+0.85,1.7+\y);
			}
		}
		\foreach \y in {2.6} {
			\foreach \x in {-1.5,5.} { 
				\draw [fill=myblue] (\x +0.15, 0.2+\y) rectangle (\x+0.85,0.8+\y);
			}
		}
		\foreach \y in {2.6} {
			\foreach \x in {-0.5,4} { 
				\draw [fill=myblue] (\x +0.15,1.1+\y) rectangle (\x+0.85,1.7+\y);
			}
		}
		\draw[thick, fill=black] (-1.35,2.25) circle (0.1cm); 
		\node at (-0.89,2.2) {$\mathcal{O}_i$};
		\draw[thick, fill=black] (5.85,2.25) circle (0.1cm); 
		\node at (5.45,2.2) {$\mathcal{O}_{j}$};
		\end{tikzpicture}
	\end{center}
	\caption{Tensor network corresponding to the two-point correlation functions $\braket{\Psi^L_t(\mathcal{N})|\mathcal{O}_i\mathcal{O}_j|\Psi^L_t(\mathcal{N})}$ for a solvable initial MPS and $t=2$. We see that almost all the elementary unitary operators have canceled with each other, after repeated application of the graphical identities in Eq.~\eqref{eq:unitarity_pic}, \eqref{eq:dual_unitarity_pic}, and~\eqref{eq:solvability_pic}. The resulting tensor network can be efficiently contracted, as explained in Sec.~\ref{sec:time_ev}.}
	\label{fig:two-point_functions}
\end{figure}
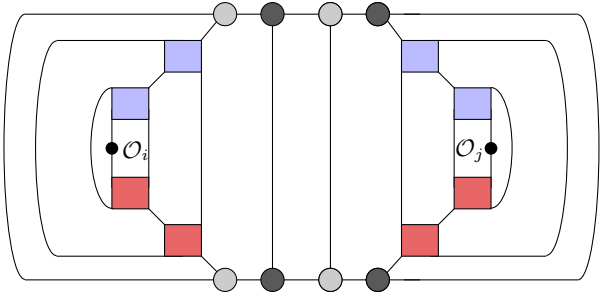

\section{The exact quantum dynamics}
\label{sec:time_ev}

In this section we finally explore the dynamics arising from the solvable MPSs. We focus in particular on three aspects: the thermalization time of local observables, the two-point correlation functions, and the entanglement growth.

\subsection{Local thermalization and the quasi-particle picture}
\label{sec:local_therm}

As we have already discussed, the knowledge of the left and right eigenvectors of the transfer matrix $E(t)$ allows us to compute the expectation values of observables that are supported over finite regions of space. However, it follows from Eq.~\eqref{eq:transverse_contraction} that the dynamics of observables localized at one site is quite trivial. Precisely, for a solvable MPS (evolved at time $t$) $\ket{\Psi_{t}^L(\mathcal{N})}$, we have 
\be
\lim_{L\to\infty}\braket{\Psi^L_t(\mathcal{N})|\mathcal{O}_j|\Psi^L_t(\mathcal{N})}={\rm tr}[\mathcal{O}_j],\quad \forall t,
\ee
where we used Eq.~\eqref{eq:right_left_eigenstate}. Namely, one-point functions remain constant at the infinite-temperature value. 

\begin{figure*}
	\begin{tabular}{lll}
		\includegraphics[width=0.325\textwidth]{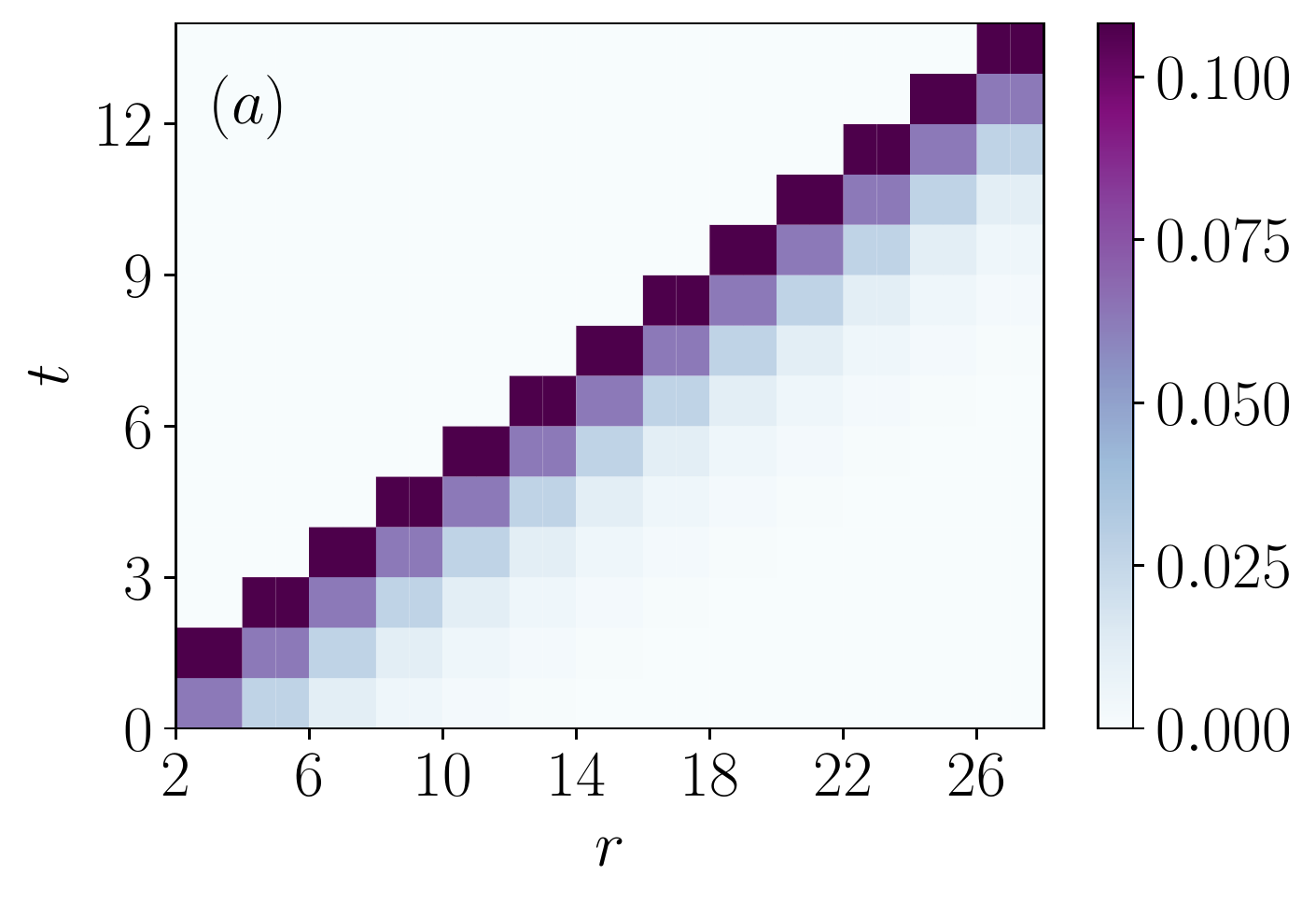} & \includegraphics[width=0.325\textwidth]{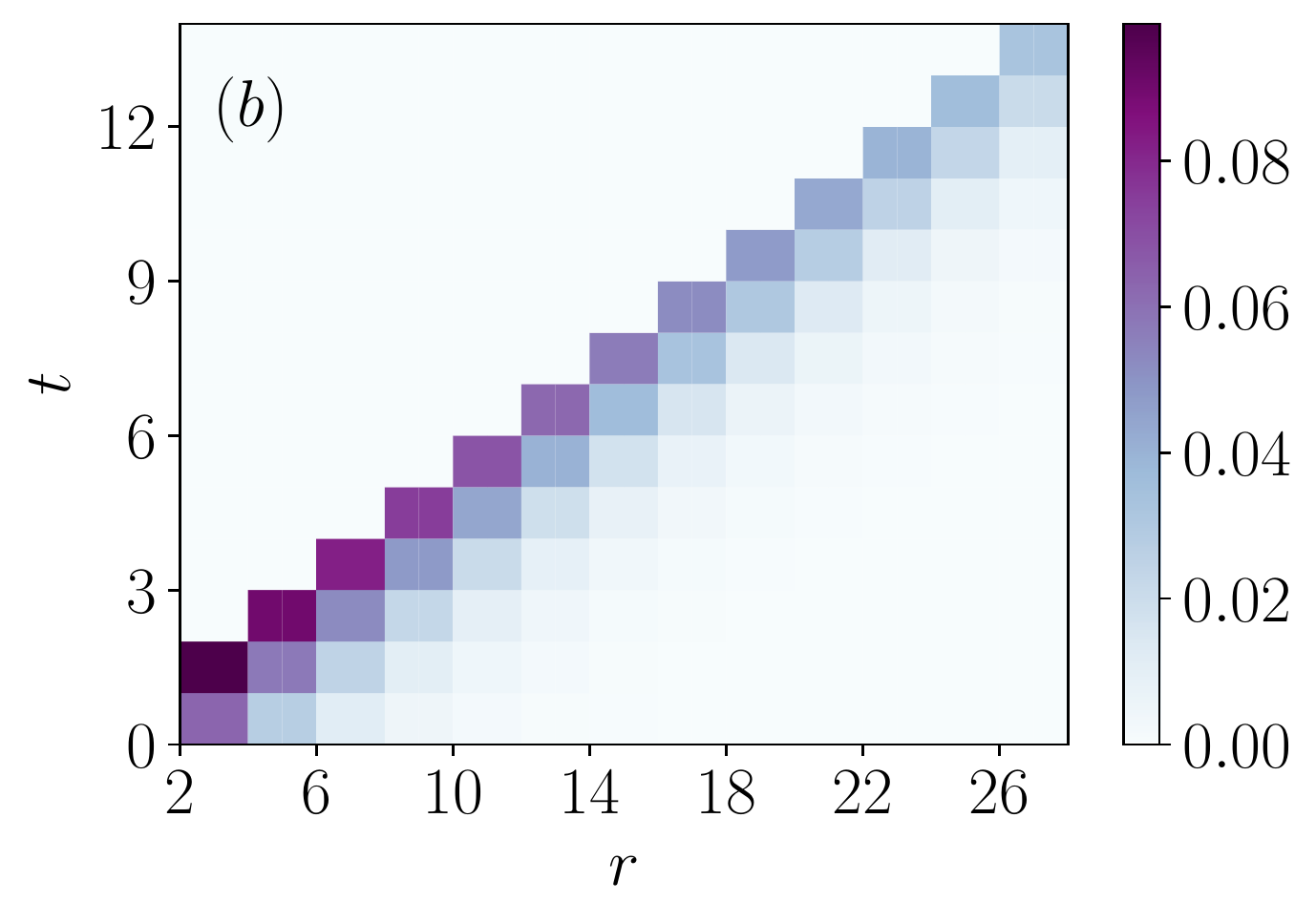}&
		\includegraphics[width=0.325\textwidth]{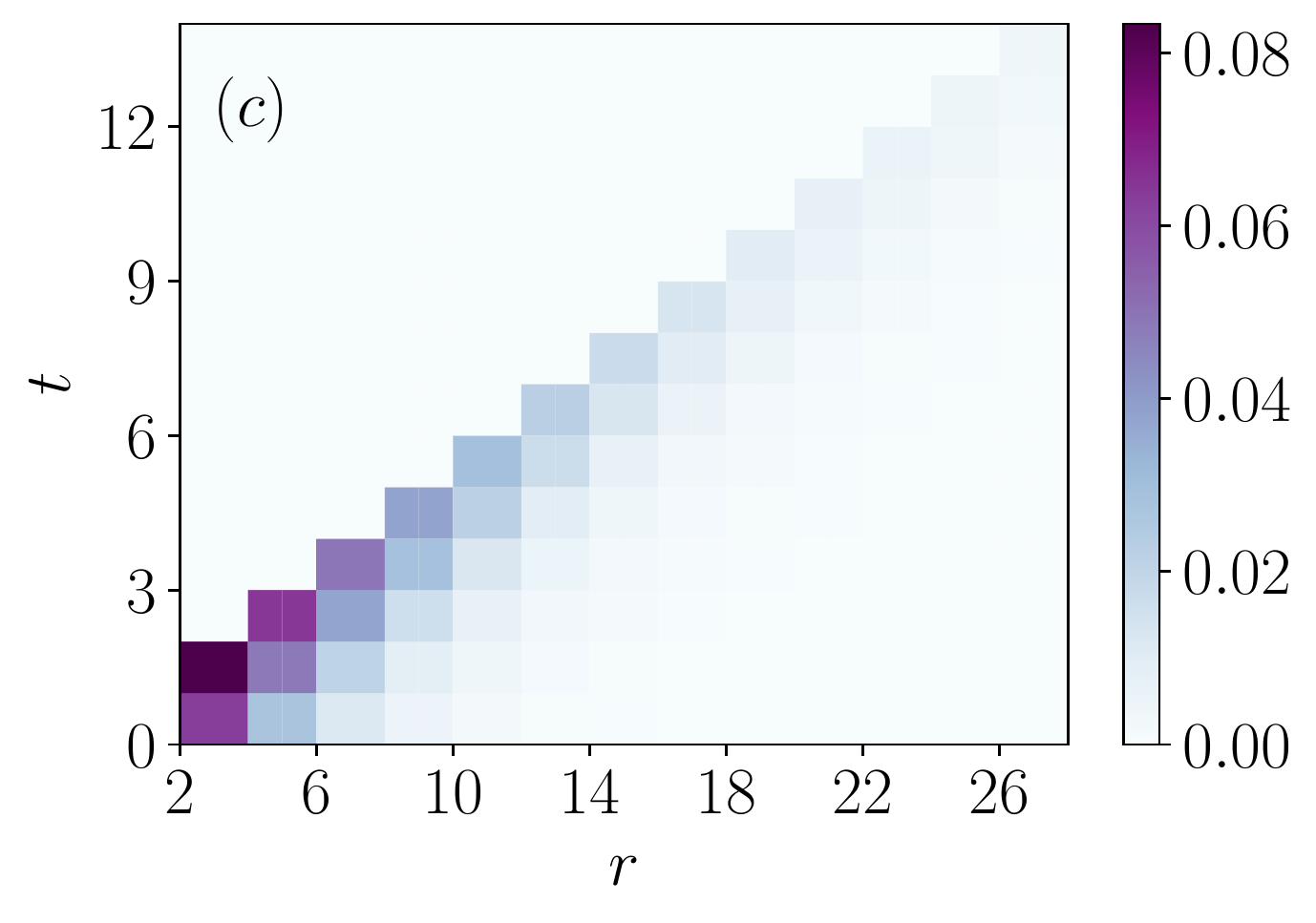}
	\end{tabular}
	\caption{Two-point correlation function $C^{x,x}_0(r,t)$ [as defined in Eq.~\eqref{eq:averaged_c}] for a solvable MPS $\ket{\Psi^L_0(\mathcal{N})}$ with bond dimension $\chi=2$. The initial state corresponds to choosing $u=v=\mathbb{1}$ and $\{K_i\}_{i=1}^3=(0.3,0.5,1.25)$ in Eqs.~\eqref{eq:n_11}--\eqref{eq:n_22}. The dynamics is driven by the quantum circuit corresponding to the self-dual kicked Ising chain, namely we chose $U_{\rm SDKI}$ as defined in Eq.~\eqref{eq:gate_SKDI}, and set the magnetic field to $(a)$: $h=0.0$, $(b)$: $h=0.15$, $(c)$: $h=0.25$.}
	\label{fig:lightcone}
\end{figure*}

\begin{figure*}
	\begin{tabular}{lll}
		\includegraphics[width=0.325\textwidth]{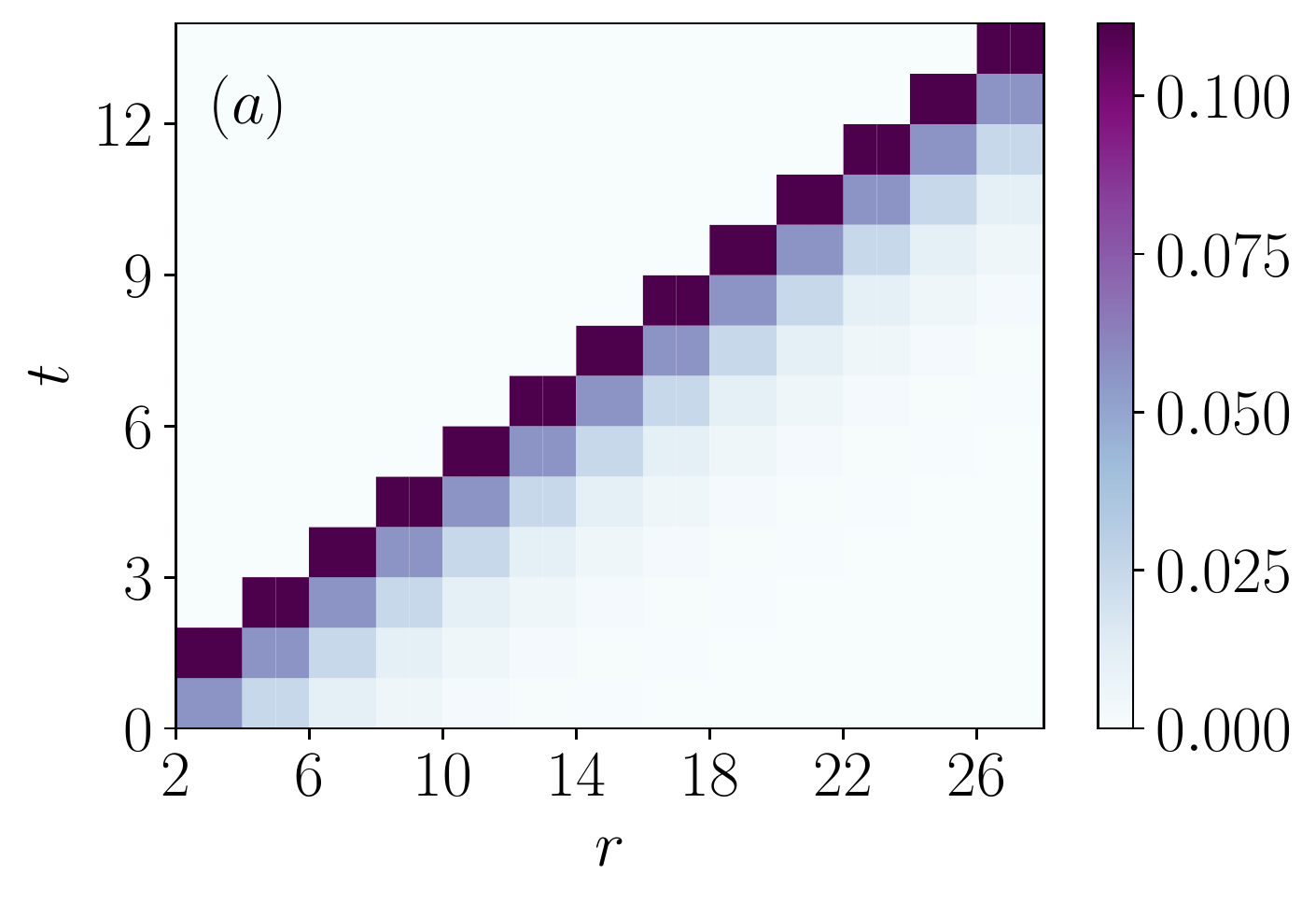} & \includegraphics[width=0.325\textwidth]{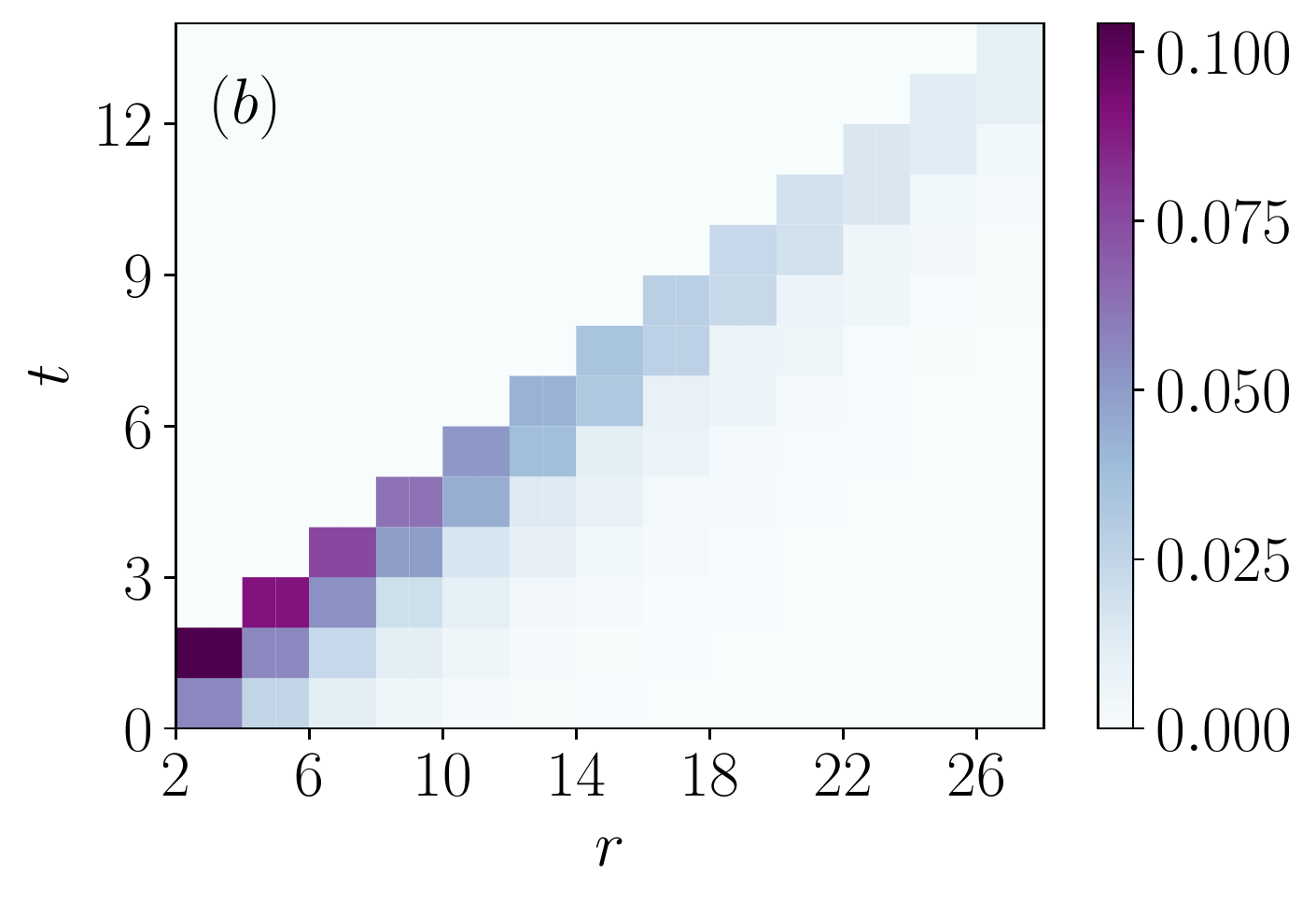}&
		\includegraphics[width=0.325\textwidth]{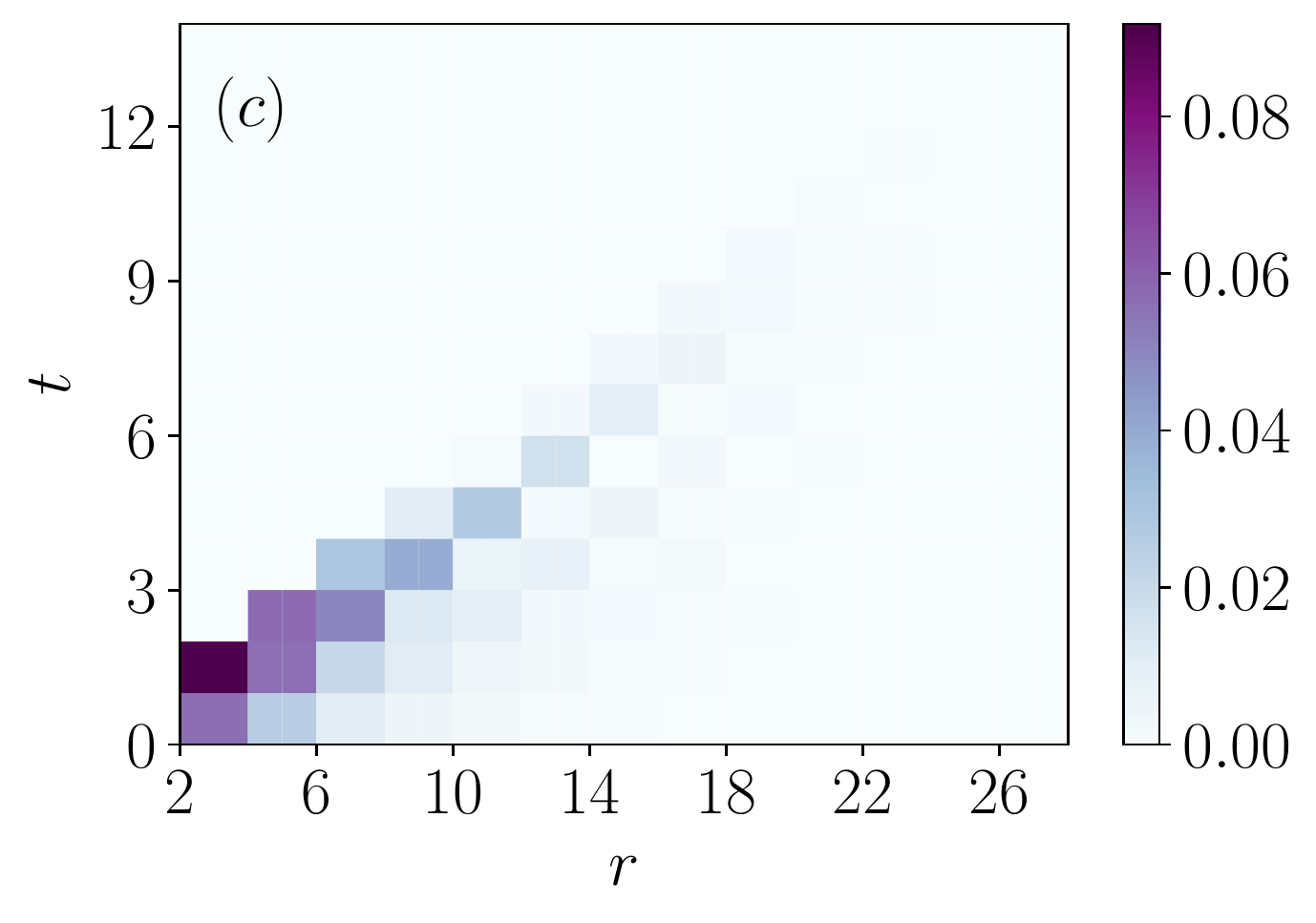}
	\end{tabular}
	\caption{Two-point correlation function $C^{z,z}_0(r,t)$ [as defined in Eq.~\eqref{eq:averaged_c}] for a solvable MPS $\ket{\Psi^L_0(\mathcal{N})}$  with bond dimension $\chi=2$. The initial state corresponds to choosing $u=v=\mathbb{1}$ and $\{K_i\}_{i=1}^3=(0.3,0.5,1.25)$ in Eqs.~\eqref{eq:n_11}--\eqref{eq:n_22}.  The dynamics is given by the dual-unitary gate $U=R(x\vartheta_1,x\vartheta_2) V[J] R(x\vartheta_3,x\vartheta_4)$ where $V[J]$ is defined in Eq.~\eqref{eq:v_mat_dual}, while $R(\alpha,\beta)=r(\alpha)\otimes r(\beta)$ and $r(\alpha)=[(\cos \alpha,\sin \alpha),(-\sin \alpha, \cos \alpha)]$. 
	In the plots, we chose $J=0.3$, and $\{\vartheta_j\}_{j=1}^4=\{0.2,0.3,0.4,0.5\}$. Finally, different subfigures corresponds to $(a)$: $x=0$, $(b)$: $x=0.15$, $(c)$: $x=0.25$.}
	\label{fig:int_lightcone}
\end{figure*}

In general, we can extend this result to local observables that are supported over finite regions of space containing more than one site. Consider the generic operator
\be
\mathcal{O}^{\{\alpha_1:\alpha_R\}}_{\{j_{1}:j_{R}\}}=\mathcal{O}^{\alpha_1}_{j_1}\otimes \cdots \otimes \mathcal{O}^{\alpha_R}_{j_R}\,,
\ee
where $\mathcal{O}^\alpha_{j}\in {\rm End}\left(h_{j}\right)$ and where ${\rm tr}\left[\mathcal{O}^\alpha_{j}\right]=0$. First note that due to even-odd effects, its expectation value on time-evolved solvable MPSs will be always zero unless $j_R- j_1\equiv 1$ (mod $2$). Second, one can prove
\bea
\lim_{L\to\infty}\braket{\Psi^L_t(\mathcal{N})|\mathcal{O}^{\{\alpha_1:\alpha_R\}}_{\{j_{1}:j_{R}\}}|\Psi^L_t(\mathcal{N})}\nonumber\\
={\rm tr}\left[\mathcal{O}^{\{\alpha_1:\alpha_R\}}_{\{j_{1}:j_{R}\}}\right]\,, \quad t\geq t^\ast\,,
\label{eq:thermalization_time}
\eea
where
\be
t^\ast=\frac{j_R-j_1+1}{2}\,.
\label{eq:thermalization_time_star}
\ee
The proof of this can be easily performed graphically. Indeed, for a given observable $\mathcal{O}^{\{\alpha_1:\alpha_R\}}_{\{j_{1}:j_{R}\}}$, one can write down the tensor network corresponding to its expectation value. By a repeated use of the graphical identities in Eqs.~\eqref{eq:unitarity_pic}, \eqref{eq:dual_unitarity_pic} and \eqref{eq:solvability_pic} it is then straightforward to see that if $t>t^\ast$ the tensor network simplifies to a new one, where all the ``upper legs'' of $\mathcal{O}^{\{\alpha_1:\alpha_R\}}_{\{j_{1}:j_{R}\}}$ are connected to its ``lower legs''. In turn, this is exactly the tensor network corresponding to the infinite-temperature expectation value in the r.h.s. of Eq.~\eqref{eq:thermalization_time}. From the graphical representation it is also straightforward to see that the correlation function in Eq.~\eqref{eq:thermalization_time} displays an even-odd effect in time, \emph{i.e.}, it is zero for all even (odd) times if $j_1$ is even (odd) (where we are again considering traceless operators).   

Eq.~\eqref{eq:thermalization_time} implies that, while the time evolution of solvable MPSs is non-trivial, finite regions reach infinite temperature in a time which is proportional to their sizes. It is interesting to observe how this behavior is \emph{exactly} predicted by the standard conformal quasi-particle picture~\cite{calabrese2005evolution}. Indeed, $t^\ast$ in Eq.~\eqref{eq:thermalization_time_star} is exactly the time needed for a pair of quasi-particles produced at the center of the region $R$ and moving at maximal speed (which is $1$ for our choice of units) to exit from it. Finally, we see that local thermalization to infinite temperature takes place irrespective of the presence of local conserved quantities (namely, of the integrability of the unitary dynamics) and that $t^\ast$ does not depend on the unitary gates chosen. In the next section, we will see that qualitative differences emerge in the study of the light-cone spreading of two-point correlation functions for integrable and non-integrable circuits.

\subsection{The two-point correlation functions}
\label{sec:two_point_calculations}

We now move on to examine two-point correlation functions. For concreteness we focus on the case of qubits (although our treatment is valid for general physical local dimension $d$), and consider
\be
\mathcal{C}^{\alpha,\beta}(j,r,t)=\braket{\Psi^L_t(\mathcal{N})|\sigma^{\alpha}_{j}\sigma^{\beta}_{j+r}|\Psi^L_t(\mathcal{N})}\,,
\ee
where $\sigma_j^\alpha$, $\alpha=x,y,z$ are Pauli matrices acting at site $j$, while $\ket{\Psi^L_t(\mathcal{N})}$ is a solvable (injective) initial MPS satisfying Eq.~\eqref{eq:solvability_identity}, evolved at time $t$. From the results of the previous subsection, we have that for fixed $j$, $r$ the function $\mathcal{C}^{\alpha,\beta}(j,r,t)$ will become zero after a time $t^\ast={(r +1)}/{2}$. Here, however, we are interested in its full time evolution, which is non-trivial for generic unitaries~$U$.

It turns out that it is possible to compute exactly $\mathcal{C}^{\alpha,\beta}(j,r,t)$ for arbitrary values of $j$, $r$ and $t$, following an approach similar to the one developed in Ref.~[\onlinecite{bertini2019exact}] for infinite-temperature dynamical two-point functions. The first step consists in simplifying the graphical representation for $\mathcal{C}^{\alpha,\beta}(j,r,t)$  by means of the dual unitarity conditions~\eqref{eq:unitarity_pic}, \eqref{eq:dual_unitarity_pic} and the solvability relation~\eqref{eq:solvability_pic}. By doing this, we obtain the formula
\be
\mathcal{C}^{\alpha,\beta}(j,r,t)=\delta_{r({\rm mod\ }2),1}\delta_{j-t({\rm mod\ }2),1}\widetilde{\mathcal{C}}^{\alpha,\beta}(r,t)\,,
\ee
where
\be
\widetilde{\mathcal{C}}^{\alpha,\beta}(r,t)=
\begin{cases}
	0 & r< 2t+1\,, \\
	\mathcal{D}_1^{\alpha,\beta}(t) & r=2t+1\,,\\
	\mathcal{D}_2^{\alpha,\beta}(r,t) & r>2t+1\,.
\end{cases}
\label{eq:final_result}
\ee
Here $\mathcal{D}_1^{\alpha,\beta}(t)$ and 	$\mathcal{D}_2^{\alpha,\beta}(r,t)$ are functions which admit a simple graphical representation. As an example, $\mathcal{D}_2^{\alpha,\beta}(r,t)$ is depicted in Fig.~\ref{fig:two-point_functions} for $r=7$ and $t=2$. 

The simplified tensor networks associated with $\mathcal{D}_1^{\alpha,\beta}(t)$ and $\mathcal{D}_2^{\alpha,\beta}(r,t)$ can be contracted efficiently, by slightly generalizing the method employed in Ref.~[\onlinecite{bertini2019exact}]. In particular, based on its graphical representation, one can derive the following formula for the function $\mathcal{D}_1^{\alpha,\beta}(t)$
\bea
\mathcal{D}_1^{\alpha,\beta}(t)&=&\frac{1}{\chi }{\rm tr}\left\{\mathcal{P}\left[\widetilde{\mathcal{F}}^{t}\left(\sigma^\alpha\right)\right]\left[\mathcal{F}^{t}\left(\sigma^\beta\right)\right]^T\right\}\,,
\label{eq:d1_function}
\eea
where $\chi$ is the bond dimension of the initial solvable MPS $\ket{\Psi^L_0(\mathcal{N})}$ and $(\cdot)^T$ denotes matrix transposition. Here we introduced the following definitions. First, the functions $\mathcal{F}$, $\widetilde{\mathcal{F}}$ are maps acting on the space of linear operators ${\rm End}(\mathbb{C}^2)$, reading
\be
\mathcal{F}\left[a\right]=\frac{1}{d}\operatorname{tr}_{2}\left[U^{\dagger}( \mathbb{1} \otimes a) U\right]
=\frac{1}{d}
\begin{tikzpicture}[baseline={([yshift=-.5ex]current bounding box.center)},vertex/.style={anchor=base,
	circle,fill=black!25,minimum size=18pt,inner sep=2pt}, scale=0.7]
\draw[=latex] (5.15,0.65)  -- (5.15,1.45);
\draw[=latex] (5.15,3.9)  -- (5.15,3.15);
\draw[=latex] (5.15,1.5)  -- (5.15,3);
\draw[=latex] (5.85,1.5)  -- (5.85,3);
\draw[=latex, ] (5.85,1.115) arc(-90:90: 0.4 and 1.145);
\foreach \y in {0} {
	\foreach \x in {5.} { 
		\draw [fill=myred] (\x +0.15,1.1+\y) rectangle (\x+0.85,1.7+\y);
	}
}
\foreach \y in {2.6} {
	\foreach \x in {5.} { 
		\draw [fill=myblue] (\x +0.15, 0.2+\y) rectangle (\x+0.85,0.8+\y);
	}
}
\draw[thick, fill=black] (5.85,2.25) circle (0.1cm); 
\node at (5.45,2.2) {$a$};
\node at (6.5,2.2) {$\,,$};
\end{tikzpicture}
\label{eq:f_function}
\ee
\be
\widetilde{\mathcal{F}}\left[a\right]=\frac{1}{d}\operatorname{tr}_{1}\left[U^{\dagger}(a \otimes \mathbb{1}) U\right]
=\frac{1}{d}
\begin{tikzpicture}[baseline={([yshift=-.5ex]current bounding box.center)},vertex/.style={anchor=base,
	circle,fill=black!25,minimum size=18pt,inner sep=2pt}, scale=0.7]
\draw[=latex] (-0.65,0.65)  -- (-0.65,1.45);
\draw[=latex] (-0.65,3.9)  -- (-0.65,3.15);
\draw[=latex] (-1.35,1.5)  -- (-1.35,3);
\draw[=latex] (-0.65,1.5)  -- (-0.65,3);
\draw[=latex, ] (-1.35,3.4) arc(90:270: 0.4 and 1.145);
\foreach \y in {0} {
	\foreach \x in {-1.5} { 
		\draw [fill=myred] (\x +0.15,1.1+\y) rectangle (\x+0.85,1.7+\y);
	}
}
\foreach \y in {2.6} {
	\foreach \x in {-1.5} { 
		\draw [fill=myblue] (\x +0.15, 0.2+\y) rectangle (\x+0.85,0.8+\y);
	}
}
\draw[thick, fill=black] (-1.35,2.25) circle (0.1cm); 
\node at (-0.89,2.2) {$a$};
\node at (-0.2,2.2) {$\,,$};
\label{eq:tilde_f_function}
\end{tikzpicture}
\ee
where $d=2$ for qubits. Second, the function $\mathcal{P}$ is a map on the same space which is defined in coordinate space as 
\be
\left(\mathcal{P}\left[a\right]\right)_{m,n}
=
a_{i,j}{\mathcal{N}^{i,m}_{\alpha,\beta}}^*\mathcal{N}^{j,n}_{\alpha,\beta}
=
\begin{tikzpicture}[baseline={([yshift=-.5ex]current bounding box.center)},vertex/.style={anchor=base,
	circle,fill=black!25,minimum size=18pt,inner sep=2pt}, scale=0.7]
		\draw[=latex, ] (2.2,2.0) arc(90:270: 0.4 and 1.125);
		\draw[thick, fill=black] (2.8,0.85) circle (0.1cm); 
		\node at (2.4,0.85) {$a$};
		\draw[=latex] (2.2,-0.25)  -- (4.3,-0.25);
		\draw[=latex] (2.2,2.8-0.8)  -- (4.3,2.8-0.8);
		\draw[=latex] (3.75,-0.25)  -- (4.2,0.25);
		\draw[=latex] (3.75,4.85-2.8)  -- (4.2,4.25-2.8);
		\draw[=latex] (2.8,-0.25)  -- (2.8,2.);
		\foreach \y in {-0.25,2.} {
			\draw[fill=white!80!black] (2.8,+\y) circle (.22);
			\draw[fill=white!35!black] (3.7,+\y) circle (.22);
		}
		\draw[=latex, ] (4.3,-0.25) arc(-90:90: 0.4 and 1.125);
		\node at (4.3,0.4) {$n$};
		\node at (4.3,1.25) {$m$};
		\node at (5.,0.85) {$\,,$};
\end{tikzpicture}
\ee
where repeated indices are summed over. 

From the above definitions, it is clear that Eq.~\eqref{eq:d1_function} can be easily evaluated numerically, with a little computational effort. An expression with a similar structure can be derived for the function $\mathcal{D}_2^{\alpha,\beta}(r,t)$. In particular, we have
\bea
\mathcal{D}^{\alpha,\beta}_2(r,t)=\frac{1}{\chi}
{\rm tr}\left\{\widetilde{\mathcal{G}}\left[\widetilde{\mathcal{F}}^{t}\left(\sigma^\alpha\right)\right]\right.\nonumber \\
\left. \mathcal{E}_\mathcal{N}^{(r-2t-3)/2}\left[\mathcal{G}\left[\mathcal{F}^{t}\left(\sigma^\beta\right)\right]^T\right]\right\}\,.
\label{eq:d_2_function}
\eea
Here $\mathcal{F}$, $\widetilde{\mathcal{F}}$ are given in Eq.~\eqref{eq:f_function} and \eqref{eq:tilde_f_function}, while $\mathcal{E}_\mathcal{N}$ reads
\be
\mathcal{E}_{\mathcal{M}}(X)=\sum_{j,k=1}^d \mathcal{M}^{(j,k)} X \left(\mathcal{M}^{(j,k)}\right)^\dagger\,.
\ee
Next, the functions $\mathcal{G}$ and $\widetilde{\mathcal{G}}$ are maps acting on the space of linear operators ${\rm End}(\mathbb{C}^2)$, and are defined in coordinate space as
\be
\left(\mathcal{G}\left[a\right]\right)_{m,n}
=
a_{i,j}\left[\mathcal{N}^{b,i}\right]^\ast_{m,\alpha}\mathcal{N}^{b,j}_{n,\alpha}
=
\begin{tikzpicture}[baseline={([yshift=-.5ex]current bounding box.center)},vertex/.style={anchor=base,
	circle,fill=black!25,minimum size=18pt,inner sep=2pt}, scale=0.7]
\draw[thick, fill=black] (3.7,0.85) circle (0.1cm); 
\node at (4.,0.85) {$a$};
\draw[=latex] (2.2,-0.25)  -- (4.3,-0.25);
\draw[=latex] (2.2,2.8-0.8)  -- (4.3,2.8-0.8);
\draw[=latex] (2.8,-0.25)  -- (2.8,2.);
\draw[=latex] (3.7,-0.25)  -- (3.7,2.);
\foreach \y in {-0.25,2.} {
	\draw[fill=white!80!black] (2.8,+\y) circle (.22);
	\draw[fill=white!35!black] (3.7,+\y) circle (.22);
}
\draw[=latex, ] (4.3,-0.25) arc(-90:90: 0.4 and 1.125);
\node at (1.8,-0.2) {$n$};
\node at (1.8,2) {$m$};
\node at (5.,0.85) {$\,,$};
\end{tikzpicture}
\ee
and
\be
\left(\widetilde{\mathcal{G}}\left[a\right]\right)_{m,n}
=
a_{i,j}\left[\mathcal{N}^{i,b}\right]^\ast_{\alpha,m}\mathcal{N}^{j,b}_{\alpha,n}
=
\begin{tikzpicture}[baseline={([yshift=-.5ex]current bounding box.center)},vertex/.style={anchor=base,
	circle,fill=black!25,minimum size=18pt,inner sep=2pt}, scale=0.7]
\draw[=latex, ] (2.2,2.0) arc(90:270: 0.4 and 1.125);
\draw[thick, fill=black] (2.8,0.85) circle (0.1cm); 
\node at (2.4,0.85) {$a$};
\draw[=latex] (2.2,-0.25)  -- (4.3,-0.25);
\draw[=latex] (2.2,2.8-0.8)  -- (4.3,2.8-0.8);
\draw[=latex] (3.7,-0.25)  -- (3.7,2.);
\draw[=latex] (2.8,-0.25)  -- (2.8,2.);
\foreach \y in {-0.25,2.} {
	\draw[fill=white!80!black] (2.8,+\y) circle (.22);
	\draw[fill=white!35!black] (3.7,+\y) circle (.22);
}
\node at (4.65,-0.3) {$n$};
\node at (4.65,2.) {$m$};
\node at (5.,0.85) {$\,,$};
\end{tikzpicture}
\ee
where repeated indices as summed over.  Note that also the function $\mathcal{E}_\mathcal{N}$ admits a convenient graphical representation, which reads
\be
\left(\mathcal{E}_\mathcal{N}\left[a\right]\right)_{m,n}=
\begin{tikzpicture}[baseline={([yshift=-.5ex]current bounding box.center)},vertex/.style={anchor=base,
	circle,fill=black!25,minimum size=18pt,inner sep=2pt}, scale=0.7]
\draw[thick, fill=black] (4.7,0.85) circle (0.1cm); 
\node at (4.4,0.85) {$a$};
\draw[=latex] (2.2,-0.25)  -- (4.3,-0.25);
\draw[=latex] (2.2,2.8-0.8)  -- (4.3,2.8-0.8);
\draw[=latex] (2.8,-0.25)  -- (2.8,2.);
\draw[=latex] (3.7,-0.25)  -- (3.7,2.);
\foreach \y in {-0.25,2.} {
	\draw[fill=white!80!black] (2.8,+\y) circle (.22);
	\draw[fill=white!35!black] (3.7,+\y) circle (.22);
}
\draw[=latex, ] (4.3,-0.25) arc(-90:90: 0.4 and 1.125);
\node at (1.8,-0.2) {$n$};
\node at (1.8,2) {$m$};
\node at (5.,0.85) {$\,.$};
\end{tikzpicture}
\ee
Once again, the proof of Eq.~\eqref{eq:d_2_function} can be easily obtained by graphical inspection, and since it presents no difficulty we omit it here.

Putting all together, Eq.~\eqref{eq:final_result} provides an efficient formula for the computation of two-point correlation functions, which allowed us to produce numerical results for different solvable initial states $\ket{\Psi^L_0(\mathcal{N})}$ and dual-unitary operators $U$. Examples of our findings are displayed in Fig.~\ref{fig:lightcone} and \ref{fig:int_lightcone}, where we report data for different choices of the latter, and the same solvable MPS. Note that in order to remove even-odd effects, we have plotted correlations averaged over neighboring sites, namely
\be
\mathcal{C}_{0}^{\alpha,\beta}(r,t)=\frac{1}{4}\sum_{x,y=\pm1} \mathcal{C}^{\alpha,\beta}(j+x,r+y,t)\,,
\label{eq:averaged_c}
\ee
which only depend on the relative distance $r$ of the two Pauli matrices.

In Fig.~\ref{fig:lightcone}$(a)$, we report the two-point correlation function $\mathcal{C}_{0}^{x,x}(r,t)$ for the integrable point of the self-dual kicked Ising chain, \emph{cf.} Eq.~\eqref{eq:gate_SKDI}: we see that there is no decay along the main light-cone.  Conversely, in Fig.~\ref{fig:lightcone}$(b)$,$(c)$ we report data for unitary gates where integrability is broken by increasing the value of the magnetic field $h$ and in this case the plots clearly display an exponential decay in time. In fact, from the explicit formula~\eqref{eq:d1_function} and the results of Ref.~[\onlinecite{bertini2019exact}] this behavior is expected. Indeed, in Ref.~[\onlinecite{bertini2019exact}] unitary gates were classified in four classes of increasing level of ergodicity, depending on the behavior of dynamical infinite-temperature correlation functions. Even though the formula for $\mathcal{C}_{0}^{\alpha,\beta}(r,t)$ along the main light-cone is different from the one found in Ref.~[\onlinecite{bertini2019exact}], both of them involve a repeated application of the function ($1$-qudit channel) $\mathcal{F}$ defined in Eq.~\eqref{eq:f_function}, and hence depend essentially on the spectrum of ${\cal F}$. Accordingly, the classification of unitary gates reported in Ref.~[\onlinecite{bertini2019exact}] not only distinguishes the behavior of dynamical correlations at infinite temperature, but also applies to predict the qualitative features of two-point functions during the quantum dynamics arising from solvable initial MPSs . 

In Fig.~\ref{fig:int_lightcone} similar data are reported for $\mathcal{C}_{0}^{z,z}(r,t)$ and different dual-unitary quantum circuits. In particular, Fig.~\ref{fig:int_lightcone}$(a)$ displays the case of the integrable evolution corresponding to the XXZ gate in Eq.~\eqref{eq:gate_xxz}, while in Fig.~\ref{fig:int_lightcone}$(b)$ and Fig.~\ref{fig:int_lightcone}$(c)$ integrability is broken by choosing non-trivial matrices $u_{\pm}, v_{\pm}$ in Eq.~\eqref{eq:dualunitaryU}. As for Fig.~\ref{fig:lightcone}, we see that a breaking of integrability corresponds to an exponential decay of the correlations along the light-cone. 

Finally, it is important to stress that the analytic formula in Eq.~\eqref{eq:final_result} does not predict a broadening of the light-cone during the time evolution, as we also observe from Figs.~\ref{fig:lightcone} and \ref{fig:int_lightcone}. Once again, this is consistent with an exact CFT picture of non-interacting quasi-particles that are created in pairs at time $t=0$ and spread ballistically for $t>0$~[\onlinecite{calabrese2005evolution}]. 

\subsection{The entanglement growth}
\label{sec:entanglement_growth}

As a final aspect of the dynamics arising from solvable MPSs, we discuss the spreading of entanglement. In particular, we consider the setting already studied in Ref.~[\onlinecite{bertini2019entanglement}] for the special case of the self-dual kicked Ising chain. We focus on a system of size $2L$ with periodic boundaries, and compute the time evolution of the entanglement of the subsystem $A$ --- composed by a connected block of $\ell$ sites -- and the rest, in the limit $L \to \infty$ (see also Ref.~[\onlinecite{gopalakrishnan2019unitary}], where a similar analysis was carried out for the entanglement of half chain). The initial state is given by a solvable MPS $\ket{\Psi^L_0(\mathcal{N})}$ and the system is evolved using a dual-unitary quantum circuit. As it is customary, in order to measure the spreading of entanglement we study the growth of the R\'enyi entropies
\be
S_{A}^{(\alpha)}(t)=\frac{1}{1-\alpha} \log \operatorname{tr}\left[\left(\rho_{A}(t)\right)^{\alpha}\right], \quad \alpha>0\,,
\label{eq:renyi}
\ee
where $\rho_{A}(t)$ is the density matrix reduced to the subsystem $A=[j_1\,,j_1+\ell-1]$ containing $\ell$ neighboring sites. At this point we note that the  entanglement growth displays an even/odd effect in space and time. In order to simplify our discussion, in the rest of this section we will consider the case of $t$ even, and choose $j_1$ to be odd, although a very similar treatment can be carried out, with minor modifications, when either $t$ is odd or $j_1$ is even. In the case of $t$ even and $j_1$ odd, we can further assume, without loss of generality, $\ell$ to be even. Indeed, if $\ell=2k+1$, it is easy to see that $\rho_{A_{2k+1}}(t)=\rho_{A_{2k}}(t)\otimes (\mathbb{1}_{2k+1}/d)$, where $A_k$ is the connected region containing sites from $j_1$ to $j_1+k-1$, while $\mathbb{1}_{k}$ is the identity acting on site $j_1+k-1$. Using \eqref{eq:solvability_pic} and the unitarity of the gates we can write the thermodynamic limit of the reduced density matrix as 
\be
\!\!\!\!\!\rho_A(t)\!\!=\!\!\frac{1}{\chi}\,
\begin{tikzpicture}[baseline=(current  bounding  box.center), scale=0.35]
\def\eps{4.75};
\def\elle{2};
\def\tmax{2};
\draw[=latex] (-10.25,3.9)  -- (5.25,3.9);
\draw[=latex] (-10,3.9)  -- (-10,-3.9+\eps);
\draw[=latex] (5,3.9)  -- (5,-3.9+\eps);
\draw[=latex, rounded corners] (-8.5,3.9)  -- (-10.25,3.9) -- (-10.25,-3.9+\eps) -- (-8.5,-3.9+\eps);
\draw[=latex, rounded corners] (3.5,3.9)  -- (5.25,3.9) -- (5.25,-3.9+\eps) -- (3.5,-3.9+\eps); 
\draw[=latex, rounded corners] (-9.5+0.5,5.1)  -- (-0.2-10.3,5.1) -- (-0.2-10.3,-5.1+\eps) -- (-9.5+0.5,-5.1+\eps);
\draw[=latex, rounded corners] (4.5-0.5,5.1)  -- (0.2+5.3,5.1) -- (0.2+5.3,-5.1+\eps) -- (4.5-0.5,-5.1+\eps);
\draw[=latex, rounded corners] (-9.5+1.5,6.1)  -- (-0.4-10.3,6.1) -- (-0.4-10.3,-6.1+\eps) -- (-9.5+1.5,-6.1+\eps);
\draw[=latex, rounded corners] (4.5-1.5,6.1)  -- (0.4+5.3,6.1) -- (0.4+5.3,-6.1+\eps) -- (4.5-1.5,-6.1+\eps);
\draw[=latex, rounded corners] (-9.5+2.5,7.1)  -- (-0.6-10.3,7.1) -- (-0.6-10.3,-7.1+\eps) -- (-9.5+2.5,-7.1+\eps);
\draw[=latex, rounded corners] (4.5-2.5,7.1)  -- (0.6+5.3,7.1) -- (0.6+5.3,-7.1+\eps) -- (4.5-2.5,-7.1+\eps);

\foreach \x in {-9,-7,-5,-3,-1,1,3,5} {
	\draw[fill=white!80!black] (\x-0.025-1,3.9) circle (.22);
	\draw[fill=white!35!black] (\x+1.025-1,3.9) circle (.22);
}
\foreach \yy[evaluate=\yy  as \aa using int(-\tmax-\elle/2+\yy), evaluate=\yy as \bb using int(\tmax+\elle/2-\yy)] in {0,...,1}
\foreach \x in {\aa,...,\bb} {
	\draw[=latex] (2*\x-3+0.85,4.3+2*\yy+0.6)  -- (2*\x-3+0.85+0.15,4.3+2*\yy+0.6+0.2);
	\draw[=latex] (2*\x-3 +0.15, 4.9+2*\yy)  -- (2*\x-3, 4.9+2*\yy+0.2);
	\draw[=latex] (2*\x-3 +0.15, 4.3+2*\yy)  -- (2*\x-3, 4.3+2*\yy-0.2);
	\draw[=latex] (2*\x-3+0.85,4.3+2*\yy)  -- (2*\x-3+1,4.3+2*\yy-0.2);
	\draw [fill=myred] (2*\x-3 +0.15, 4.3+2*\yy) rectangle (2*\x-3+0.85,4.3+2*\yy+0.6);
}
\foreach \yy[evaluate=\yy  as \aa using int(-\tmax-\elle/2+\yy), evaluate=\yy as \bb using int(\tmax+\elle/2-\yy+1)] in {1,...,\tmax}
\foreach \x in {\aa,...,\bb} {
	\draw[=latex] (2*\x+0.85-4,4.9+2*\yy-1)  -- (2*\x+0.85-4+0.15,4.9+2*\yy-1+0.2);
	\draw[=latex] (2*\x +0.15-4, 4.9+2*\yy-1)  -- (2*\x-4, 4.9+2*\yy-1+0.2);
	\draw[=latex] (2*\x +0.15-4, 4.3+2*\yy-1)  -- (2*\x -4, 4.3+2*\yy-1-0.2);
	\draw[=latex] (2*\x+0.85-4,4.3+2*\yy-1)  -- (2*\x+1-4,4.3+2*\yy-1-0.2);
	\draw [fill=myred] (2*\x +0.15-4, 4.3+2*\yy-1) rectangle (2*\x+0.85-4,4.9+2*\yy-1);
}

\draw[=latex] (-10.25,-3.9+\eps)  -- (5.25,-3.9+\eps);
\foreach \x in {-9,-7,-5,-3,-1,1,3,5} {
	\draw[fill=white!80!black] (\x-0.025-1,-3.9+\eps) circle (.22);
	\draw[fill=white!35!black] (\x+1.025-1,-3.9+\eps) circle (.22);
}
\foreach \yy[evaluate=\yy  as \aa using int(-\tmax-\elle/2+\yy), evaluate=\yy as \bb using int(\tmax+\elle/2-\yy)] in {0,...,1}
\foreach \x in {\aa,...,\bb} {
	\draw[=latex] (2*\x-3+0.85,-4.3-2*\yy-0.6+\eps)  -- (2*\x-3+0.85+0.15,-4.3-2*\yy-0.6-0.2+\eps);
	\draw[=latex] (2*\x-3 +0.15, -4.9-2*\yy+\eps)  -- (2*\x-3, -4.9-2*\yy-0.2+\eps);
	\draw[=latex] (2*\x-3 +0.15, -4.3-2*\yy+\eps)  -- (2*\x-3, -4.3-2*\yy+0.2+\eps);
	\draw[=latex] (2*\x-3+0.85,-4.3-2*\yy+\eps)  -- (2*\x-3+1,-4.3-2*\yy+0.2+\eps);
	\draw [fill=myblue] (2*\x-3 +0.15, -4.3-2*\yy+\eps) rectangle (2*\x-3+0.85,-4.3-2*\yy-0.6+\eps);
}
\foreach \yy[evaluate=\yy  as \aa using int(-\tmax-\elle/2+\yy), evaluate=\yy as \bb using int(\tmax+\elle/2-\yy+1)] in {1,...,\tmax}
\foreach \x in {\aa,...,\bb} {
	\draw[=latex] (2*\x+0.85-4,-4.9-2*\yy+1+\eps)  -- (2*\x+0.85-4+0.15,-4.9-2*\yy+1-0.2+\eps);
	\draw[=latex] (2*\x +0.15-4, -4.9-2*\yy+1+\eps)  -- (2*\x-4, -4.9-2*\yy+1-0.2+\eps);
	\draw[=latex] (2*\x +0.15-4, -4.3-2*\yy+1+\eps)  -- (2*\x -4, -4.3-2*\yy+1+0.2+\eps);
	\draw[=latex] (2*\x+0.85-4,-4.3-2*\yy+1+\eps)  -- (2*\x+1-4,-4.3-2*\yy+1+0.2+\eps);
	\draw [fill=myblue] (2*\x +0.15-4, -4.3-2*\yy+1+\eps) rectangle (2*\x+0.85-4,-4.9-2*\yy+1+\eps);
}
\draw [thick, decorate, decoration={brace,amplitude=5pt,raise=-4pt},yshift=0pt]
(-6,\tmax+\eps+2) -- (1,\tmax+\eps+2) node [black,midway,yshift=0.25cm] {$\ell$};
\draw [thick, decorate, decoration={brace,amplitude=5pt,mirror,raise=4pt},yshift=0pt]
(5.9,\tmax+2) -- (5.9,\tmax+\eps+1) node [black,midway,xshift=0.45cm] {$t$};
\draw [thick, decorate, decoration={brace,amplitude=5pt,mirror,raise=4pt},yshift=0pt]
(-10,\tmax+\eps-2.75) -- (5,\tmax+\eps-2.75) node [black,midway,yshift=-0.5cm] {$\ell+2t$};
\node at (0,-6.25) {};
\end{tikzpicture}
\!\!\!\!\!\!\!\!\!,
\ee
where we used that, thanks to Theorem~\ref{th:classification}, the matrix $S$ in \eqref{eq:solvability_pic} can be chosen equal to the identity. This circuit can be further contracted by repeated use of the initial state's solvability condition~\eqref{eq:solvability_pic} combined with the dual-unitarity property~\eqref{eq:dual_unitarity_pic}. Two different results are obtained depending on whether or not $2t$ is larger than $\ell$. Specifically, for $2t\leq \ell$ we find 
\be
\!\!\!\!\!\rho_A(t)\!=\frac{1}{d^{2t}\chi}
\begin{tikzpicture}[baseline=(current  bounding  box.center), scale=0.35]
\def\eps{4.75};
\def\elle{4};
\def\tmax{1};
\draw[=latex] (-4.25,3.9)  -- (-0.75,3.9);
\draw[=latex, rounded corners] (-4.25,3.9)  -- (-4.75,3.9) -- (-4.75,-3.9+\eps) -- (-4.25,-3.9+\eps);
\draw[=latex, rounded corners] (-0.75,3.9)  -- (-.25,3.9) -- (-.25,-3.9+\eps) -- (-.75,-3.9+\eps);
\draw[=latex, rounded corners] (-5,4.1) -- (-5,-4.1+\eps);
\draw[=latex, rounded corners] (0,4.1) -- (0,-4.1+\eps);	
\draw[=latex, rounded corners] (-6,5.1) -- (-6,-5.1+\eps);
\draw[=latex, rounded corners] (1,5.1) -- (1,-5.1+\eps);

\foreach \x in {-3,-1} {
	\draw[fill=white!80!black] (\x-0.025-1,3.9) circle (.22);
	\draw[fill=white!35!black] (\x+1.025-1,3.9) circle (.22);
}
\foreach \yy[evaluate=\yy  as \aa using int(-\tmax+\elle/2+\yy), evaluate=\yy as \bb using int(\tmax-\elle/2-\yy)] in {0,...,0}
\foreach \x in {\aa,...,\bb} {
	\draw[=latex] (2*\x-3+0.85,4.3+2*\yy+0.6)  -- (2*\x-3+0.85+0.15,4.3+2*\yy+0.6+0.2);
	\draw[=latex] (2*\x-3 +0.15, 4.9+2*\yy)  -- (2*\x-3, 4.9+2*\yy+0.2);
	\draw[=latex] (2*\x-3 +0.15, 4.3+2*\yy)  -- (2*\x-3, 4.3+2*\yy-0.2);
	\draw[=latex] (2*\x-3+0.85,4.3+2*\yy)  -- (2*\x-3+1,4.3+2*\yy-0.2);
	\draw [fill=myred] (2*\x-3 +0.15, 4.3+2*\yy) rectangle (2*\x-3+0.85,4.3+2*\yy+0.6);
}
\foreach \yy[evaluate=\yy  as \aa using int(-\tmax+\elle/2+\yy), evaluate=\yy as \bb using int(\tmax-\elle/2-\yy+1)] in {1,...,\tmax}
\foreach \x in {\aa,...,\bb} {
	\draw[=latex] (2*\x+0.85-4,4.9+2*\yy-1)  -- (2*\x+0.85-4+0.15,4.9+2*\yy-1+0.2);
	\draw[=latex] (2*\x +0.15-4, 4.9+2*\yy-1)  -- (2*\x-4, 4.9+2*\yy-1+0.2);
	\draw[=latex] (2*\x +0.15-4, 4.3+2*\yy-1)  -- (2*\x -4, 4.3+2*\yy-1-0.2);
	\draw[=latex] (2*\x+0.85-4,4.3+2*\yy-1)  -- (2*\x+1-4,4.3+2*\yy-1-0.2);
	\draw [fill=myred] (2*\x +0.15-4, 4.3+2*\yy-1) rectangle (2*\x+0.85-4,4.9+2*\yy-1);
}

\draw[=latex] (-4.5,-3.9+\eps)  -- (-0.5,-3.9+\eps);
\foreach \x in {-3,-1} {
	\draw[fill=white!80!black] (\x-0.025-1,-3.9+\eps) circle (.22);
	\draw[fill=white!35!black] (\x+1.025-1,-3.9+\eps) circle (.22);
}
\foreach \yy[evaluate=\yy  as \aa using int(-\tmax+\elle/2+\yy), evaluate=\yy as \bb using int(\tmax-\elle/2-\yy)] in {0,...,0}
\foreach \x in {\aa,...,\bb} {
	\draw[=latex] (2*\x-3+0.85,-4.3-2*\yy-0.6+\eps)  -- (2*\x-3+0.85+0.15,-4.3-2*\yy-0.6-0.2+\eps);
	\draw[=latex] (2*\x-3 +0.15, -4.9-2*\yy+\eps)  -- (2*\x-3, -4.9-2*\yy-0.2+\eps);
	\draw[=latex] (2*\x-3 +0.15, -4.3-2*\yy+\eps)  -- (2*\x-3, -4.3-2*\yy+0.2+\eps);
	\draw[=latex] (2*\x-3+0.85,-4.3-2*\yy+\eps)  -- (2*\x-3+1,-4.3-2*\yy+0.2+\eps);
	\draw [fill=myblue] (2*\x-3 +0.15, -4.3-2*\yy+\eps) rectangle (2*\x-3+0.85,-4.3-2*\yy-0.6+\eps);
}
\foreach \yy[evaluate=\yy  as \aa using int(-\tmax+\elle/2+\yy), evaluate=\yy as \bb using int(\tmax-\elle/2-\yy+1)] in {1,...,\tmax}
\foreach \x in {\aa,...,\bb} {
	\draw[=latex] (2*\x+0.85-4,-4.9-2*\yy+1+\eps)  -- (2*\x+0.85-4+0.15,-4.9-2*\yy+1-0.2+\eps);
	\draw[=latex] (2*\x +0.15-4, -4.9-2*\yy+1+\eps)  -- (2*\x-4, -4.9-2*\yy+1-0.2+\eps);
	\draw[=latex] (2*\x +0.15-4, -4.3-2*\yy+1+\eps)  -- (2*\x -4, -4.3-2*\yy+1+0.2+\eps);
	\draw[=latex] (2*\x+0.85-4,-4.3-2*\yy+1+\eps)  -- (2*\x+1-4,-4.3-2*\yy+1+0.2+\eps);
	\draw [fill=myblue] (2*\x +0.15-4, -4.3-2*\yy+1+\eps) rectangle (2*\x+0.85-4,-4.9-2*\yy+1+\eps);
}
\draw [thick, decorate, decoration={brace,amplitude=5pt,raise=-2pt},yshift=0pt]
(-6,\tmax+\eps+1) -- (1,\tmax+\eps+1) node [black,midway,yshift=0.4cm] {$\ell$};
\draw [thick, decorate, decoration={brace,amplitude=5pt,mirror,raise=0pt},yshift=0pt]
(1.75,\tmax+3) -- (1.75,\tmax+\eps+0.25) node [black,midway,xshift=0.35cm] {$t$};
\draw [thick, decorate, decoration={brace,amplitude=5pt,mirror,raise=4pt},yshift=0pt]
(-4.25,\tmax+\eps-1.75) -- (-0.75,\tmax+\eps-1.75) node [black,midway,yshift=-0.5cm] {$\ell-2t$};
\node at (0,-4.25) {};
\end{tikzpicture}
\!\!\!,
\label{eq:rhol>2t}
\ee
while for $2t>\ell$ the circuit simplifies to 
\be
\rho_A(t)=\frac{1}{d^\ell} \mathbbm{1}_\ell.
\label{eq:rho2t>l}
\ee 
The explicit form \eqref{eq:rhol>2t}--\eqref{eq:rho2t>l} of the reduced density matrix directly gives  
\begin{align}
	\!\!S_{A}^{(\alpha)}\!(t)\!=& {\rm min}(2t,\ell)\log d+\frac{\log[\operatorname{tr}\left[ \mathbb O^\alpha_{\ell-2t}\right]]}{1-\alpha} \,\theta_{\rm H}[\ell-2t],
	\label{eq:entropyresult}
\end{align}
where $\theta_{\rm H}[x]$ is the step function (with $\theta_{\rm H}[0]=0$) and we introduced the $d^x$-dimensional matrix 
\be
\mathbb O_{x} =\frac{1}{\chi}\,
\begin{tikzpicture}[baseline=(current  bounding  box.center), scale=0.5]
\def\eps{4.75};
\def\elle{4};
\def\tmax{1};
\draw[=latex] (-4.25,3.9)  -- (-0.75,3.9);
\draw[=latex, rounded corners] (-4.25,3.9)  -- (-4.75,3.9) -- (-4.75,-3.9+\eps) -- (-4.25,-3.9+\eps);
\draw[=latex, rounded corners] (-0.75,3.9)  -- (-.25,3.9) -- (-.25,-3.9+\eps) -- (-.75,-3.9+\eps);
\foreach \x in {-3,-1} {
	\draw[=latex] (\x-0.025-1,3.9)  -- (\x-0.025-1,3.9+0.5);
	\draw[fill=white!80!black] (\x-0.025-1,3.9) circle (.22);
	\draw[=latex] (\x+1.025-1,3.9)  -- (\x+1.025-1,3.9+0.5);
	\draw[fill=white!35!black] (\x+1.025-1,3.9) circle (.22);
}
\draw[=latex] (-4.5,-3.9+\eps)  -- (-0.5,-3.9+\eps);
\foreach \x in {-3,-1} {
	\draw[=latex] (\x-0.025-1,-3.9+\eps)  -- (\x-0.025-1,-3.9+\eps-0.5);
	\draw[fill=white!80!black] (\x-0.025-1,-3.9+\eps) circle (.22);
	\draw[=latex] (\x+1.025-1,-3.9+\eps)  -- (\x+1.025-1,-3.9+\eps-0.5);
	\draw[fill=white!35!black] (\x+1.025-1,-3.9+\eps) circle (.22);
}
\draw [thick, decorate, decoration={brace,amplitude=5pt,raise=4pt},yshift=0pt]
(-4.25,\tmax+\eps-1.6) -- (-0.75,\tmax+\eps-1.6) node [black,midway,yshift=0.5cm] {$x$};
\node at (0,-1.25) {};
\end{tikzpicture}.
\ee
The result \eqref{eq:entropyresult} follows directly from the observation that the reduced density matrix  \eqref{eq:rhol>2t} is unitary equivalent to the tensor product of the matrices $\mathbb O_{\ell-2t}$ and $\mathbbm{1}_{2t}/d^{2t}$. 

To analyze \eqref{eq:entropyresult}, it is instructive to consider its prediction for $S_{A}^{(\alpha)}(t)/\ell$ in the scaling limit $\ell,t\to \infty$ where the ratio $\ell/t=\zeta$ is kept fixed. Let us start considering the second term on the r.h.s. of \eqref{eq:entropyresult}. This term can be written in terms of the state transfer matrix [\emph{cf.} \eqref{eq:statetransfermatrix}]
\be
\tau(\mathcal N)=
\begin{tikzpicture}[baseline={([yshift=-.5ex]current bounding box.center)},vertex/.style={anchor=base,
	circle,fill=black!25,minimum size=18pt,inner sep=2pt}, scale=0.7]
\draw[=latex] (2.2,-0.25)  -- (4.3,-0.25);
\draw[=latex] (2.2,2.8-0.8)  -- (4.3,2.8-0.8);
\draw[=latex] (2.8,-0.25)  -- (2.8,2.);
\draw[=latex] (3.7,-0.25)  -- (3.7,2.);
\foreach \y in {-0.25,2.} {
	\draw[fill=white!80!black] (2.8,+\y) circle (.22);
	\draw[fill=white!35!black] (3.7,+\y) circle (.22);
}
%
%
\end{tikzpicture}=
\begin{tikzpicture}[baseline={([yshift=-.5ex]current bounding box.center)},vertex/.style={anchor=base,
	circle,fill=black!25,minimum size=18pt,inner sep=2pt}, scale=0.7]
\draw[=latex] (2,0)  -- (4,0);
\draw[=latex] (2,2.25)  -- (4,2.25);
\draw[fill=white!50!black] (3,1.125) ellipse (.4 and 1.3);
\end{tikzpicture},
\ee
as follows 
\be
\operatorname{tr}\left[ \mathbb O^n_{2x}\right]=\frac{1}{\chi^n} \begin{tikzpicture}[baseline={([yshift=-.5ex]current bounding box.center)},vertex/.style={anchor=base,
	circle,fill=black!25,minimum size=18pt,inner sep=2pt}, scale=0.7]
\draw[=latex, ] (4.3,-0.25) arc(-90:0: 0.2 and .25);
\draw[=latex, ] (4.3,-1) arc(-90:90: 0.2 and .25);
\draw[=latex, ] (4.3,-2.75) arc(-90:90: 0.2 and .25);
\draw[=latex, ] (4.3,-3) arc(90:0: 0.2 and .25);
\draw[=latex] (4.3,-.25)  -- (3.5,-.25);
\draw[=latex] (1.8,-.25)  -- (2.5,-.25);
\draw[=latex] (4.3,-.5)  -- (3.5,-.5);
\draw[=latex] (1.8,-.5)  -- (2.5,-.5);
\draw[=latex,dotted] (2.5,-.25)  -- (3.5,-.25);
\draw[=latex,dotted] (2.5,-.5)  -- (3.5,-.5);
\draw[fill=white!50!black] (2.5,-.375) ellipse (.1 and .325);
\draw[fill=white!50!black] (3.5,-.375) ellipse (.1 and .325);
\draw[fill=white!50!black] (2,-.375) ellipse (.1 and .325);
\draw[fill=white!50!black] (4,-.375) ellipse (.1 and .325);
\draw[=latex] (4.3,-1.25)  -- (3.5,-1.25);
\draw[=latex] (1.8,-1.25)  -- (2.5,-1.25);
\draw[=latex] (4.3,-1)  -- (3.5,-1);
\draw[=latex] (1.8,-1)  -- (2.5,-1);
\draw[=latex,dotted] (2.5,-1.25)  -- (3.5,-1.25);
\draw[=latex,dotted] (2.5,-1)  -- (3.5,-1);
\draw[fill=white!50!black] (2.5,-1.125) ellipse (.1 and .325);
\draw[fill=white!50!black] (3.5,-1.125) ellipse (.1 and .325);
\draw[fill=white!50!black] (2,-1.125) ellipse (.1 and .325);
\draw[fill=white!50!black] (4,-1.125) ellipse (.1 and .325);
\draw[=latex] (4.3,-2)  -- (3.5,-2);
\draw[=latex] (1.8,-2)  -- (2.5,-2);
\draw[=latex] (4.3,-2.25)  -- (3.5,-2.25);
\draw[=latex] (1.8,-2.25)  -- (2.5,-2.25);
\draw[=latex,dotted] (2.5,-2.25)  -- (3.5,-2.25);
\draw[=latex,dotted] (2.5,-2)  -- (3.5,-2);
\draw[fill=white!50!black] (2.5,-2.125) ellipse (.1 and .325);
\draw[fill=white!50!black] (3.5,-2.125) ellipse (.1 and .325);
\draw[fill=white!50!black] (2,-2.125) ellipse (.1 and .325);
\draw[fill=white!50!black] (4,-2.125) ellipse (.1 and .325);
\draw[=latex] (4.3,-3)  -- (3.5,-3);
\draw[=latex] (1.8,-3)  -- (2.5,-3);
\draw[=latex] (4.3,-2.75)  -- (3.5,-2.75);
\draw[=latex] (1.8,-2.75)  -- (2.5,-2.75);
\draw[=latex,dotted] (2.5,-2.75)  -- (3.5,-2.75);
\draw[=latex,dotted] (2.5,-3)  -- (3.5,-3);
\draw[fill=white!50!black] (2.5,-2.875) ellipse (.1 and .325);
\draw[fill=white!50!black] (3.5,-2.875) ellipse (.1 and .325);
\draw[fill=white!50!black] (2,-2.875) ellipse (.1 and .325);
\draw[fill=white!50!black] (4,-2.875) ellipse (.1 and .325);

\draw[=latex, ] (1.8,-0.25) arc(-90:-180: 0.2 and .25);
\draw[=latex, ] (1.8,-0.5) arc(90:270: 0.2 and .25);
\draw[=latex, ] (1.8,-2.25) arc(90:270: 0.2 and .25);
\draw[=latex, ] (1.8,-3) arc(90:180: 0.2 and .25);
\node at (4.4,-1.45) {$\vdots$};
\node at (1.7,-1.45) {$\vdots$};
\draw [thick, decorate, decoration={brace,amplitude=5pt,raise=4pt},yshift=0pt]
(1.9,-0.2) -- (4.2,-0.2) node [black,midway,yshift=0.5cm] {$x$};
\draw [thick, decorate, decoration={brace,amplitude=5pt,raise=4pt},yshift=0pt]
(4.5,-0.25) -- (4.5,-3) node [black,midway,xshift=0.5cm] {$n$};
\draw node at (4,-4) {};
\end{tikzpicture}.
\label{eq:piccorr}
\ee
In the scaling limit, for any $\zeta<2$ (the term is trivially zero in the opposite case), the number of matrices $\tau(\mathcal N)$ in \eqref{eq:piccorr} becomes infinite and we can make the following replacement  
\be
\begin{tikzpicture}[baseline={([yshift=-.5ex]current bounding box.center)},vertex/.style={anchor=base,
	circle,fill=black!25,minimum size=18pt,inner sep=2pt}, scale=0.7]
\draw[=latex] (2,0)  -- (4,0);
\draw[=latex] (2,2.25)  -- (4,2.25);
\draw[fill=white!50!black] (3,1.125) ellipse (.4 and 1.3);
\end{tikzpicture} = \frac{1}{\chi}\,\,
\begin{tikzpicture}[baseline={([yshift=-.5ex]current bounding box.center)},vertex/.style={anchor=base,
	circle,fill=black!25,minimum size=18pt,inner sep=2pt}, scale=0.7]
\draw[=latex, ] (2.25,2.25) arc(90:-90: 0.5 and 1.125);
\draw[=latex, ] (3.75,2.25) arc(90:270: 0.5 and 1.125);
\end{tikzpicture},
\label{eq:replacement}
\ee
where we used that, thanks to Theorem~\ref{th:classification}, the unique eigenvector of $\tau(\mathcal N)$ corresponding to eigenvalue $1$ is a maximally entangled pair of qudits. The replacement \eqref{eq:replacement} leads to 
\be
\lim_{\substack{\ell,t\to\infty \\ \ell/t=\zeta}} \frac{\log[\operatorname{tr}\left[ \mathbb O^n_{\ell-2t}\right]]}{1-n} \,\theta_{\rm H}[\ell-2t]= 2\log\chi\,\, \theta_{\rm H}[\zeta-2]\,.
\label{eq:limitcorrection}
\ee
In particular, note that the r.h.s. of \eqref{eq:limitcorrection} is finite: this means that only the first term on the r.h.s. of \eqref{eq:entropyresult} contributes to the scaling limit when dividing by $\ell$. As a consequence, $S_{A}^{(\alpha)}(t)/\ell$ takes the following universal form
\be
\lim_{\substack{\ell,t\to\infty \\ \ell/t=\zeta}}S_{A}^{(\alpha)}(t)/\ell= {\rm min}(\zeta,2)\log d\,.
\label{eq:entropyleading}
\ee
In the last few years this form has been observed in conformal invariant models~\cite{calabrese2005evolution,tsunami}, in generic isolated systems~\cite{kimhuse,Mishra_protocol_2015,pal_entangling_2018}, and in local random unitary circuits~\cite{nahum2017quantum, vonKeyserlingk2018operator, nahum2018operator,chan2018solution}. The ubiquitousness of Eq.~\eqref{eq:entropyleading} has been recently explained by introducing the so called ``minimal membrane picture"~\cite{nahum2017quantum}. In essence, the idea is to estimate the entanglement between a subsystem $A$ and the rest by measuring the length of the minimal membrane in space-time that separates the subsystem from the rest. 

Ref.~[\onlinecite{bertini2019entanglement}] showed that \eqref{eq:entropyleading} is exact at any time in the self-dual kicked Ising model evolving from ``separating states''. Considering the special case $\chi=1$ of \eqref{eq:entropyresult} we see that this statement carries over for generic dual-unitary circuits. Indeed, in this case we have  ${\rm tr}[\mathbb O_{x}^\alpha]=1$ for all $\alpha$. From \eqref{eq:entropyresult}, however, we also see that for more general initial ``solvable" MPSs there appear some corrections that can make the entanglement spectrum non-trivial. These corrections are encoded by the matrix $\mathbb O_{x}$ and depend solely on the initial state. 

We remark that the corrections contained in \eqref{eq:entropyresult} can be generically calculated numerically with high efficiency because the rank of the matrix that encodes them is constant and equal to $\chi$. A simple analytically tractable limit is that of infinite length of the block $A$. Indeed, proceeding as we did to obtain \eqref{eq:limitcorrection}, we readily find 
\be
\lim_{\ell\to\infty} S_{A}^{(\alpha)}(t)= 2\log\chi+ 2t \log d\,.
\ee
We see that in this case the entanglement spectrum is again completely flat. 

Finally, we note that the form \eqref{eq:rhol>2t}--\eqref{eq:rho2t>l} of the reduced density matrix can also be used to compute the evolution of the entanglement entropy of disjoint blocks. In this case one has to connect some of the central uppermost and lowermost lines of the circuits \eqref{eq:rhol>2t}--\eqref{eq:rho2t>l} to divide $A$ into two disconnected parts. The resulting quantum circuit is again very simple in the case \eqref{eq:rho2t>l} and implies that for $2t>\ell$ the reduced density matrix is again proportional to the identity. The case \eqref{eq:rhol>2t}, however, becomes more complicated: the density matrix is not unitary equivalent to $\mathbb O_{\ell-2t}\otimes \mathbbm{1}_{2t}/d^{2t}$ anymore, and the result depends on the specific dual-unitary gate considered.

\section{Non-solvable initial states}
\label{sec:non_solvable_states}

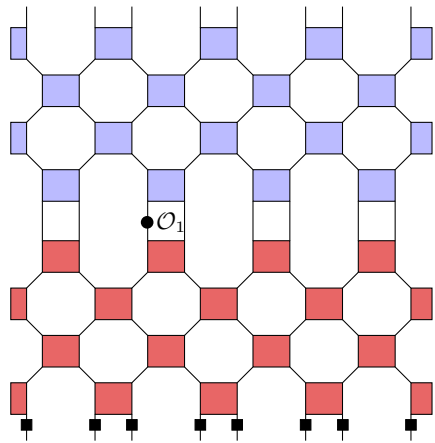
\begin{figure}
	\begin{center}
		\begin{tikzpicture}[scale=0.7]
		\foreach \x in {0.85,2.85,4.85,6.85} { 
			\draw[=latex] (\x+0.,-0.3)  -- (\x+0.,0.2);
			\draw[=latex] (\x,7.75-0.2)  -- (\x+0.,7.75+0.2);
		}
		\foreach \x in {2.85-0.7,4.85-0.7,6.85-0.7,8.85-0.7} { 
			\draw[=latex] (\x,-0.3)  -- (\x,0.2);
			\draw[=latex] (\x,7.75-0.2)  -- (\x,7.75+0.2);
		}
		\foreach \x in {2.85-0.7,4.85-0.7,6.85-0.7,8.85-0.7} { 
			\draw [fill=black] (\x -0.1,-0.1) rectangle (\x+0.1,0.1);
		}
		\foreach \x in {0.85,2.85,4.85,6.857} { 
			\draw [fill=black] (\x -0.1,-0.1) rectangle (\x+0.1,0.1);
		}
		\foreach \y in {0,1.8} {
			\foreach \x in {2,4.,...,6.} { 
				\draw [fill=myred] (\x +0.15, 0.2+\y) rectangle (\x+0.85,0.8+\y);
			}
			\draw [fill=myred] (0.55, 0.2+\y) rectangle (0.85,0.8+\y);
			\draw [fill=myred] (8.15, 0.2+\y) rectangle (8.55,0.8+\y);
		}
		\foreach \y in {0,1.8} {
			\foreach \x in {1.,3.,...,7.} { 
				\draw [fill=myred] (\x +0.15,1.1+\y) rectangle (\x+0.85,1.7+\y);
			}
		}
		\foreach \y in {4.5+0.45,6.3+0.45} {
			\foreach \x in {2.,4.,...,6.} { 
				\draw [fill=myblue] (\x +0.15, 0.2+\y) rectangle (\x+0.85,0.8+\y);
			}
			\draw [fill=myblue] (0.55, 0.2+\y) rectangle (0.85,0.8+\y);
			\draw [fill=myblue] (8.15, 0.2+\y) rectangle (8.55,0.8+\y);
		}
		\foreach \y in {2.7+0.45,4.5+0.45} {
			\foreach \x in {1.,3.,...,7.} { 
				\draw [fill=myblue] (\x +0.15,1.1+\y) rectangle (\x+0.85,1.7+\y);
			}
		}
		\foreach \y in {1.8,4.5+0.45,6.3+0.45}{
			\foreach \x in {2.,4.,6.,8.} { 
				\draw[=latex] (\x-0.15,-0.1+\y)  -- (\x+0.15,0.2+\y);
			}
		}
		\foreach \y in {0.9,2.7,5.4+0.45}{
			\foreach \x in {2.,4.,6.,8.} { 
				\draw[=latex] (\x-0.15,0.2+\y)  -- (\x+0.15,-0.1+\y);
			}
		}
		\foreach \y in {1.8,4.5+0.45,6.3+0.45}{
			\foreach \x in {1,3.,5.,7.} { 
				\draw[=latex] (\x-0.15,0.2+\y)  -- (\x+0.15,-0.1+\y);
			}
		}
		\foreach \y in {0.9,2.7,5.4+0.45}{
			\foreach \x in {1,3.,5.,7.} { 
				\draw[=latex] (\x-0.15,-0.1+\y)  -- (\x+0.15,0.2+\y);
			}
		}
		\foreach \x in {1,3.,5.,7.} { 
			\draw[=latex] (\x+0.15,3.5)  -- (\x+0.15,3.8+0.45);
		}
		\foreach \x in {2,4.,6.,8.} { 
			\draw[=latex] (\x-0.15,3.5)  -- (\x-0.15,3.8+0.45);
		}
		\draw[thick, fill=black] (3.22-0.08,3.85) circle (0.1cm); 
		\node at (3.6,3.85) {$\mathcal{O}_1$};
		\end{tikzpicture}
	\end{center}
	\caption{Pictorial representation of a time-dependent one-point function, where the initial state for the quantum dynamics is chosen as in Eq.~\eqref{eq:Gibbs_initial}. In the figure, $\tau=2$ (namely, $t=4$) and periodic boundary conditions are assumed in the ``time direction'', \emph{i.e.} the vertical lines at the bottom and at the top are understood to be joined together. Small black rectangles denote the operators $\rho^{(j)}(\beta,a)$ defined in Eq.~\eqref{eq:local_density_matrix}.}
	\label{fig:product_states}
\end{figure}

It is natural to wonder how the features of the quantum dynamics studied in the previous section depend on the solvability of the MPSs chosen as initial states. In general, for instance, one may expect that while the system will still locally approach 
an infinite temperature density matrix for generic initial states and unitary gates, local expectation values will display an exponential decay to zero, rather than approaching zero in a finite number of time steps.

The dynamics arising from generic states can be studied using numerical MPS techniques~\cite{schollwock2011density}. Here, in order to compare the solvable and non-solvable cases, we follow a different approach, and consider a family of initial states which depend on one real parameter $\beta$, interpolating between the infinite temperature density matrix and arbitrary initial pure product states, as $\beta$ is varied from zero to infinity. This allows us to study analytically the dynamics for small values of $\beta$, highlighting some qualitative differences that arise with respect to the solvable dynamics.

For concreteness, we focus once again on the case of a qubit system and consider the initial mixed state
\be
\rho_0(\beta,a)=\rho^{(1)}(\beta,a)\otimes \rho^{(2)}(\beta,a)\otimes \cdots \otimes \rho^{(2L)}(\beta,a)\,,
\label{eq:Gibbs_initial}
\ee
where we introduced the single-site density matrix
\be
\rho^{(j)}(\beta,a)=\frac{1}{\mathcal{Z}(\beta,a)}\exp\left[-\beta a\right] \in \mathrm{End}(h_j) \,.
\label{eq:local_density_matrix}
\ee
Here ${a\in \mathrm{End}(\mathbb{C}^2)}$, with ${\rm tr} (a)=0$ and $a^2=\mathbb{1}$, while $\mathcal{Z}(\beta,a)={\rm tr}\left[\exp\left(-\beta a\right)\right]$, so that $\rho^{(j)}(\beta, a)$ admits the expansion
\be
\rho^{(j)}(\beta, a)=\frac{1}{2}\mathbb{1}-\frac{1}{2}\tanh(\beta)a\,.
\label{eq:expansion}
\ee
In this section, we will focus on the computation of time-dependent one-point correlation functions. First, note that according to Eqs.~\eqref{eq:ev_rule_even} and \eqref{eq:ev_rule_odd}, the initial density matrix~\eqref{eq:Gibbs_initial} is time-evolved as
\bea
\rho_{2t+1}(\beta, a)&=&\mathcal{U}^{\dagger}_-\rho_{2t}(\beta, a)\mathcal{U}_-\,,\\
\rho_{2t+2}(\beta, a)&=&\mathcal{U}_+^\dagger \rho_{2t+1}(\beta, a) \mathcal{U}_+\,.
\eea
Next, making use of Eq.~\eqref{eq:expansion}, one can formally write
\be
\braket{\mathcal{O}_j\rho_t(\beta,a)}=\sum_{n=0}^\infty c^a_n(t)\left[\tanh\left(\beta\right)\right]^n\,.
\label{eq:gibbs_exp}
\ee
In the following, we show how dual unitarity allows us to compute the coefficients $c^a_n(t)$ exactly up to $n=2$, providing a perturbative knowledge of the one-point correlation functions.

As for the solvable initial states, one-point functions for the density matrix~\eqref{eq:Gibbs_initial} display an even/odd effect in time, due to the discrete nature of the dynamics. In order to simplify the following discussion, in analogy to Sec.~\ref{sec:entanglement_growth}, also in this section we restrict for concreteness to even values of times
\be
t=2\tau\,,\qquad \tau\in \mathbb{N}\,,
\label{eq:tau}
\ee
and choose an operator placed at an odd site $j=2k+1$.

\begin{figure}
	\begin{center}
		\begin{tikzpicture}[scale=0.6]
		\draw[=latex] (6.85,-0.3)  -- (6.85,7.75+0.2);
		\draw[=latex] (6.85-0.7,7.75-0.3)  -- (6.85-0.7,7.75+0.2);
		\draw[=latex] (6.85-0.7,-0.3)  -- (6.85-0.7,0.2);
		\draw[=latex] (5.85-0.7,7.75-0.3-1)  -- (5.85-0.7,7.75+0.2);
		\draw[=latex] (5.85-0.7,-0.3)  -- (5.85-0.7,0+1.2);
		\draw[=latex] (4.85-0.7,7.75-0.3-2)  -- (4.85-0.7,7.75+0.2);
		\draw[=latex] (4.85-0.7,-0.3)  -- (4.85-0.7,0+2);
		\draw[=latex] (3.85-0.7,7.75-0.3-3)  -- (3.85-0.7,7.75+0.2);
		\draw[=latex] (3.85-0.7,-0.3)  -- (3.85-0.7,0+3);
		\draw[=latex] (4.85,2.)  -- (4.85,5.5);
		\draw[=latex] (5.85,1.25)  -- (5.85,6.5);
		\foreach \x in {6.857} { 
			\draw [fill=green!60!black] (\x -0.1,-0.1) rectangle (\x+0.1,0.1);
		}
		\foreach \y in {0} {
			\foreach \x in {6.} { 
				\draw [fill=myred] (\x +0.15, 0.2+\y) rectangle (\x+0.85,0.8+\y);
			}
		}
		\foreach \y in {1.8} {
			\foreach \x in {4.} { 
				\draw [fill=myred] (\x +0.15, 0.2+\y) rectangle (\x+0.85,0.8+\y);
			}
		}
		\foreach \y in {0} {
			\foreach \x in {5.} { 
				\draw [fill=myred] (\x +0.15,1.1+\y) rectangle (\x+0.85,1.7+\y);
			}
		}
		\foreach \y in {1.8} {
			\foreach \x in {3.} { 
				\draw [fill=myred] (\x +0.15,1.1+\y) rectangle (\x+0.85,1.7+\y);
			}
		}
		\foreach \y in {4.5+0.45} {
			\foreach \x in {4.} { 
				\draw [fill=myblue] (\x +0.15, 0.2+\y) rectangle (\x+0.85,0.8+\y);
			}
		}
		\foreach \y in {6.3+0.45} {
			\foreach \x in {6.} { 
				\draw [fill=myblue] (\x +0.15, 0.2+\y) rectangle (\x+0.85,0.8+\y);
			}
		}
		\foreach \y in {2.7+0.45} {
			\foreach \x in {3} { 
				\draw [fill=myblue] (\x +0.15,1.1+\y) rectangle (\x+0.85,1.7+\y);
			}
		}
		\foreach \y in {4.5+0.45} {
			\foreach \x in {5} { 
				\draw [fill=myblue] (\x +0.15,1.1+\y) rectangle (\x+0.85,1.7+\y);
			}
		}
		\foreach \y in {4.5+0.45}{
			\foreach \x in {4.} { 
				\draw[=latex] (\x-0.15,-0.1+\y)  -- (\x+0.15,0.2+\y);
			}
		}
		\foreach \y in {6.3+0.45}{
			\foreach \x in {6.} { 
				\draw[=latex] (\x-0.15,-0.1+\y)  -- (\x+0.15,0.2+\y);
			}
		}
		\foreach \y in {0.9}{
			\foreach \x in {6.} { 
				\draw[=latex] (\x-0.15,0.2+\y)  -- (\x+0.15,-0.1+\y);
			}
		}
		\foreach \y in {2.7}{
			\foreach \x in {4.} { 
				\draw[=latex] (\x-0.15,0.2+\y)  -- (\x+0.15,-0.1+\y);
			}
		}
		\foreach \y in {1.8}{
			\foreach \x in {5.} { 
				\draw[=latex] (\x-0.15,0.2+\y)  -- (\x+0.15,-0.1+\y);
			}
		}
		\foreach \y in {5.4+0.45}{
			\foreach \x in {5.} { 
				\draw[=latex] (\x-0.15,-0.1+\y)  -- (\x+0.15,0.2+\y);
			}
		}
		\foreach \x in {3.} { 
			\draw[=latex] (\x+0.15,3.5)  -- (\x+0.15,3.8+0.45);
		}
		\foreach \x in {4.} { 
			\draw[=latex] (\x-0.15,3.5)  -- (\x-0.15,3.8+0.45);
		}
		\draw[thick, fill=black] (3.22-0.08,3.85) circle (0.1cm); 
		\node at (2.5,3.85) {$\mathcal{O}_1$};
		\node at (0.5,3.85) {$c^a_1(\tau)=\frac{1}{2^{2\tau}}\times $};
		\end{tikzpicture}
	\end{center}
	\caption{Pictorial representation for the first coefficient $c_1^a(t)$, for $\tau=2$ (namely, $t=4$). From the picture, it is clear that it can be computed using the analytic formula for the infinite-temperature dynamical two-point functions derived in Ref.~[\onlinecite{bertini2019exact}]. The small green rectangle corresponds to the operator $-1/2 \tanh(\beta) a$}
	\label{fig:first_coefficient}
\end{figure}
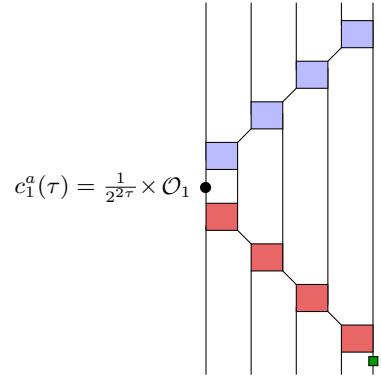

We begin by noting that Eq.~\eqref{eq:gibbs_exp} can be pictorially represented as in Fig.~\ref{fig:product_states}, where a small black rectangle placed at site $j$ denotes the single-site density matrix $\rho^{(j)}(\beta,a)$ defined in Eq.~\eqref{eq:local_density_matrix}. Next, from Eq.~\eqref{eq:expansion} it is clear that each operator $a$ bears a factor $\tanh(\beta)$, so that at the first order in $x=\tanh(\beta)$ Fig.~\ref{fig:product_states} simplifies in a sum of tensor networks, each one corresponding to setting all operators $\rho^{(j)}(\beta,a)$ equal to $\mathbb{1}/2$, except for one. By means of the usual graphical identities, it is straightforward to see that only one term in this sum is non-zero. In the case of Fig.~\ref{fig:product_states}, for instance, this corresponds to the tensor network displayed in Fig.~\ref{fig:first_coefficient}, where a small green rectangle now corresponds to the operator $-1/2 \tanh(\beta) a$. We recognize that the latter is proportional to an infinite-temperature two-point correlation function, and can thus be computed using the results of Ref.~[\onlinecite{bertini2019exact}], thus obtaining
\bea
c^a_{1}(\tau)=-\frac{1}{2}{\rm tr}\left\{\widetilde{\mathcal{F}}^{2\tau}\left[\mathcal{O}\right]a^T\right\}\,.
\eea
Here $\widetilde{\mathcal{F}}[a]$ is the map defined in Eq.~\eqref{eq:tilde_f_function}, $(\cdot)^T$ denotes, as before, matrix transposition, while $\tau$ was introduced in Eq.~\eqref{eq:tau}.

The same logic can be followed to compute the second-order coefficient $c_2^a(\tau)$, which can be expressed as a sum of some particular three-point correlation functions. The computation of the latter is slightly more involved, but one can once again exploit dual unitarity to evaluate them in an efficient way, in complete analogy with what we did in Sec.~\ref{sec:two_point_calculations} for the case of time-dependent two-point functions. Since the computation does not present additional difficulties, we omit the intermediate steps and in the following only present the final result, which reads
\be
\!\!\!\!\!\!c^a_{2}(\tau)=\!\frac{1}{4}\sum_{r=1}^{2\tau}{\rm tr}\left\{\!\widetilde{\mathcal{F}}^{r-1}\left[\mathcal{O}\right]\! \mathcal{S}\left[\widetilde{\mathcal{L}}^{2\tau-r}(a),\mathcal{L}^{2\tau-r}(a)\right]^T\right\}\!.
\ee
Here $\mathcal{L}[a]$ and $\widetilde{\mathcal{L}}[a]$ are defined similarly to $\mathcal{F}[a]$ and $\widetilde{\mathcal{F}}[a]$ [introduced in Eqs.~\eqref{eq:f_function} and \eqref{eq:tilde_f_function}], but with $U$ and $U^\dagger$ exchanged, namely
\bea
\widetilde{\mathcal{L}}\left[a\right]=\frac{1}{d}\operatorname{tr}_{1}\left[U(a \otimes \mathbb{1}) U^\dagger\right]\,,\\
\mathcal{L}\left[a\right]=\frac{1}{d} \operatorname{tr}_{2}\left[U( \mathbb{1} \otimes a) U^\dagger\right]\,,
\eea
with $d=2$ for qubits. Finally, the function $\mathcal{S}$ is a map acting on the space of linear operators ${\rm End}(\mathbb{C}^2\otimes \mathbb{C}^2)$
\be
S:{\rm End}(\mathbb{C}^2\otimes \mathbb{C}^2)\to {\rm End}(\mathbb{C}^2)
\ee
which is defined in coordinate space as
\bea
\left(\mathcal{S}\left[a,b\right]\right)_{m,n}&=&
\begin{tikzpicture}[baseline={([yshift=-.5ex]current bounding box.center)},vertex/.style={anchor=base,
	circle,fill=black!25,minimum size=18pt,inner sep=2pt}, scale=0.9]
\draw[=latex] (5.15,1.5)  -- (5.15,2);
\draw[=latex] (5.15,2.5)  -- (5.15,3);
\draw[=latex] (5.85,1.5)  -- (5.85,3);
\draw[=latex, ] (5.85,1.115) arc(-90:90: 0.6 and 1.145);
\draw[=latex, ] (5.15,3.4) arc(90:270: 0.6 and 1.145);
\foreach \y in {0} {
	\foreach \x in {5.} { 
		\draw [fill=myred] (\x +0.15,1.1+\y) rectangle (\x+0.85,1.7+\y);
	}
}
\foreach \y in {2.6} {
	\foreach \x in {5.} { 
		\draw [fill=myblue] (\x +0.15, 0.2+\y) rectangle (\x+0.85,0.8+\y);
	}
}
\draw[thick, fill=black] (4.55,2.25) circle (0.1cm); 
\draw[thick, fill=black] (6.45,2.25) circle (0.1cm); 
\node at (4.25,2.2) {\scalebox{0.8}{$a^T$}};
\node at (6.8,2.2) {\scalebox{0.8}{$b^T$}};
\node at (5.3,2) {\scalebox{0.8}{$n$}};
\node at (5.3,2.5) {\scalebox{0.8}{$m$}};
\end{tikzpicture}
\nonumber\\
&=&a_{j,i}b_{p,l}U_{j,p}^{n,q}\left[U^{\dagger}\right]_{m,q}^{i,l}\,.
\label{eq:s_function}
\eea

\begin{figure}
	\includegraphics[width=0.45\textwidth]{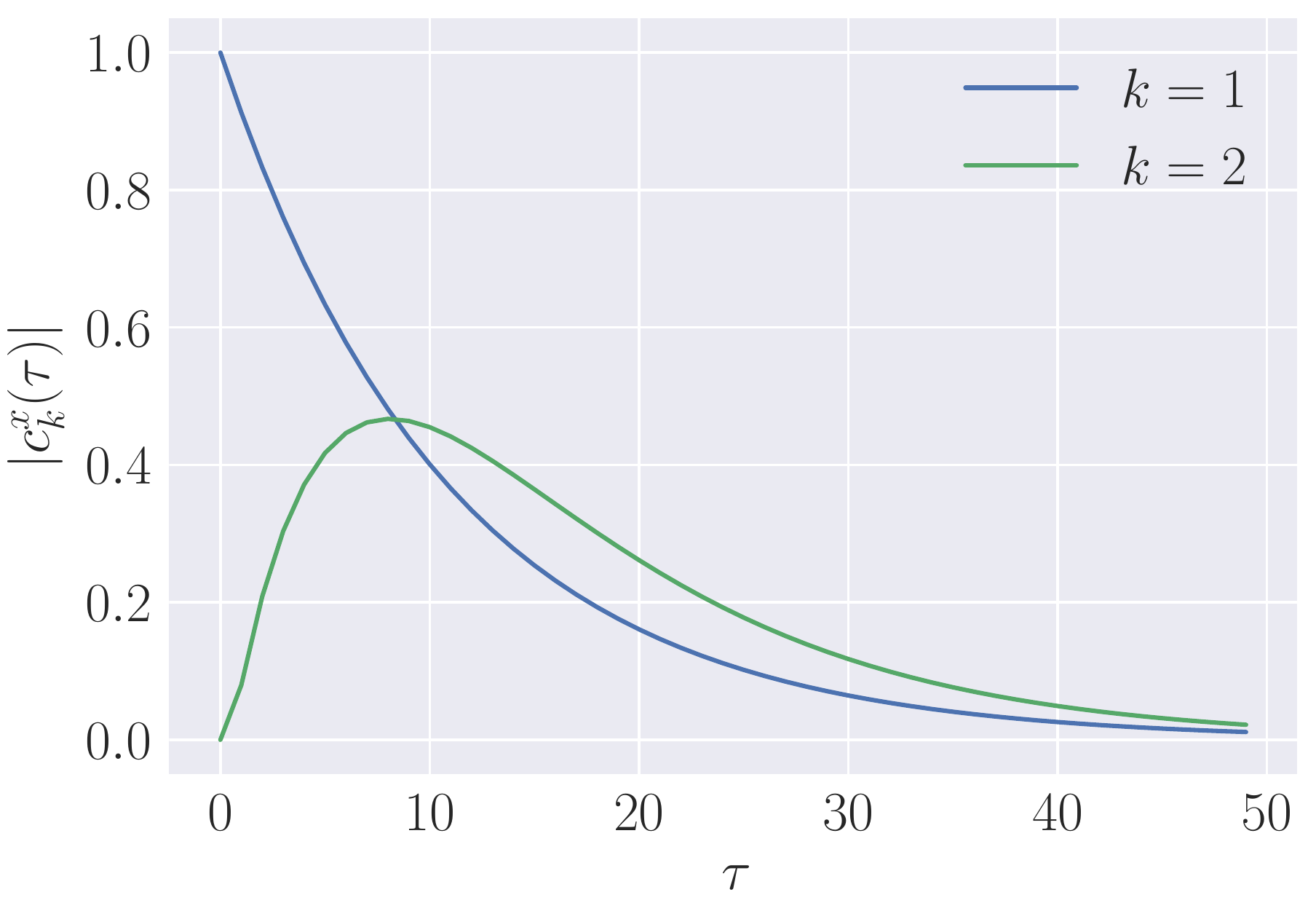} 
	\caption{Coefficients $c^x_k(\tau)$ defined in Eq.~\eqref{eq:gibbs_exp}. Here we chose the local operator $\mathcal{O}_j=\sigma_j^x$, while for the initial Gibbs state we set $a=\sigma^x$. The dynamics is driven by the quantum circuit corresponding to the self-dual kicked Ising chain, namely we chose $U_{\rm SDKI}$ as defined in Eq.~\eqref{eq:gate_SKDI}, and set the magnetic field to $h=0.15$.}
	\label{fig:coefficients}
\end{figure}

Dual unitarity does not seem to provide a substantial simplification in the computation of higher-order coefficients $c^a_{k>2}(t)$ when compared to a generic evolution. Still, the cases $k=1,2$ already allow us to highlight some of the expected differences with respect to the dynamics arising from solvable MPSs. In Fig.~\ref{fig:coefficients} we present results for the evolution of the expectation value of $\mathcal{O}_j=\sigma^x_j$ in a chaotic quantum circuit corresponding to the self-dual kicked Ising chain (we chose $U_{\rm SDKI}$ as defined in Eq.~\eqref{eq:gate_SKDI}, and set the magnetic field to $h=0.15$). We initialized the system in the state \eqref{eq:Gibbs_initial} with $a=\sigma^x$. We see that, as expected, both coefficients $c^x_1(\tau)$, $c^x_2(\tau)$ approach zero, although in an exponential fashion rather than in a finite number of time steps. In general, based on this result, we also expect that our analytic formulae for two-point correlation functions, valid for solvable MPSs, will acquire some ``exponential corrections'' in the case of generic initial states and dual-unitary gates.

\section{Conclusions}
\label{sec:conclusions}

In this work we have considered the dynamics of the recently introduced class of dual-unitary quantum circuits, and exhibited a family of initial states for which different physical quantities can be computed exactly. We have characterized these states in terms of a particular ``solvability'' condition, and provided a complete classification of them. In particular, we have shown that this family includes MPSs of arbitrary bond dimension $\chi$, and provided explicit examples for $\chi=1\,,2$. (\emph{cf.} Sec.~\ref{sec:explicit_examples}).

 For the dynamics arising from these initial states, we have derived three main results. First, we were able to compute explicitly the time it takes for an observable supported on a finite region $\ell$ to approach its infinite-temperature value, and shown that $t\propto \ell$ (Sec.~\ref{sec:local_therm}). Second, we have provided an exact formula, efficient to evaluate, to compute two-point equal-time correlation functions of local observables, and we have shown that they display different qualitative features depending on the ergodicity of the quantum circuit (Sec.~\ref{sec:two_point_calculations}). Third, we have derived a closed formula for the time evolution of the entanglement entropy of a connected block of length $\ell$, computing explicitly the limit $\ell\to\infty$ (Sec.~\ref{sec:entanglement_growth}). This generalizes the exact result of Ref.~[\onlinecite{bertini2019entanglement}] to all dual-unitary quantum circuits, extending it for more general initial states. Remarkably, we showed that solvable MPSs with bond dimension larger than one produce non-trivial finite-time corrections. 

Finally, we have also considered a family of non-solvable initial mixed states depending on one real parameter $\beta$, which interpolate between the infinite temperature density matrix and arbitrary initial pure product states, \emph{cf.} Sec.~\ref{sec:non_solvable_states}. We have studied analytically their dynamics for small values of $\beta$, and highlighted the main differences from the case of solvable MPSs. 

In the light of our results, there are several natural directions to be explored. For example, the class of solvable MPSs introduced in this paper was defined by the property that the leading eigenstate of the transverse transfer matrix is a product state in the  ``folded picture'' of Refs.~[\onlinecite{banuls2009matrix},\onlinecite{muller-hermes_tensor_2012}]. One can wonder whether it is possible to find different types of initial states or unitary gates for which such leading eigenstates are written instead in the form of non-trivial MPSs (with fixed bond dimension). In this case, the dynamics would still be solvable, but a richer phenomenology would emerge for the evolution of local observables.

Finally, it is certainly interesting to wonder whether some aspects of the present paper can be generalized to the case of higher spatial dimensions, where the application of numerical techniques is known to be much harder with respect to the one-dimensional case. A successful description of the dual-unitary dynamics in higher spatial dimensions would provide a rare benchmark, for instance, for the development of numerical computational methods.

\section*{Acknowledgments} 
LP acknowledges support from the Alexander von Humboldt foundation.  BB and TP acknowledge support by the EU Horizon 2020 program through the ERC Advanced Grant OMNES No. 694544, and by the Slovenian Research Agency (ARRS) under the Programme P1-0402. JIC acknowledges support by the EU Horizon 2020 program through the ERC Advanced Grant QENOCOBA No. 742102, and from the DFG (German Research Foundation) under Germany’s Excellence Strategy - EXC-2111 - 390814868. Part of this work has been carried out during a visit of LP to the University of Ljubljana, whose hospitality is kindly acknowledged.

\appendix

\section{Comparison between the solvability and separating conditions}
\label{sec:solvability_vs_separability}

In this section we discuss the relation between the ``separating'' initial states of Ref.~[\onlinecite{bertini2019entanglement}] and the solvable MPSs introduced in this paper. First, we observe that the work~[\onlinecite{bertini2019entanglement}] focused on one particular realization of a dual-unitary quantum circuit, corresponding to the self-dual kicked Ising model~\cite{bertiniSDKI}. In this specific case, the separating property was also introduced as a technical condition on initial \emph{product} states, allowing for an analytic determination of the right eigenstate of an appropriate transverse transfer matrix. 

There are two main differences between the present work and Ref.~[\onlinecite{bertini2019entanglement}]: first, here we allow for more general initial states in the form of MPSs, and do not restrict to product states; second, we look for initial conditions that are analytically tractable for any dual-unitary circuit, and not just for one specific case. In fact, we show in the following that the separating states of Ref.~[\onlinecite{bertini2019entanglement}] are not solvable MPSs for a generic dual-unitary evolution, but become so after applying one layer of unitary operators, which corresponds to the first time step of the self-dual kicked Ising Floquet dynamics.

We start by recalling that the Floquet evolution associated with the self-dual kicked Ising chain is defined by~\cite{bertini2019exact, gopalakrishnan2019unitary}
\bea
\mathcal U_{\rm SDKI}^{2t}&=& \mathbb U^{\phantom{\rm e}}_{\rm K}\mathbb U^{\rm e}_{\rm I}  \mathbb U^{\rm o}_{\rm SDKI} \mathbb U_{\rm SDKI}^t \mathbb U^{\rm e}_{\rm I}, \label{eq:even_ev}\\
\mathcal U_{\rm SDKI}^{2t+1}&=& \mathbb U_{\rm K}\mathbb U^{\rm o}_{\rm I} \mathbb U^{\rm o}_{\rm SDKI}\mathbb U_{\rm SDKI}^{t-1} \mathbb U^{\rm e}_{\rm SDKI}\mathbb U^{\rm e}_{\rm I}\,,\label{eq:odd_ev}
\eea
where $t\in\mathbb N$. Here $\mathbb{U}^{\rm e}_{\rm SDKI}$ ($\mathbb{U}^{\rm o}_{\rm SDKI}$) is the transfer matrix defined by applying the local gate~\eqref{eq:gate_SKDI} to each pair of neighboring qubits, where the first one is chosen at an even (odd) position, while $\mathbb{U}_{\rm SDKI}=\mathbb{U}^{\rm e}_{\rm SDKI}\mathbb{U}^{\rm o}_{\rm SDKI}$. The operators $\mathbb{U}^{\rm e/o}_{\rm I}$ and $\mathbb{U}_{\rm I}$ are defined analogously, with two-site gates given by
\bea
U_{\rm I} &=& e^{- i (\pi/4) \sigma^{z}\otimes\sigma^z} (e^{- i h \sigma^{z}}\otimes \mathbb{1})\,.
\eea
Finally $\mathbb{U}_{\rm K}=U_{\rm K}^{\otimes 2L}$ with
\be
U_{\rm K} = e^{- i (\pi/4) \sigma^{x}}\,.
\ee

Next, the separating states were defined in Ref.~[\onlinecite{bertini2019entanglement}] as the two subclasses of product states
\be
\left|\psi_{\boldsymbol{\theta}, \phi}\right\rangle=\bigotimes_{j=1}^{2L}\left[\cos \left(\frac{\theta_{j}}{2}\right)|\uparrow\rangle+\sin \left(\frac{\theta_{j}}{2}\right) e^{i \phi_{j}}|\downarrow\rangle\right]\,,
\ee
defined by
\bea
\mathcal{T}&=&\left\{\left|\psi_{(\pi / 2) \mathbf{1}, \phi}\right\rangle, \quad \phi_{j} \in[0,2 \pi]\right\}\,,\label{eq:t_class}\\
\mathcal{L}&=&\left\{\left|\psi_{\bar{\theta}, \phi}\right\rangle, \quad \bar{\theta}_{j} \in\{0, \pi\}\right\}\,,\label{eq:s_class}
\eea
where $\mathbf{1}$ denotes a vector of length $2L$ with all entries equal to 1. Note that we do not need to specify the value of $\phi$ in Eq.~\eqref{eq:s_class}. In order to compare with the present article, where we restricted to two-site shift invariant initial states, we can assume 
\bea
\phi_{2j}&=&\phi_e \qquad \phi_{2j-1}=\phi_o\,,\qquad  j=1\,,\ldots L\,, \\
\bar{\theta}_{2j}&=&\bar{\theta}_e\qquad \bar{\theta}_{2j-1}=\bar{\theta}_o\,, \qquad  j=1\,,\ldots L\,.
\eea

We see that the above class of states does not belong to the family of solvable MPSs of dimension $\chi=1$, which was characterized in Sec.~\ref{sec:bond_chi1}. However, from Eq.~\eqref{eq:even_ev}, \eqref{eq:odd_ev}, we have that the Floquet dynamics of the self-dual kicked Ising model is made up of different steps. In particular, given an initial state $\ket{\psi}$, the first step consists in multiplying it by either $\mathbb U^{\rm e}_{\rm I}$ or $ \mathbb U^{\rm e}_{\rm SDKI}\mathbb U^{\rm e}_{\rm I}$: after that the evolution is dictated by the dual-unitary quantum circuit encoded in $\mathbb U_{\rm SDKI}^t$. Accordingly, one should check that if $\ket{\psi}\in\mathcal{T}\cup \mathcal{L}$, then $\mathbb U^{\rm e}_{\rm I}\ket{\psi}$ and $\mathbb U^{\rm e}_{\rm SDKI} \mathbb U^{\rm e}_{\rm I}\ket{\psi}$ are solvable. This is indeed the case, as one can verify by direct computation.

In conclusion, the Floquet dynamics of the separating product states introduced in Ref.~[\onlinecite{bertini2019entanglement}] can be described in terms of a dual-unitary circuit where the initial configuration is indeed a solvable MPS with bond dimension $\chi=1$.

\section{Proof of Theorem~\ref{th:classification}}
\label{sec:proof_theorem_1}

In this appendix we provide a full proof of Theorem~\ref{th:classification}. Let us start by proving the first part of the statement, namely that each solvable MPS is equivalent in the thermodynamic limit to an injective MPS satisfying Eq.~\eqref{eq:solvability_identity}. 

Consider the following transfer matrix associated with the MPS $|\Psi^L_0\left(\mathcal{M}\right)\rangle$
\be
\tau(\mathcal{M})=\sum_{j,k=1}^d \mathcal{M}^{(j,k)}\otimes \left[\mathcal{M}^{(j,k)}\right]^\ast\,,
\label{eq:statetransfermatrix}
\ee
which acts on the tensor product $\mathbb{C}^\chi\otimes \mathbb{C}^\chi$ of two auxiliary spaces. It is easy to show that condition $C1$ is equivalent to requiring that the transfer matrix $\tau(\mathcal{M})$ has a unique eigenvalue $\lambda_0$ with largest absolute value, with $\lambda_0=1$ and algebraic multiplicity equal to $1$. This follows from $\braket{\Psi^L_t|\Psi^L_t}=\braket{\Psi^L_0|\Psi^L_0}$, and the identity
\be
\braket{\Psi^L_0(\mathcal{M})|\Psi^L_0(\mathcal{M})}= {\rm tr}\left[\tau(\mathcal{M})^{L}\right]\,,
\ee
where we used that the system size is $2L$. Defining now the state
\be
\ket{R_\tau}=\sum_{i,j=1}^\chi V_{i,j}\ket{i}\ket{j}\in \mathbb{C}^{\chi}\otimes \mathbb{C}^{\chi}\,,
\ee 
we see that the condition $\tau(\mathcal{M})\ket{R_\tau}=\ket{R_\tau}$ is verified if and only if
\be
\sum_{j,k=1}^d \mathcal{M}^{(j,k)} V \left(\mathcal{M}^{(j,k)}\right)^\dagger =V\,,
\label{eq:to_be_verified}
\ee
where $V$ is the  $\chi$-dimensional matrix with entries $V_{i,j}$. Hence, the transfer matrix $\tau(\mathcal{M})$ has a unique largest eigenvalue $\lambda_0=1$ if and only if the linear map
\be
\mathcal{E}_{\mathcal{M}}(X)=\sum_{j,k=1}^d \mathcal{M}^{(j,k)} X \left(\mathcal{M}^{(j,k)}\right)^\dagger\,,
\label{eq:quantum_channel}
\ee
has a unique largest eigenvalue $\lambda_0=1$. On the other hand, setting $i=j$ in Eq.~\eqref{eq:solvability}, and summing over $j$, we obtain
\be
\sum_{j,k=1}^d \mathcal{M}^{(j,k)} S \left(\mathcal{M}^{(j,k)}\right)^\dagger =S\,.
\label{eq:fixed_point}
\ee
Namely, combining conditions $C1$ and $C2$, we obtain that if $\ket{\Psi^L_0(\mathcal{M})}$ is a solvable MPS, then $S$ is the only fixed point for the linear map $\mathcal{E}_{\mathcal{M}}(X)$ in Eq.~\eqref{eq:quantum_channel}. 

It is important to observe now that $\mathcal{E}_{\mathcal{M}}(X)$ is a completely positive, linear map on the space of matrices ${\rm End}(\mathbb{C}^\chi)$ and that its spectral radius (defined as the largest among the absolute values of its eigenvalues) is equal to $1$. This allows us to exploit known results in the theory of positive maps on $C^\ast$-algebras. In particular, using Theorem $2.5$ in Ref.~[\onlinecite{evans_spectral_1978}], it follows that $\mathcal{E}_{\mathcal{M}}(X)$ has a \emph{positive} fixed point~\cite{perez2006matrix}, namely there exists a Hermitian matrix $T$ with a non-negative spectrum such that $\mathcal{E}_{\mathcal{M}}(T)=T$. However, due to condition $C1$, it follows that $T\propto S$, since $\mathcal{E}_{\mathcal{M}}(X)$ has a unique fixed point. Hence, up to a proportionality factor, we just showed that for a solvable MPS, the matrix $S$ in condition $C2$ must be positive.

Suppose now that the matrix $S$ is invertible. Then $S$ must be strictly positive and we can make use of the following lemma to conclude that the MPS $\ket{\Psi^L_0(\mathcal{M})}$ is injective. 
\begin{Lemma}
	\label{th:strong_irr}
	\emph{The (normalized) MPS $\ket{\Psi^L_0(\mathcal{M})}$ is injective for $L$ sufficiently large if and only if the linear map $\mathcal{E}_\mathcal{M}(X)$ defined in Eq.~\eqref{eq:quantum_channel} satisfies the following two conditions}
	\begin{enumerate}
		\item \emph{ $\mathcal{E}_\mathcal{M}(X)$ has a unique maximum eigenvalue $\lambda_0$ with $|\lambda_0|=1$\,};
		\item \emph{The corresponding eigenvector $\Lambda$ is a strictly positive operator, namely it is Hermitian with strictly positive spectrum.}
	\end{enumerate}
\end{Lemma}
The proof of this Lemma is non-trivial, as it requires some technical tools in the theory of quantum channels. The basic idea underlying  the Lemma is to exploit the  fact that injectivity for sufficiently large $L$ implies that the channel $\mathcal{E}_\mathcal{M}(X)$ is \emph{primitive}~\cite{sanz2010quantum}. In turn, one can show that this property is equivalent to the above conditions $1$ and $2$. We omit further details of the proof here, and refer the reader to Ref.~[\onlinecite{sanz2010quantum}] where it is explicitly worked out, \emph{cf.} Prop.~$3$ therein. For a pedagogical introduction of the relevant quantum information tools, see instead Ref.~[\onlinecite{sanz_thesis}].

Conversely, suppose that the matrix $S$ is not invertible. Then, we can assume that there exists a basis of $\mathbb{C}^{\chi}$ in which the matrices $\mathcal{M}^{(i,j)}$ can be written in block diagonal form, so that without loss of generality we can assume
\be
\mathcal{M}^{(i,j)}=
\begin{pmatrix}
	\mathcal{M}^{(i,j)}_1 & 0\\
	0 & \mathcal{M}^{(i,j)}_2
\end{pmatrix}\,,
\label{eq:block_diagonal_form}
\ee
where $\{\mathcal{M}^{(i,j)}_{\alpha}\}_{i,j=1}^{d}$ are  $\chi_{\alpha}$-dimensional matrices with $\chi_{\alpha}<\chi$, $\alpha=1,2$. In order to see this, we proceed as follows.

Since $S$ is positive, we can write $S=\sum_{\alpha=1}^{\chi^\prime} \mu_\alpha \ket{\alpha}\bra{\alpha}$, where $\chi^\prime<\chi$ and $\mu_\alpha>0$. Following [\onlinecite{perez2006matrix}], and defining $P_\mathcal{R}$ to be the projector onto the space $\mathcal{R}$ spanned by $\ket{\alpha}$'s, we can prove that $\mathcal{M}^{(i,j)}P_\mathcal{R}=P_\mathcal{R}\mathcal{M}^{(i,j)}P_\mathcal{R}$, namely $\mathcal{M}^{(i,j)}\ket{\alpha}\in \mathcal{R}\,, \forall (i,j), \alpha$. Indeed, suppose that this is not true. Then, there exists $(m,n), \beta$ such that
\be
\sum_{\alpha} \mu_{\alpha}\ket{\alpha}\bra{\alpha}-\mu_{\beta} \mathcal{M}^{(m,n)}\ket{ \beta}\langle\beta|\left[\mathcal{M}^{(m,n)}\right]^\dagger \not\geq 0\,.
\ee
But since 
\be
\sum_{\alpha} \mu_{\alpha}|\alpha\rangle\langle\alpha|=\sum_{i,j}\sum_{\alpha}\mu_{\alpha} \mathcal{M}^{(i,j)}\ket{ \alpha}\langle\alpha|\left[\mathcal{M}^{(i,j)}\right]^\dagger\,,
\ee
we obtain
\be
\sum_{\substack{(i,j)\neq (m,n)\\ \alpha\neq \beta }}\mu_{\alpha} \mathcal{M}^{(i,j)}\ket{ \alpha}\langle\alpha|\left[\mathcal{M}^{(i,j)}\right]^\dagger \not\geq 0\,,
\ee
which is a contradiction. Thus, the matrices $\mathcal{M}^{(i,j)}$ must be block-triangular. However, it is now immediate to show that we can write the same state as an MPS $\ket{\Psi_0^L(\widetilde{\mathcal{M}})}$ with tensors $\widetilde{\mathcal{M}}^{(i,j)}$ that are block-diagonal and obtained from $\mathcal{M}^{(i,j)}$ by setting to zero  off-diagonal terms. Putting all together, we conclude that we can indeed assume Eq.~\eqref{eq:block_diagonal_form} without loss of generality, with $\chi_1$ being the rank of the matrix $S$.

We note now that Eq.~\eqref{eq:block_diagonal_form} implies
\be
\ket{\Psi^L_{0}(\mathcal{M})}=\ket{\Psi^L_{0}(\mathcal{M}_1)}+\ket{\Psi^L_{0}(\mathcal{M}_2)}\,,
\ee
namely $\ket{\Psi^L_{0}(\mathcal{M})}$ can be written as the sum of two distinct MPSs. Next, we observe that the norm of $\ket{\Psi^L_{0}(\mathcal{M}_2)}$ must be vanishing in the thermodynamic limit, since the map $\mathcal{E}_{\mathcal{M}_2}(X)$ has spectral radius $r<1$. Indeed, suppose that this is not the case and there exists $\tilde{S}$ such that $\mathcal{E}_{\mathcal{M}_2}(\tilde{S})=\tilde{\lambda}_0 \tilde{S}$ with $|\tilde{\lambda}_0|\geq 1$. Then, 
\be
S_1=
\begin{pmatrix}
	S & 0\\
	0 & 0
\end{pmatrix}\,,
\quad 
S_2=
\begin{pmatrix}
	0 & 0\\
	0 & \tilde{S}
\end{pmatrix}\,,
\ee
are both eigenstates for $\mathcal{E}_{\mathcal{M}}(X)$ with eigenvalue $|\lambda_0|,|\tilde{\lambda}_0|\geq 1$, which contradicts the hypothesis. In conclusion, one can find an injective MPS $\ket{\Psi^L_0(\mathcal{M}_1)}$, with bond dimension $\chi_1<\chi$, which is equivalent to $\ket{\Psi^L_0(\mathcal{M})}$ in the thermodynamic limit. 

Putting all together, up to equivalence in the thermodynamic limit, we can always assume that the fixed point $S$ of $\mathcal{E}_{\mathcal{M}}(X)$ is strictly positive, and define
\be
\mathcal{N}^{(i,j)}= S^{-1/2} \mathcal{M}^{(i,j)} S^{1/2}\,,
\ee
where we also have $\left[S^{1/2}\right]^{\dagger}=S^{1/2}$. It is now straightforward to verify that $\ket{\Psi^L_0(\mathcal{N})}$ is equivalent to $\ket{\Psi_0(\mathcal{M})}$ in the thermodynamic limit, is injective and satisfies Eq.~\eqref{eq:solvability_identity}. This proves the first part of the statement of Theorem~\ref{th:classification}.

Finally, let us prove the second part of Theorem~\ref{th:classification}, namely that if a state is equivalent to an injective MPS satisfying~\eqref{eq:solvability_identity}, then it is also equivalent to a solvable MPS satisfying conditions $C1$ and $C2$. Clearly, using Lemma~\ref{th:strong_irr}, we only need to prove that the algebraic multiplicity of the leading eigenvalue is equal to $1$. This can be done once again by invoking a general result in the theory of quantum channels~\cite{wolf2012quantum}: if $\mathcal{E}_\mathcal{M}(X)$ is a positive linear map with spectral radius $r=1$ satisfying Eq.~\eqref{eq:solvability_identity} (\emph{i.e.} it is \emph{unital}) and $\lambda_0$ is an eigenvalue with $|\lambda_0|=1$, then its algebraic and geometric multiplicities coincide, namely all blocks in the Jordan form corresponding to $\lambda_0$ are one-dimensional. Here we omit the proof of this statement, for which we refer the reader to Ref.~[\onlinecite{wolf2012quantum}] (see Proposition 6.2 therein). This concludes the proof of Theorem~\ref{th:classification}.~\qed

\bibliography{./bibliography}

\end{document}